\documentclass[conference]{IEEEtran} % NDSS 2020 and IEEE S&P
\pdfoutput=1

\usepackage{multirow}
\usepackage{pifont}
\usepackage{graphicx}
\usepackage{color}
\usepackage[table,dvipsnames]{xcolor}
\usepackage[breaklinks]{hyperref}
\usepackage{subcaption}
\usepackage{eurosym}
\usepackage{makecell}
\usepackage{multirow}
\usepackage{enumerate} 
\usepackage{booktabs}

\usepackage{nag} % complain when using outdated commands

\usepackage{stfloats} % fix double column float problem

\newcommand{\one}{\ding{202}}
\newcommand{\two}{\ding{203}}
\newcommand{\three}{\ding{204}}
\newcommand{\four}{\ding{205}}

\def\cmp{\texttt{\_\_cmp()}}
\def\nba{28~257}
\def\nbasuccess{22~949}
\def\nbbanner{1~426}
\def\nbsa{560}
\def\nbsarefusal{508}
\def\cmplocator{\texttt{\_\_cmpLocator}}
\def\preaction{\textit{Consent stored before choice}}
\def\nonrespect{\textit{Non-respect of choice}}
\def\nooption{\textit{No way to opt out}}
\def\preticked{\textit{Pre-selected choices}}

\def\scname{\textit{Cookinspect}}
\def\extname{\textit{Cookie Glasses}}
\newif\ifanon
\newif\ifextended
\extendedfalse

\newif\ifediting
\def\REMARK#1#2{        % 1 = reviewer (1, 2, ...); 2 = comment
  \ifediting
  \begin{center}
  \noindent\fbox{
  \begin{minipage}[b]{0.4\textwidth}       
  \textbf{[#1]: #2}
  \end{minipage}}
  \end{center}
  \fi}

\def\NB#1{\REMARK{Nataliia}{#1}}
\def\nc#1{\REMARK{Celestin}{#1}}

\def\SHORTEN{\vspace{-0.5cm}}
\def\SHORTENTAB{\vspace{-0.3cm}} % \SHORTEN is a bit violent for tables

\newcolumntype{C}[1]{>{\centering\arraybackslash}p{#1}}
\newcolumntype{R}[1]{>{\raggedleft\arraybackslash}p{#1}}
\newcolumntype{L}[1]{>{\raggedright\arraybackslash}p{#1}}

\def\BibTeX{{\rm B\kern-.05em{\sc i\kern-.025em b}\kern-.08em
    T\kern-.1667em\lower.7ex\hbox{E}\kern-.125emX}}

\begin{document}

%\title{Are Cookie Banners GDPR-Compliant? \\
\title{Do Cookie Banners Respect my Choice? \\
%\title{I can opt out... But do they care?\\
%  \LARGE \textbf{Measuring and evaluating violations in IAB Europe's Transparency and Consent Framework}}
  \LARGE {%Measuring GDPR and ePrivacy violations within the \\%in cookie banners of\\ 
  Measuring Legal Compliance of Banners from\\
  IAB Europe's Transparency and Consent Framework}
  }

\ifanon
\author{
Anonymous
}
\else
\author{\IEEEauthorblockN{C\'elestin Matte}
\IEEEauthorblockA{Universit\'e C\^ote d'Azur, Inria \\
France\\
celestin.matte@inria.fr}
\and
\IEEEauthorblockN{Nataliia Bielova}
\IEEEauthorblockA{Universit\'e C\^ote d'Azur, Inria\\
France\\
nataliia.bielova@inria.fr}
\and
\IEEEauthorblockN{Cristiana Santos}
%\IEEEauthorblockA{School of Law, University\\Toulouse 1-Capitole, SIRIUS Chair\\
%cristiana.santos@ut-capitole.fr}
\IEEEauthorblockA{%Independent Legal Researcher\\
Research Centre for Justice and Governance \\
School of Law, University of Minho\\
cristianasantos@protonmail.com}
}
\fi

% NDSS 2020
% \IEEEoverridecommandlockouts
% \makeatletter\def\@IEEEpubidpullup{6.5\baselineskip}\makeatother
% \IEEEpubid{\parbox{\columnwidth}{
%     Network and Distributed Systems Security (NDSS) Symposium 2020\\
%     23-26 February 2020, San Diego, CA, USA\\
%     ISBN 1-891562-61-4\\
%     https://dx.doi.org/10.14722/ndss.2020.23xxx\\
%     www.ndss-symposium.org
% }
% \hspace{\columnsep}\makebox[\columnwidth]{}}

\maketitle

%==========================================================

% Todo:
% - Fix country numbers

\begin{abstract}
As a result of the GDPR and the ePrivacy Directive, European users encounter cookie banners on almost every website. Many of such banners are implemented by Consent Management Providers (CMPs), who respect IAB Europe's Transparency and Consent Framework (TCF). Via cookie banners, CMPs collect and disseminate user consent to third parties.
In this work, we systematically study IAB Europe's TCF and analyze consent stored behind the user interface of TCF cookie banners. We analyze the GDPR and the ePrivacy Directive to identify potential legal violations in implementations of cookie banners based on the storage of consent and detect such suspected violations by crawling \nbbanner{} websites that contains TCF banners, found among \nba{} crawled European websites.
With two automatic and semi-automatic crawl campaigns, we detect suspected violations, and we find that:
141 websites register positive consent even if the user has not made their choice; 
236 websites nudge the users towards accepting consent by pre-selecting options; and
27 websites store a positive consent even if the user has explicitly opted out. 
Performing extensive tests on 560 websites, we find at least one suspected violation in 54\% of them.
Finally, we provide a browser extension to facilitate manual detection of suspected violations for regular users and Data Protection Authorities.
% 1 426
% 28 257
\end{abstract}
%\category{K.4}{Public Policy Issues}{Privacy}

%\terms{Security; Privacy; Web; GDPR; Consent;}

%\nc{Not sure if I should add keywords, it's not clear}

\begin{IEEEkeywords}
Privacy; GDPR; Consent; Web measurement
\end{IEEEkeywords}

\iffalse
\nc{Policy:
  \begin{itemize}
  \item We call the TCF ``The TCF'', not ``IAB Europe TCF'', ``TCF framework'' etc.
  \item We explain abbreviations (e.g. CMP) in the abstract, at the beginning of introduction + background, not later.
  \item ``IAB'', not ``the IAB'' (+ IAB Europe when necessary)
  \item CMP ID (in upper cases)
  \item We can use the term ``TCF banner'' (defined in the introduction)
  \item Gender-neutral pronouns: they/them
  \item We mention websites using \textbackslash{}texttt\{domain.tld\} (no \textbackslash{}url)
  \item ``Advertisers'' instead of ``vendors''
  \end{itemize}
}

\nc{Todo: check numbers everywhere}
\fi
%==========================================================
\section{Introduction}
\label{sec:intro}

\iffalse
\NB{Submitted papers may include up to 13 pages of text and up to 5 pages for references and appendices, 
totaling no more than 18 pages. }
\NB{Deadline: October 1st, at 3:00 PM (UTC-7, i.e., PDT). }
\fi

Today's web advertising ecosystem heavily relies on continuous data collection 
and tracking that allows advertising companies as well as data brokers to continuously 
profit from collecting a vast amount of data associated to the users. 
Adopted in April 2016 and implemented in May 2018, the General Data Protection Regulation (GDPR)~\cite{gdpr} changed the rules on consent, 
shaking the tracking and advertisement industry in its practices. 
The ePrivacy Directive, amended in 2009 
(ePD, also known as ``cookie law'') %\footnote{The upgrade of the ePD into a regulation is currently under discussion.}
~\cite{epd} 
made it mandatory to collect user's consent before any access or storage of non-mandatory data 
 (not strictly necessary for the service requested by the user).
%on their 
%%terminal equipment. 
%browsers. 
In case of websites, the %such 
consent is usually presented in the form of \emph{cookie banners}, 
or cookie notices that inform the user of data collection and should 
provide a meaningful choice on whether to accept or reject such collection. 
The website visitors in the European Union observe such banners on many  
websites they visit today. 
%However, the implementations of %such 
%consent required by the ePD 
%%data collection, its implementations 
%varied 
%across European countries, and sanctions that Data Protection Authorities (DPAs) could give to 
%companies were not high enough to be sufficiently dissuasive. On the contrary, the GDPR, as 
%a European regulation, was immediately enforceable in every European country, 
%setting uniformed requirements on the implementation of consent. 
%Moreover, under the GDPR, the possible sanctions increased up to a possible maximum of \euro 20 million or 4\% of the worldwide revenue of the company.

Various research studies looked into detection and measurement of web tracking technologies that 
perform silent data collection without user's explicit consent~\cite{Roes-etal-12-NSDI,Olej-etal-14-NDSS,englehardt2016online,Acar-etal-13-CCS,Niki-etal-13-SP,Lern-etal-16-USENIX,papa-etal-19-www}. 
Several recent works~\cite{libert2018changes,degeling2018we,trevisan2019years,sanchez2019can}  
have been measuring the impact of GDPR 
%The GDPR had an impact 
on the web tracking and advertising ecosystem.  
Libert et al.
%Researchers 
\cite{libert2018changes} 
%researchers 
observed a 22\% drop in the amount of third-party cookies before %(April 2018) 
and after 
%(July 2018) 
the GDPR, but only a 2\% drop in third-party content. %~\cite{libert2018changes}.
%Other works 
Degeling et al.~\cite{degeling2018we} recently measured the prevalence of cookie banners and showed that 
the amount of banners increased over time %~\cite{degeling2018we} 
after the GDPR.
Legal scholars, %and 
authorities and computer science researchers 
independently noticed that some banners do not allow users to refuse data collection, and raised this in various 
studies~\cite{degeling2018we,leenes2015taming,wp292015sweep,vallina2019tales}. 
%Researchers 
Several recent works~\cite{traverso2017benchmark,trevisan2019years,sanchez2019can} measured the impact of choices set in cookie banners on tracking: 
upon accepting and rejecting the consent proposed in a cookie banner, 
researchers evaluated the number of cookies set in the browser 
and the number of third-party tracking requests across websites.  
%Recent works also measured the prevalence of cookie banners and showed that 
%the amount of banners increased over time~\cite{degeling2018we} after the GDPR.
%Others 
Latest works~\cite{utz2019uninformed,nouwens2020dark} evaluated whether the design of cookie banners made an impact on how users 
would interact with them.  

%and usefulness of cookie banners that are supposed 
%to help Web users express their consent to tracking and data collection. 
%Some recently studied design of cookie banners and measured how users will interact with 
%different designs~\cite{utz2019uninformed}, others 
% shows that %Observation that 
%the number of cookie banners asking for user consent increased over time~\cite{degeling2018we}, 
%and the design of cookie banners impact how people will interact with it~\cite{utz2019uninformed}. 

Although many research efforts took place after the GDPR to detect and analyze cookie banners 
and their impact on tracking technologies and on the users, 
no study has analyzed
%Even though the GDPR has been in force for over a year and a half, no research study so far has 
%analyzed 
what actually happens behind the user interface of cookie banners yet. 
It is unclear how to meaningfully compare the interface of the banners 
shown to the users to the actual consent that banners store and 
transmit to the third parties present on the website. Our work is 
motivated by the following questions:

{\em
Do banners actually respect user's choice made in the user interface?
Do banners silently register a positive consent even if the user has not made their choice?
%Do banners comply with GDPR and ePD and systematically give the user an option to refuse? 
% the previous question was analyzed by the EURECOM team~\cite{sanchez2019can}. 
% So let's not mention it here.
Do they nudge the user to accept everything by pre-choosing a positive consent?}

Answering such questions, %responsibilities for 
%Ensuring
ensuring a proper functionality and legal compliance of a cookie banner 
is usually left to %be handled by 
the website publisher and 
is completely obscure for the website visitor. 

\begin{figure}[t]
  \center
  \includegraphics[width=0.45\textwidth]{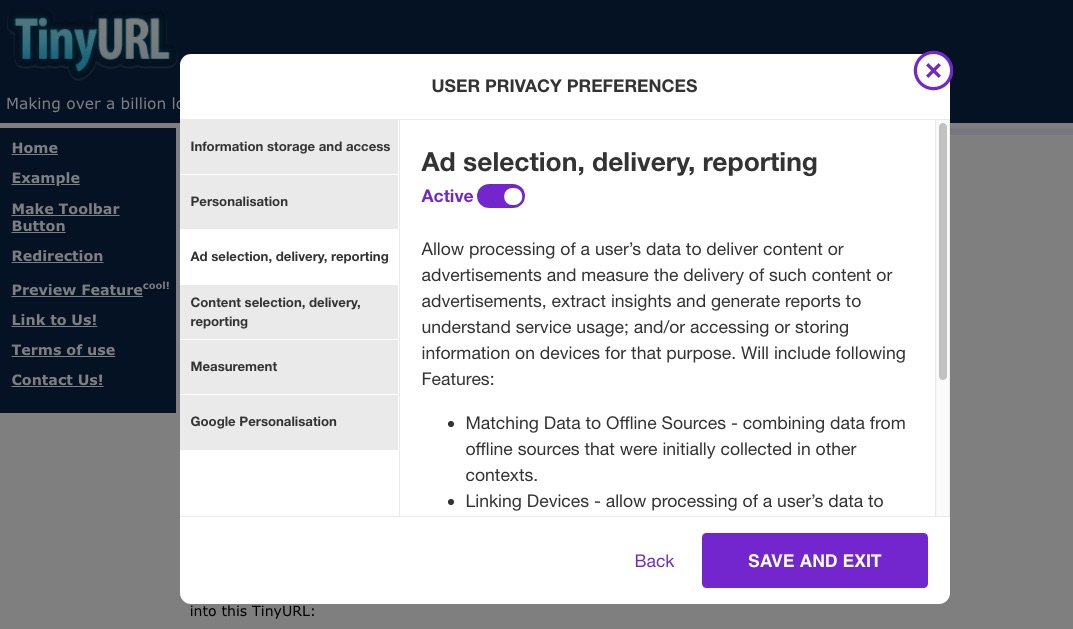}
  \caption{A cookie banner 
  %on \texttt{euronews.com}
  on \texttt{{tinyurl.com}} %that implements 
  of a Consent Management Provider %``Didomi'' 
  that implements 
  IAB Europe's Transparency \& Consent Framework (TCF).}
  \SHORTEN
  \label{fig:banner-example}
\end{figure}

In reaction to the GDPR, the European branch of the Interactive Advertising Bureau (IAB Europe), 
an advertising business organization, produced the Transparency and Consent Framework (TCF)~\cite{iab-tcf2018} 
to structure the practices of actors of the tracking and advertisement industry regarding consent collection. 
Notably, they introduced the notion of \emph{Consent Management Providers (CMPs)} -- actors in charge of 
collecting consent from the end-user, and redistributing this consent to advertisers. 
Figure~\ref{fig:banner-example} shows a typical example of a cookie banner implemented by a CMP 
%of IAB Europt TCF 
that %contains five purposes for data processing. 
allows the user to agree or disagree with five predefined purposes of data processing. 
%on a popular website
%\texttt{tinyurl.com}\footnote{The website \texttt{tinyurl.com} was accessed on 29 October 2019.}.
%All banners that implement an IAB Europe TCF are called ``TCF banners'' in our study.
%
%
%On a website where a CMP is present, end users see a cookie banner. We call such TCF-related banners ``TCF banner'' in this paper.
%\nc{Maybe an example here?}
 
\textbf{Contributions.} Thanks to the open specification of the TCF, we perform the first systematic comparison of the consent chosen by the users and the consent stored by the CMPs, which is further transmitted to third-party advertisers present on a website. 
With our analysis of consent, we are able to measure both the GDPR and the ePD compliance of cookie banners implemented in the TCF. 
We note that the responsibility for the suspected violations are joint between the publishers and the CMPs.
Our main contributions are:

\begin{enumerate}
\item We design an automatic method to detect the presence of a 
cookie banner developed by a
 Consent Management Provider (CMP) 
(Section~\ref{sec:methodo:detecting}). 
We automatically detect \nbbanner{} websites with such banners. 
\item We develop and use a methodology to intercept the 
consent stored in the browser %by the CMP 
(Section~\ref{sec:methodology:methods}).
By analyzing the content of consent, we bring transparency by %assigning the responsibility for each consent on 
revealing 
the companies behind CMPs and publishers. 
\item By collaborating with legal scholars (one of the co-authors and external ones), we thoroughly analyze the GDPR, the ePrivacy 
Directive and other legal texts to identify four potential legal violations specific to cookie 
banners:  \preaction, \nooption, \preticked\ and \nonrespect\ (Section~\ref{sec:violations}). 
\item We develop a method to evaluate regulatory compliance of websites 
%with TCF banners 
(Section~\ref{sec:methodology:violations}).  
We quantify the identified suspected violations on \nbbanner\ websites %that contain 
%TCF banners 
by automatic-, semi-automatic crawls and manual detection (Section~\ref{sec:results:violations}). 
By analyzing cookie banners' design on a subset of \nbsa\ websites (from countries 
whose language the authors speak), we find  
that 236 (47\%) websites nudge the users towards acceptance by 
pre-selecting options, while 38 (7\%) websites do not provide any means to refuse consent. 
By analyzing the consent stored in the browser, we automatically detect 
141 out of \nbbanner\ (10\%)  websites  that store a positive consent before user has made any choice in the cookie banner, 
while 27 out of \nbsa\ (5\%) websites  store an all -- accepting consent even if the user has explicitly 
opted out in the cookie banner interface.
In total, we find at least one suspected violation in 304 out of 560 websites (54\%).
We discuss the difficulty to attribute responsibility of these suspected violations in Section~\ref{sec:discussion}.
\item We measure the problem of escalation of shared consent between CMPs (Section~\ref{sec:results:shared_cookie}). 
The TCF allows different CMPs and publishers to rely on each other's consent, 
set in a shared cookie. 
We observe that 3 websites store a positive consent before user action 
in the shared cookie, while 20 websites store a positive consent in a shared cookie  
even if the user has explicitly opted out. 
Such invalid consent can be reused by any CMP and publisher 
and therefore  escalates  %non-compliant consent. 
non-compliance to other websites. 
%According to our estimations, 62 websites 
%
\item We quantify  third-party requests that transmit consent % of websites 
and 
that belong to known third-party tracking services (Section~\ref{sec:additional}). 
We observe that various third parties receive consent with third-party requests, 
where the origin of consent %is unknown. 
does not necessarily match the CMP present on the website. 
%Moreover, such originless consents are also 
Such consents are set 
before user action on 69 websites and despite user refusal on 38 websites.
We observe that the number of third-party tracking requests increases both 
after positive consent and after refusal.
 \end{enumerate}

To measure compliance, we have designed two tools. 
%We first design two tools.  
\scname{}~\cite{Cookinspect} is a Selenium- and Chromium-based crawler which automatically and semi-automatically
visits websites, %gathers information related to the transmission of consent %-passing 
logs stored consent and intercepts transmission of consent to third parties. 
%in the TCF,
%and 
\extname{}~\cite{Cookie-Glasses} is a publicly available 
browser extension for Google Chrome and Firefox 
that allows users to detect a CMP that implements a TCF banner and 
see if their choice is correctly transmitted to advertisers by CMPs.
%check whether the banner 
%stores a consent that correspond to the user's preferences. 

%\textbf{Outline.}
%%\b{Todo: update}
%\CM{\b{Todo: update}}
%Section~\ref{sec:background} gives some background information about the advertisement and 
%tracking industry, and presents IAB Europe's Transparency and Consent Framework (TCF). 
%Section~\ref{sec:methodology} presents our methodology and Section~\ref{sec:results} 
%shows results in detecting GDPR and TCF violations by TCF banners and advertisers. 
%%our preliminary results. 
%Section~\ref{sec:ethics} discusses some ethical considerations about this work. 
%Section~\ref{sec:related_work} lists related work.
%Section~\ref{sec:limitations} discusses some limitations. 
%Section~\ref{sec:future_work} list future plans for the work and concludes the paper.

%==========================================================

%\section{IAB TCF, GDPR requirements on valid consent and possible violations}
%\section{Background on IAB Europe TCF and EU Data Protection regulations}
\section{IAB Europe's Transparency and Consent Framework (TCF)}
\label{sec:background}

%In this section, we provide a technical description of IAB Europe's Transparency and Consent Framework (TCF).
%We then give an introduction to two EU data protection laws: 
%ePrivacy directive (ePD) and GDPR that pose legal %and derive 
%requirements for valid consent. 
%We then list possible violations of the GDPR and the ePD in cookie banners 
%that we further detect in this work.

%========================================================================

%\subsection{Technical background on IAB Europe's Transparency \& Consent Framework}
%\subsection{IAB Europe's Transparency \& Consent Framework (TCF)}

The third-party advertising and tracking ecosystem contains different actors. \emph{Publishers} provide websites to users and include third-party advertising content. \emph{Advertisers} and trackers collect users' data and display ads. Finally, \emph{users} consume content. % (possibly free). 
With the arrival of the GDPR, it became evident that the different actors of this ecosystem were not equipped 
to properly collect and exchange user's consent. % to tracking.
%missed entities responsible for gathering user's consent and providing third parties and advertisers access to consent provided by the user.

\begin{figure}[t]
  \center
  \includegraphics[width=0.35\textwidth]{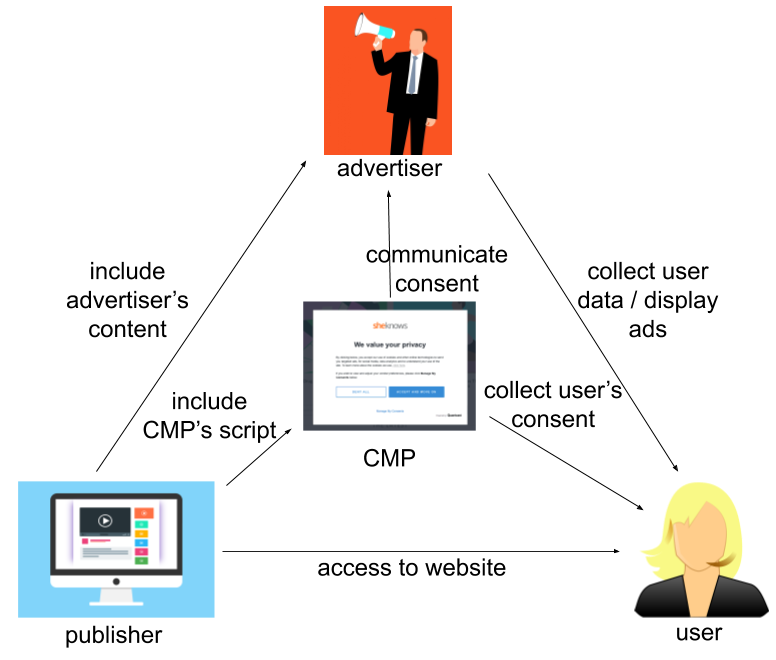}
  \caption{Consent Management Providers (CMPs) under IAB Europe's TCF.}
  \SHORTEN
  \label{fig:ecosystem}
\end{figure}

%In April 2018 (one month before GDPR), 
%One month before GDPR came in force, in 
In April 2018, 
IAB (Interactive Advertising Bureau) Europe
%the European division of the Interactive Advertising Bureau (IAB) 
published the %TCF 
Transparency \& Consent Framework (TCF) -- a technical specification that allows third-parties 
and publishers to collect and exchange user's consent to data collection and the use of cookies\footnote{In this work, we study version 1.1 of the TCF. 
Even though IAB Europe published TCF version 2 on August 21\textsuperscript{st} 2019, we have not %. Because that version was released one day before the %crawling period, we did not see 
observed its application in the wild, and therefore did not address it in this work.}.
The TCF was presented as a way to help digital advertising industry to ``interpret and comply with EU rules on data protection and privacy -- notably the GDPR''~\cite{iab-tcf2018}.
%
%\begin{figure}[t]
%  \center
%  \includegraphics[width=0.35\textwidth]{figures/business_model}
%  \caption{Consent Management Providers (CMPs) under IAB Europe's TCF.}
%  \SHORTEN
%  \label{fig:ecosystem}
%\end{figure}
%
%In the TCF, 
IAB Europe introduced new actors, % to manage users' consent 
called 
\emph{Consent Management Providers (CMPs)}, 
who 
%CMPs 
are responsible for collecting the end user's consent, 
storing it in the user's browser and implementing methods to respond 
to advertisers' queries regarding this consent. 
%
%\begin{figure}[t]
%  \center
%  \includegraphics[width=0.45\textwidth]{figures/tinyurl-banner}
%  \caption{A cookie banner on \texttt{{tinyurl.com}} that implements 
%  IAB Europe's Transparency \& Consent Framework (TCF).}
%  \label{fig:tinyurl-banner}
%\end{figure}
%
%Figure~\ref{fig:tinyurl-banner} shows an example of a cookie banner on a popular website 
%\texttt{tinyurl.com} that is implemented by a CMP
%under IAB Europe TCF\footnote{The website \texttt{tinyurl.com} was accessed on 29 October 2019.}. 

Figure~\ref{fig:ecosystem} summarizes 
the updated interaction of CMPs with publishers, users and advertisers.
%
%\begin{figure}[t]
%  \center
%  \includegraphics[width=0.4\textwidth]{figures/business_model}
%  \caption{Consent Management Providers (CMPs) under IAB Europe TCF.}
%  \SHORTEN
%  \label{fig:ecosystem}
%\end{figure}
%
%	\subsubsection{Registration process of CMPs and advertisers in the TCF}
%\label{sec:TCFregistration}
%
To become a part of the TCF, each CMP and advertiser must register with IAB Europe.
%
%{\bf Consent Management Platforms (CMPs)}
%To become a part of the TCF, each %Consent Management Platform (CMP) 
%CMP must register at the IAB Europe and 
%pay a registration fee of \euro 1200 to participate in the framework\footnote{\url{https://register.consensu.org/CMP}, 
%accessed on September 5th, 2019.}. 
As a result, IAB Europe maintains: (1) a \emph{public list of CMPs}~\cite{iab-cmplist} that participate in the framework, 
%(alongside identifiers, called CMP IDs), 
and (2) a \emph{Global Vendor List} (GVL)~\cite{iab-gvl} -- a public list of registered advertisers (called ``vendors'').
%\begin{enumerate}[(i)]
%\item a \emph{public list of CMPs}~\cite{iab-cmplist}, who participate in the frameworks alongside their unique identifiers to be used in the framework, and 
%\item a  \emph{Global Vendor List} (GVL)~\cite{iab-gvl} -- a public list of registered advertisers (called ``vendors'').
%\end{enumerate}
%
As of October 25\textsuperscript{th} 2019, there are 117 CMPs in the CMP list % contains 117 CMPs and the GVL
and 550 advertisers in the GVL list.
%
%{\bf Advertisers}
%
%To participate in TCF, each advertiser, called ``vendor'' in the TCF, must register at the IAB Europe (and pay a fee of 
%\euro 1200\footnote{\url{https://register.consensu.org/}, accessed on September 5th, 2019.}), called 
%the  \emph{Global Vendor List} (GVL)\footnote{Figure~\ref{fig:GVL-registration} of the Appendix 
%shows a screenshot of an IAB Europe vendor registration process.}. 
%The GVL  is a public list~\cite{iab-gvl}, where each registered advertiser 
%(called \emph{vendor}) is present. 
%
%
While registering in the GVL, among other information, 
each advertiser must declare one or more of 
the \emph{five pre-defined purposes} for which the data is collected 
and for which of them consent will be used -- 
these are the purposes the user usually sees in the interface of the cookie banner 
(see Figure~\ref{fig:banner-example}). 
Table~\ref{tab:purposes} in Appendix~\ref{app:purposes} shows the full list and description of all the
predefined purposes.

%\subsubsection{Consent format and storage}
%The TCF proposes a specification for expressing and storing user's consent and methods that allow the consent to be shared across different actors.

\subsection{Consent String}
The TCF defines a standard format for consent~\cite{iab-tcf2018}, called \emph{consent string}.
This string contains 
(1) advertisers for whom the user consented their data to be sent to, 
(2) purposes of data processing the user consented to, and 
(3) the CMP identifier, along with other information.
This format is a slightly-modified version of base64 of an array of values. %replacing some characters and adding the 
We use a script provided by IAB~\cite{iab-cssdk} to decode this format. % that we use in our work.

%For example, Figure~\ref{fig:consent-string} shows a decoding of 
%the consent string ``\texttt{BOX5uluOX5uluCLAAAENB6-AAAAizAAA}'', obtained on \texttt{telerama.fr}.
%, is decoded as follows (trimmed from unnecessary information):
%\NB{We need an example wit only one purpose: purpose 4, and few vendors, including vendor 25. Can we have it? Then it's easy to explain the problem of verification of consent that I refer to in the end of this section}

\begin{figure*}[t]
  \center
  \includegraphics[width=0.89\textwidth]{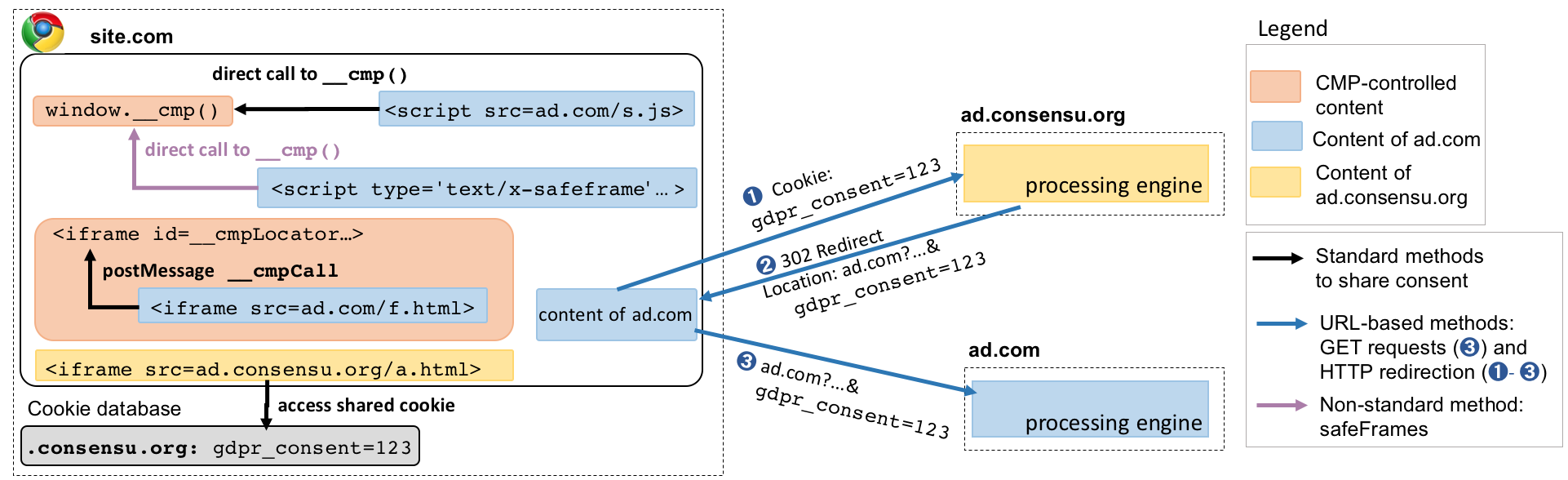}
  \caption{The methods to share consent in IAB Europe's TCF.}
  \vspace{-0.7cm}
  \label{fig:consent-sharing}
\end{figure*}

\begin{figure}[t]
%\fbox{
  %\begin{minipage}[b]{0.4\textwidth}
    \{"\texttt{created}": "2018-11-27 17:24:14",\\
    %\hspace*{1mm} "\texttt{vendorListVersion}": 122,\\
    \hspace*{1mm} "\texttt{cmpId}": 139,\\
    \hspace*{1mm} "\texttt{allowedPurposeIds}": [1,2,3,4,5],\\
    \hspace*{1mm} "\texttt{allowedVendorIds}": [1,2,3 ... ,554,555,556], ...\}
  %\end{minipage}
%}
\caption{Example of a decoding of a consent string (only fields relevant for this paper are shown).}
\SHORTEN
\label{fig:consent-string}
\end{figure}

Figure~\ref{fig:consent-string} shows a decoding of 
the consent string ``\texttt{BOX5uluOX5uluCLAAAENB6-AAAAizAAA}'', obtained on \texttt{telerama.fr}.
%We analyze the following parameters of the consent strings in our study. 
The \texttt{cmpId} is an identifier of a CMP from the CMP list~\cite{iab-cmplist} 
responsible for storing a consent string; 
\texttt{allowedPurposeIds} are the identifiers of the five pre-defined TCF purposes 
of data processing; %(normally, these are 
%the purposes the user has agreed upon while setting their choices in a cookie banner). 
%Table~\ref{tab:purposes} in Appendix lists all the five pre-defined 
%purposes for data processing that are specified by the TCF.
and 
\texttt{allowedVendorIds} are the identifiers of advertisers %(called ``vendors'') 
from the  %Global Vendor List~\cite{iab-gvl}. % that are allowed by the consent. 
GVL~\cite{iab-gvl}. % that are allowed by the consent. 
%
%Note that by 
%By analyzing a consent string, any %advertiser (or more generally, any third party) 
%third party can %also 
%%extract information about 
%identify 
%the CMP that registered consent by matching the \texttt{cmpId} field of the 
%consent string against the %publicly available list of CMP Ids and CMP entities in the TCF~\cite{iab-cmplist}). 
%public list of CMPs~\cite{iab-cmplist}.
Note that any %advertiser (or more generally, any third party) 
third party can %also 
%extract information about 
identify 
the CMP that registered the consent string by comparing the \texttt{cmpId} field of the 
consent string to the %publicly available list of CMP Ids and CMP entities in the TCF~\cite{iab-cmplist}). 
public list of CMPs~\cite{iab-cmplist}.
%By comparing the \texttt{cmpId} field of the 
%consent string to the public list of CMPs~\cite{iab-cmplist}, any 
%third party can identify the CMP that registered consent.

%
%\begin{figure*}[t]
%  \center
%  \includegraphics[width=0.99\textwidth]{figures/consent_sharing}
%  \caption{The methods to share consent in IAB Europe TCF.}
%  \SHORTEN
%  \label{fig:consent-sharing}
%\end{figure*}

\subsection{Consent Storage} 
\label{sec:background:consentstorage}
The TCF does not impose any particular mechanism for storing user's consent in the browser. It only suggests that CMPs use a ``first-party service-specific cookie'', without further details~\cite{iab-jsapi}.

%However, as 
As one way to implement consent storage, the TCF proposes CMPs to 
 store a consent string in a cookie, named \texttt{euconsent} in the \texttt{consensu.org} domain owned by IAB, reachable by CMPs through DNS subdomains delegation 
%to store a consent string in a cookie, named ``euconsent'', having \texttt{.consensu.org} as its origin 
%(we call it ``shared cookie'' in the rest of the paper)\footnote{The website-specific cookie has priority over the shared cookie, when both are defined (i.e., the website-specific cookie will be used).}.
(we call it ``shared cookie'' in the rest of the paper)\footnote{TCF mentions that when website-specific cookie 
and shared cookies are both defined, the website-specific cookie will be used.}. This mechanism allows CMPs to share consent information despite the Same-Origin-Policy.
%
%Therefore, any script executed in an iframe from a subdomain of \texttt{consensu.org} origin can access and modify %this 
%the shared ``euconsent'' cookie.
%
Since each CMP registered in the TCF has access to its own subdomain (e.g. %\texttt{example.consensu.org}), so 
\texttt{ad.consensu.org}), it can host scripts in it to read and modify the shared cookie. 

%Notice that even though CMPs are responsible for implementing the consent storage, publishers that host them can also customize the implementation of consent storage. 

\subsection{Consent Sharing}
\label{sec:background:consentsharing}

Once consent is stored in the user's browser, any advertiser (or, more generally, any third party) present on a page
can query a CMP to obtain the consent that was given by the user. In the TCF, consent sharing can be done via: 
%\begin{enumerate}
%\item 
%(1) standard methods for first-party scripts and third-party iframes,
%%: a direct call to a function called \cmp{}, or sending a postMessage to an iframe called \cmplocator{},
%%\item 
%(2) a shared cookie set on  \texttt{.consensu.org}, % domain,
%%\item 
%(3) URL-based methods: GET queries and redirection mechanism, 
%%\item 
%(4) a non-standard method: safeFrames. %or OpenRTB extension fields (excluded from this study).
%%\end{enumerate}
standard APIs,
%: a direct call to a function called \cmp{}, or sending a postMessage to an iframe called \cmplocator{},
%\item 
 a shared cookie, 
%\item 
 URL-based methods, 
%\item 
and a non-standard method (safeFrames). %or OpenRTB extension fields (excluded from this study).
%\end{enumerate}
We present each consent sharing mechanism in more details below. 
Figure~\ref{fig:consent-sharing} graphically presents how each method obtains a consent string.

%\begin{figure*}[htpb]
%  \center
%  \includegraphics[width=0.99\textwidth]{figures/consent_sharing}
%  \caption{The methods to share consent in IAB Europe TCF.}
%  \SHORTEN
%  \label{fig:consent-sharing}
%\end{figure*}

%\NB{Client-side pre-bid requests are also observed in step (5). Step (5) covers redirection mechanism, direct URL-based communication and  pre-bid of OpenRTB.}

%There are several ways a third party present on a webpage can use to verify whether 
%the consent is stored by the CMP.  

{\bf Standard APIs.}
The TCF v1.1~\cite{iab-tcf2018} specifies APIs that each CMP must implement -- such APIs allow any 
third-party advertiser present on a publisher website to 
verify whether a CMP has already stored a consent on a given website.
%According to the specification, \emph{these APIs should not return the consent string before the user gives their decision on consent} 
%(or consent is retrieved from existing cookies)~\cite{iab-jsapi}.
%
In particular, each CMP %in the framework 
must implement:
%\begin{itemize}
\begin{enumerate}[(i)]
\item a javascript function called ``\cmp{}'', that scripts in a first-party position can call directly, 
% to communicate with the CMP,
\item an iframe named ``\cmplocator{}'', that iframes in a third-party position can communicate with using the \texttt{postMessage} API using a \texttt{\_\_cmpCall} parameter.
%\end{itemize}
\end{enumerate}

%
%In a preliminary test on 21~000 websites, we looked for both elements on websites. We found that CMPs using at least one of these two elements define the \cmp{} function in 100.00\% of cases (only one exception), and the \cmplocator{} in 79.24 \% of cases. We can then consider that the \cmp{} function is always defined on TCF websites.
%In a preliminary test on 21~000 websites, we analyzed how often \cmp{} and \cmplocator{} are used. 
%Among CMPs that implement the standard APIs, we have observed that 100\% of them 
%have a \cmp{} function defined and 
% %in 100.00\% of cases (only one exception), 
% % NATALIIA: 100% OR 1 EXCEPTION?
% 79.24\% of CMPs  support  \cmplocator{}. 
% Therefore we can safely use a hypothesis that the \cmp{} function is defined, 
% then a CMP is present on a website. 
% In our large scale experiments, we rely on a presence of a \cmp{} function 
%to conclude that a CMP is present on a website (see \red{Section~\ref{sec:INSERT}}). 

{\bf Shared cookie.}
%First, if 
As explained above, CMPs can store consent in a shared cookie named \texttt{euconsent} 
on the \texttt{.consensu.org} domain, 
that any other CMP can read. 

{\bf URL-based methods.}
%Additionally to the standard methods to access the consent string,
The TCF specifies~\cite{iab-urlbased} other methods to send consent to third parties devoid of the possibility 
to execute JavaScript, such as tracking pixels. 
First, consent can be passed:
%\begin{itemize}
%\item 
through GET requests, in a predefined ``\texttt{gdpr\_consent}'' parameter. 
Second, 
%\item 
using an HTTP redirecting mechanism that we show in steps \one-\three\ in Figure~\ref{fig:consent-sharing}. 
%third-party scripts willing to read to shared cookie 
Third parties can issue a %query 
request to a subdomain of \texttt{.consensu.org}, such as \texttt{ad.consensu.org}, %giving a URL as a parameter. 
providing a final destination URL as a parameter of the request, such as \texttt{ad.com}. 
Because browsers automatically %insert %the shared 
attach the cookie of \texttt{.consensu.org} to the request, 
the CMP owning \texttt{ad.consensu.org} obtains a shared cookie (step \one{}). 
The CMP then responds with the ``302 Redirect'' status, indicating 
 \texttt{ad.com} as a destination URL and attaching the cookie value in the  
parameters (step \two{}). Finally, \texttt{ad.com} receives the shared cookie (step \three{}).  
%of the 
%CMPs can implement redirection %scripts hosted on their 
%mechanism on \texttt{.consensu.org} subdomain, 
%which will redirect the original %query 
%request to the provided URL, adding the shared cookie to this %query.
%URL. 
%\end{itemize}

{\bf Non-standard method: SafeFrames}
Finally, the TCF proposes an additional non-standard method to share consent: {\em safeFrames}.
A SafeFrame~\cite{safeFrames} is an API-enabled iframe (implemented via specific first-party scripts) that controls the communication between the webpage content and third-party ads. The SafeFrame proxy obtains consent by calling to the standard \cmp{} function. 
%Because safeFrames is a very specific implementation of iframes and anyway reuses the calls to \cmp{} function, we have excluded it form our study.
%we already detect consent strings set via safeFrames by 
%analyzing the \cmp{} function.

%==========================================================
%\input{questions}
%==========================================================
%\section{GDPR and ePrivacy violations for cookie banners}
%\section{GDPR and ePrivacy Requirements for Cookie Banners}
\section{GDPR and ePrivacy Suspected Violations}
\label{sec:violations}
\label{sec:background:legal}

In our work, we focus on the European regulatory framework on data protection and privacy. 
This section provides a legal analysis of the suspected violations %foreseen 
\ifextended
both at European and at national levels
\fi
, and the limitations that this analysis entails. 
\subsection{Legal Background}
\label{sec:legal}

%While a directive (in our case, the ePrivacy Directive ``ePD’’, also known as ``cookie law’’)~\cite{ePD-09}) is left up to each member State to implement in its own national law, a regulation -- such as the GDPR~\cite{gdpr} -- is directly enforceable in every European country.
In May of 2018, the GDPR 
enforced the rules regarding the processing of personal data in any environment~\cite{EDPB-4-07}. In order to lawfully process personal data, companies need to choose a \emph{legal basis of processing}~\cite[Article 6(1)(a)]{gdpr}. One of the most well-known one is \emph{consent}. Articles 4(11) and 7 of the GDPR have set 
precise requirements on \emph{valid consent}: 
it must be freely given, specific, informed, unambiguous, 
explicit, revocable, given prior to any data collection, 
and requested in a readable and accessible manner~\cite{EDPB-4-18}. 

The ePrivacy Directive (``ePD’’, also known as ``cookie law’’)\footnote{The upgrade of the ePD into a regulation is currently under discussion.}~\cite{ePD-09} provides \emph{supplementary} 
rules to the GDPR with respect to the processing of personal data in the 
electronic communication sector, such as websites. 
While GDPR is a \emph{regulation}, and hence is  directly enforceable in every European country, 
ePD is a \emph{directive}, and hence  is left up to each member State to implement in its own national law.

%a directive is left up to each member State to implement in its own national law, a regulation -- such as the GDPR~\cite{gdpr} -- is directly enforceable in every European country.

%\footnote{{\em``In situations where the ePD “particularises” (i.e. renders more specific) the rules of the GDPR, these (specific) provisions of the ePD shall, as “lexspecialis”, take precedence over the (more general) provisions of the GDPR''}, EDPB, Opinion 5/2019~\cite{EDPB-5-19}.}. 
%
According to %this specification of 
the ePD~\cite[Article 5(3)]{ePD-09}, and pursuant to the guidance of the European Data Protection Board (EDPB) and 
Information Commissioner's Office (ICO, the UK's %DPA),  
Data Protection Authority),
website publishers have to %use only one legal basis of processing -- user consent – 
rely on \emph{user consent} when they collect and process personal data using non-mandatory 
(non strictly necessary for the service requested by the user)
cookies or other tracking technologies~\cite{EDPB-2-10,ICO-cookie-19}. 

An alternative legal basis to process personal data is the \emph{legitimate interest} pursued by the controller or by a third party (Article 6(1)(f) GDPR), e.g. freedom to conduct a business. Nevertheless, the EDPB stated that legitimate interests’ grounding is not considered to be an appropriate lawful basis for the processing of personal data in connection with purposes as tracking, profiling and advertising \cite{EDPB-3-13},\cite{EDPB-6-14}.

The following legal analysis is based in the most authoritative legal documents in this 
specific domain of privacy and data protection law. In particular, we reproduce the arguments 
already made public by the recent case decisions of the Court of Justice of the EU (CJEU, the highest court in the EU), 
by the DPAs and the EDPB guidelines, and the legal rules laid down in explicit legal provisions 
of the GDPR and the ePrivacy Directive, as well as in its recitals (recitals help legal interpretation 
of provisions in a specific context, but they are not mandatory for compliance). 
These cited expert generated legal sources are the foundational framework for data protection 
which we apply in this work to discern whether the declared practices are legally compliant.

\subsection{Legal analysis of potential violations}

As a result of a deep legal analysis by %our 
a legal scholar (a co-author in this paper), %of the relevant legal sources in this domain, 
we identify \emph{four potentially legal violations} 
specific to cookie banners that implement the IAB Europe TCF framework.

\textbf{\preaction{}}:
\emph{The CMP stores a positive consent before the user has made their choice in the banner.
Therefore, when advertisers request for consent, the CMP responds 
with the consent string even though the user has not clicked on a banner and has not made their choice.}

This practice %exposes noncompliance towards 
violates the requirement of \emph{prior consent}
which demands that  
website publishers need to request consent to users
(1) prior to any processing activity of personal data~\cite{EDPB-4-18}, and 
(2) before loading %cookie related 
tracking technologies %~\cite[Art. 5(3)]{ePD-09}.
according to Article 5(3) of the ePD~\cite{ePD-09}.
%It also violates the requirement of \emph{unambiguous consent} 
%because such practice assumes the user consent without any affirmative action of the user. 
%pursuant to Article 5(3) of the ePD. 
This requirement is further imposed in the guidance from the EDPB~\cite{EDPB-2-13}, 
the ICO~\cite{ICO-cookie-19}  and 
the CNIL (French Data Protection Authority)~\cite{CNIL-deliberation-2019}.
Moreover, the TCF's policy document explicitly states that ``a CMP will generate consent signals only on the basis of a clear affirmative action taken by a user''~\cite{iab-policy}.

\textbf{\nooption}: 
\emph{The banner does not offer a way to %give a negative 
refuse consent. The most common case is a banner simply informing the users about the site's use 
of cookies}%(Figure~\ref{fig:nooption})
\footnote{In previous works, this category has been named as "No option" in Degeling et al.~\cite{degeling2018we} and as "OnlyAccept" in Sanchez-Rola et al.\cite{sanchez2019can}}.  

This practice configures a violation of the requirement of \emph{unambiguous consent}~\cite[Art.4(11)]{gdpr}
% which is required by Article 4(11) of the GDPR 
 stipulating that in order for the user consent to be valid, 
 the user must give an “unambiguous indication” through a “clear and affirmative action” of his choice
%acceptance or refusal of cookies
~\cite[Art. 5(3)]{ePD-09}. Moreover, 
Recital 66 of the ePD is quite explicit while directing
that \emph{“the methods of offering the right to refuse should be as user-friendly as possible”}. 
In its guidelines, the {EDPB~\cite{EDPB-2-13}}
states that \emph{``consent mechanism should present the user with a real and meaningful choice
regarding cookies on the entry page''}, and that users 
\emph{``should have an opportunity to freely choose between the option to accept some or all cookies or to decline all or some cookies.''}
In this line, we posit that cookie banner design must offer users an option to 
%customize consent 
%towards accepting and rejecting the use of cookies.
either accept or refuse consent.

\textbf{\preticked}:
\emph{The banner gives users a choice between one or more purposes or vendors, 
however some of the purposes or vendors are pre-selected: pre-ticked boxes 
or sliders set to ``accept''}. %(Figure~\ref{fig:preticked}). 

Preselected choices consist in a direct violation of the requirement of 
\emph{unambiguous consent}~\cite[Article 4(11)]{gdpr}. % set forth by Article 4(11) of the GDPR, as mentioned above. 
Recital 32 of GDPR reads further that consent given in
the form of a preselected tick in a checkbox does not imply active behaviour of the user and
 that pre-ticked boxes  do not constitute consent. 
The {EDPB~\cite{EDPB-4-18}} indicates that pre-ticked boxes (or opt-out boxes) configure
ambiguous behaviours and do not render a valid consent. 
The {ICO guidance~\cite{ICO-cookie-19}} 
and the CNIL~\cite{CNIL-deliberation-2019}
observe that \emph{``pre-ticked
boxes or any equivalents, cannot be used for non-essential cookies''}. 
Finally,  {%recent 
the Court of Justice of the EU (CJEU) judgment %(the highest court in the EU) 
from October 2019}~\cite{Planet49} (also known as the “Planet49 GmbH” case),
establishes definitely that the consent which a website user must give 
%to the storage of and access to
%cookies on his equipment 
is not %validly constituted by way of a 
valid if it contains a 
pre-checked checkbox which the user must
deselect to refuse their consent. 

\textbf{\nonrespect{}}: 
\emph{The CMP stores a positive consent in the browser even though the user explicitly refused consent.}

This practice incurs in violation of the \emph{lawfulness principle} established in Articles 5(1)(a) and 6(1) of the
GDPR: %, which instructs that, 
for the processing to be lawful, it must be based on a legal ground, namely, the
user consent to the use of cookies (legal ground demanded by Article 5(3) of the ePD). The EDPB~\cite{EDPB-15-11}
 further specified that 
 \emph{``if the individual decided against consenting, any data processing that had already taken
place would be unlawful''} due to lacking legal basis for processing.

%\end{itemize}

%\begin{figure}[t]
%  \center
%  \includegraphics[width=0.45\textwidth]{figures/violations}
%  \caption{Representation of some of the third-parties-related violations (red color), plus a normal behaviour presented in section~\ref{sec:normal_behaviour}.}
%  \label{fig:venn_violations}
%\end{figure}

%\subsection{Violations at the National Level}
\ifextended
\subsection{Implementations at the National Level}

Regarding the differences among Member States' implemented national laws of the ePD, some DPAs (e.g. the Spanish and Italian ones) express in their guidance %~\cite{SpanishDPAGuidance,ItalianDPAGuidance} 
that the act of continuing to browse equates to consent, and therefore some of the websites implement this practice. 
However, since the GDPR has overridden the definition of consent, requiring among others an active behavior of the data subject, and considering that this definition has a direct effect in the legal order of each Member State, any different national implementation of the ePD regarding consent does not impact our work.
\fi

\subsection{Limitations of the Legal Analysis}

%Even though we haven't been peer-reviewed in the legal community, 
 Even if our analysis is endorsed in legislation, judicial decisions and expert-generated legal sources, this analysis is yet limited if not sustained judicially. Therefore we deliberately leave space to legal uncertainty on the assessment of the identified questionable practices and emphasize that only a judicial assessment that requires more specific fact finding of each practice could render a final appraisal of such analysis and provide legal certainty.
%==========================================================
\section{Methodology}
\label{sec:methodology}

The goal of our study is to detect the suspected violations of the GDPR and the ePrivacy Directive in cookie banners 
that implement IAB Europe's Transparency \& Consent Framework (TCF). 
We detect all suspected violations defined in Section~\ref{sec:violations} on websites that originate 
from the European Union because a TCF banner is more likely to be observed on EU websites. 
%We also test some other European and international TLDs.

In this section, we  describe the website selection and data collection processes, followed by our 
methods to detect TCF banners and  intercept the user consent string. 
We then explain how we detect suspected GDPR and ePD 
violations with our methodology. 
To detect suspected violations at scale, we built a crawler, called \scname{}~\cite{Cookinspect}, 
based on a Selenium-instrumented Chromium, 
that we use to perform large-scale automatic crawls and semi-automatic crawls (with 
certain human actions) to audit websites% and detect violations 
at scale. 
We describe the two measurement campaigns done with \scname\ in Section~\ref{sec:campaigns}. 

\subsection{Websites Selection and Reachability}
\label{sec:methodology:websites}

We used Tranco to build lists~\cite{le2019tranco}. %This work 
Tranco aggregates results from different lists over a month in order to overcome flaws inherent to these lists' creation: instability, presence of unreachable domains, possible attacks to insert domains, etc\footnote{See Appendix~\ref{sec:reproduce} for the lists and options we used.}.
From the top 1 million list of Tranco of September 20\textsuperscript{th} 2019, we extracted the top 1~000 websites of the TLD of 31 European countries and 1~000 websites from three country-independant TLDs: \texttt{.com}, \texttt{.org} and \texttt{.eu}.
Altogether, we obtained \nba{} websites 
(some countries had few websites in Tranco).
%: for example, there were only 108 websites on \texttt{.cy} (Cyprus) and 
%62 websites on \texttt{.mt} (Malta).
%The second column of Table~\ref{tab:tlds} shows the number of 
%crawled websites for each TLD. % that we have considered. 

Since our study is focused on the respect of consent, we decided to 
respect publishers’ consent regarding bots and crawling on their websites 
by checking the instructions in each website's \texttt{robots.txt} file. 
For each website $d$ in a list of \nba{} websites, we first visited the address \texttt{https://www.$d$/robots.txt}
using Python's \texttt{urllib} to verify access authorization. If access was denied, we did not crawl the website.
As a result, 3~633 (12.86\%) websites out of \nba{}
refused access to robots in their \texttt{robots.txt} file, so we removed them 
from our further analysis.

While testing authorization, we also verified reachability.
%We first open an empty browser and load the website.
If loading the \texttt{robots.txt} file failed for a network-related reason, we attempted accessing it %the same address 
through HTTP. If DNS resolution failed for \texttt{www.$d$}, we attempted accessing \texttt{$d$} instead. 
We determined if the website was loaded with a timeout of 10 seconds. 
If loading timed out 3 times, we considered access not successful. 
We applied the same criteria when visiting the main page with a Selenium-controlled browser.
%We consider websites opening alert boxes on load as not reachable (this concerns 0.04\% of websites).
In total, 1~675 (5.9\%) websites were not reachable. 

As a result, we successfully automatically crawled \nbasuccess{} websites originating 
from up to 1~000 websites from each TLD. %EU country, and up to  1~000 websites from each of 
%the country-independant TLDs: \texttt{.com}, \texttt{.org} and \texttt{.eu}.
The resulting \nbasuccess{} websites constitute the basis for the investigations that we describe in Section~\ref{sec:campaigns}.

\subsection{Detecting a TCF Cookie Banner}
\label{sec:methodo:detecting}

We first automatically detect websites with cookie banners that implement the TCF.  
\scname{} detects the presence of a TCF banner by checking whether a \cmp{} 
function is defined on a given website (each CMP
must implement a \cmp{} function according to the TCF v1.1 specification~\cite{iab-jsapi}, 
see more details in Section~\ref{sec:background:consentsharing}).  
%executing 
%a script that calls a function documented in the TCF's specification: the\cmp{} function~\cite{iab-jsapi}.

To validate our detection method in practice, we made a 
preliminary crawl on 21~000 websites, and analyzed how often \cmp{} and \cmplocator{} are used. 
We observed that all websites (except for one) that implement  \cmplocator{}  also implement the \cmp{}  function. 
%Among CMPs that implement the standard APIs, we have observed that 100\% of them 
%have a \cmp{} function defined and 
 %in 100.00\% of cases (only one exception), 
 % NATALIIA: 100% OR 1 EXCEPTION?
%We also observed that 79.24\% of websites  support  \cmplocator{}. 
% 100% - 79.24% = 
We also observed that 20.76\% of websites define a \cmp{} function but do not implement \cmplocator.
Thus, we can safely use a hypothesis that if the \cmp{} function is defined, 
 then a CMP is present on a website. 
We therefore rely on the presence of a \cmp{} function to conclude that a CMP is present on a website. 
When crawling a website, \scname\ waits for the website to 
finish all loading,  
%After a website has finished loading, \scname\ 
waits for %and an additional 
additional 3 seconds\footnote{Experimental 
tests on 200 websites showed that no more than a 2s delay is necessary.}, 
and then verifies whether the \texttt{window} object contains a \texttt{\_\_cmp} function. 
%If a \cmp{} function is not loaded upon the first page load, we do not re-load the page. 
If a \cmp{} function is not detected, \scname\ does not reload the page. 

As a result, we have automatically detected TCF banners on \nbbanner\ websites and 
we show further results on the prevalence of TCF banners and CMPs that implement 
them in Section~\ref{sec:prevalence}. 
Websites on which we did not find a cookie banner are omitted for the rest of the paper.
%Notice that our results %are not exhaustive and 
%represent a lower bound on the usage of TCF banners. 
%For example, website \texttt{notizie.it} does not load the \cmp{} function 
%on the main page, but displays the banner. We observe queries with consent string 
%that contain a positive consent before %user action. 
%user has made any choice in the cookie banner. 

\subsection{Intercepting a Consent String}
\label{sec:methodology:methods}

\begin{figure*}[t]
  \center
  \includegraphics[width=0.97\textwidth]{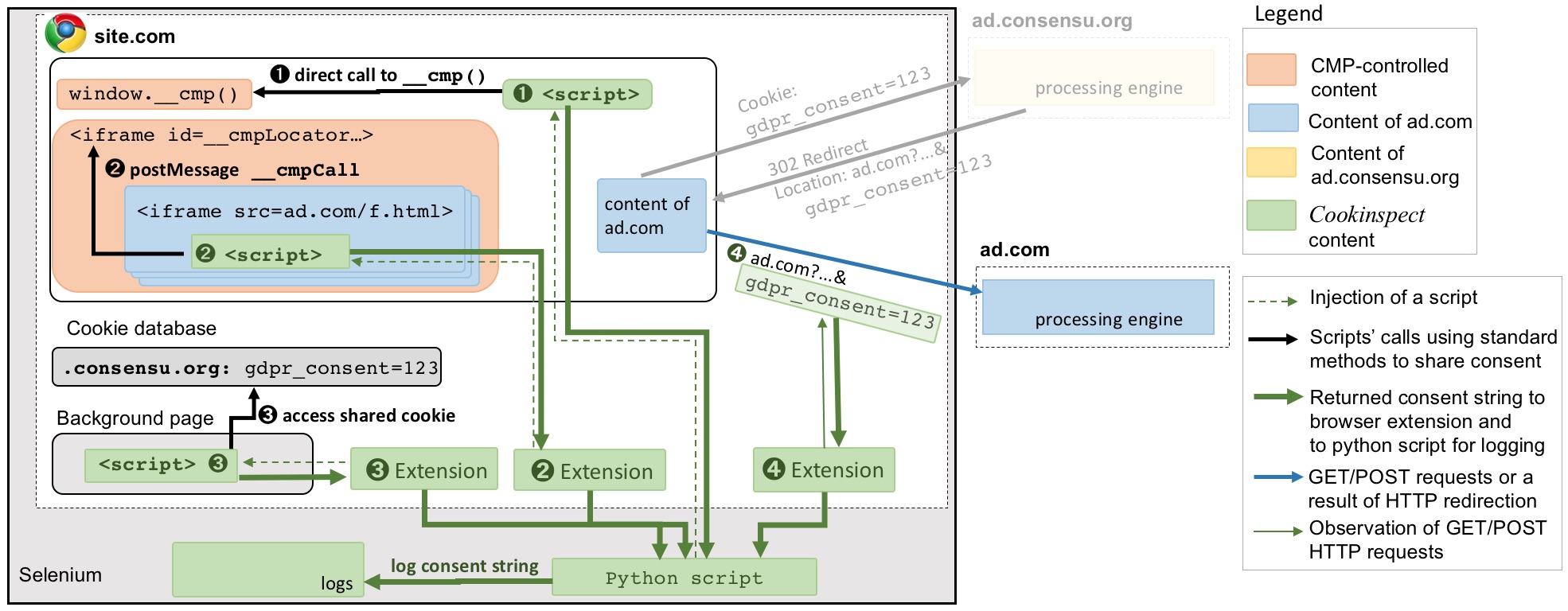}
  \caption{Detecting stored consent strings with \scname{}.}
  \label{fig:detect_consent}
  \vspace{-0.7cm}
\end{figure*}

%The IAB Europe TCF specification~\cite{iab-jsapi} describes how CMPs can 
%communicate user consent stored in a consent string to third party advertisers 
%present on the website. 
CMPs that implement a TCF banner can use a number of different methods 
to share a consent string with advertisers present on a website (see 
Section~\ref{sec:background:consentsharing}). 
%In Section~\ref{sec:background:consentsharing}, we provide a summary of all the 
%communication methods that a CMP that implements a TCF banner can use to
% share the  consent string with third party advertisers present on the website.
%
%The TCF provides several different ways to share user consent (see Section~\ref{sec:background:consentsharing}). 
%
%We described the different ways to pass consent in the TCF in section~\ref{sec:background:consentstorage}. 
In this section, we describe how %we 
\scname{}~\cite{Cookinspect} intercepts the consent strings using all available methods, 
summarized in  
%First, we actively intercept the consent string by pretending to be a third-party vendor that requests a consent from a CMP. These consent strings are directly provided by the CMP.
%Second, we passively monitor the consent string sent in HTTP requests and stored in a shared cookie.
Figure~\ref{fig:detect_consent}. 
%summarizes the four methods we use to intercept the consent string.  
%that we describe below.
\scname{} relies on several browser extensions 
that help injecting scripts and collecting the consent string using different methods.
\scname{} contains a Python script that collects all the intercepted consent strings 
and logs them in a local database. 

{\bf Standard APIs}. 
%\subsubsection{Direct calls to the \cmp{} function \one}
%We use a browser extension overriding the \cmp{} function and preventing any further modification of it. For every call to this function, we throw an exception, and extract the origin of the script responsible for this call in its stack trace. We extract the domain responsible for the call from the script's origin and transmit it to our Selenium script.
%1. Direct call: 
%We first actively 
First, \scname{} actively pretends to be an advertiser in a first party position: 
it inserts its script in the context of 
the crawled website (method \one{} in Fig.~\ref{fig:detect_consent}). The injected script is first-party because it runs in the origin 
of the crawled website (in the origin of \texttt{site.com} in Fig.~\ref{fig:detect_consent}).
The injected script makes a {direct call to the 
\cmp{} function}, and if it obtains a consent string in return, then the script transmits it back 
to the Python script that logs the consent strings. 
%With this method, we act as if we were a vendor script in a first-party position.

%\subsubsection{PostMessages to the \cmplocator{} iframe}
%We use a browser extension inserted in all frames to monitor all postMessages. When a postMessage make a TCF-related call (\texttt{\_\_cmpCall}), we extract its origin directly.
Second, \scname{} pretends to be an advertiser in a third-party position: \scname{} contains  
a custom browser extension \two{} that injects a script in all third-party iframes that 
have the \cmplocator{} iframe as a parent (only the children of an iframe with the 
\cmplocator{} identifier are able to query for the consent string).
From each such iframe, the injected script sends a postMessage \texttt{\_\_cmpCall} to the 
\cmplocator{} iframe to request the consent string (method \two{}). 
The script then transmits it back to the 
extension and further to the Python script for logging.
%With this method, we act as if we were a vendor script in a third-party position.

{\bf Shared cookie.}
%The TCF allows CMPs to store user consent in a shared cookie on the \texttt{.consensu.org} domain, which allows different CMPs to re-use the user consent.
%But due to the limitations of the Same-Origin Policy and of browser access control policy, we have no direct access to such cookie via Selenium or a browser extension. 
%We developed a
A browser extension \three{} of \scname{} attempts to read the shared cookie, and then transmits it to the extension and back to the Python script.
%injects a script in every iframe of a \texttt{*.consensu.org} origin. While executing in the correct origin, this script

{\bf URL-based methods. }
%1. GET and POST requests: 
To intercept the URL-based methods and obtain consent strings shared with 
third parties, %we develop 
a custom browser extension \four{} %that 
monitors all network requests. 
According to the TCF, a consent can be transmitted by the URL-based methods 
inside the GET \texttt{gdpr\_consent} parameter -- we therefore log all the requests 
containing such parameter (method \four). 
Although the TCF only mentions GET requests, we also monitor POST requests parameters. 
%because we see requests 
We observed that 
%requests that transmit 
%transmitting 
%a consent string using 
POST requests transmit a consent string on 399
websites (out of \nbbanner\ TCF websites, i.e. 28\%), %(in comparison to 680 (51.28\%) for the GET method)
while GET requests do so on 764 websites (53.6\%). 
%\footnote{In our experiments, we also saw other non-standard ways of transmitting the consent string: for example, on \texttt{mirror.co.uk} we observed a GET request with  \texttt{gdpr\_consent\_string} parameter instead of \texttt{gdpr\_consent}. We cannot quantify such abnormal cases at large scale and exclude them from our study.}.

%2. Consent redirecting: As 
Since the consent redirecting mechanism (detailed in section~\ref{sec:background:consentsharing}) always uses %queries 
HTTP requests to transmit the consent string in the \texttt{gdpr\_consent} parameter, 
we already detect it by intercepting all GET and POST requests that contain such parameter.
%observe the post-redirection query with the browser extension used for observing all queries described above.
%With this method, we obtain consent strings sent to third parties using queries.
%
%The specifications of the URL-based method for full-consent string passing~\cite{iab-urlbased} indicates values for ``URL parameters'', suggesting the use of GET queries to send information. In practice, we saw queries on 680 (51.28\%) websites using the GET method and on 330 (24.89\%) using the POST method. Moreover, parameters are sometimes embedded. \b{I need an example}

%\subsubsection{Shared cookie}
%We use a browser extension inserting a script in the frames of \texttt{*.consensu.org} origin. This script attempts to read the shared cookie, %(which will be possible for frames of the \texttt{consensu.org} origin)
%and transmits it to our Selenium script.

%{\bf Non-standard method: SafeFrame.}
{\bf Non-standard method.}
%safeFrames are implemented by IAB (and not by the CMPs responsible for cookie banners), and 
According to TCF specification, safeFrames act as a proxy to the \cmp{} function.  
\scname{} %therefore 
does not specifically interfere with safeFrames, because they obtain
a consent string by making a direct call to the \cmp{} function, which is already done with method \one{}. 
% expect results to be the same as calling this function directly, 
%and therefore do not use that method.
%we already observe these calls with the techniques for detecting direct calls to the \cmp{} function described above.
%However, because of this proxying via a script in a first-party position, we cannot obtain the origin of the call.

\subsection{Identifying the CMP Responsible for a TCF Banner}
\label{sec:methodo:cmpid}

To identify a CMP behind a banner,
we use the consent string that we obtain from the standard APIs and the shared cookie. 
%We extract information from the consent strings to identify the CMP on a website.
We decode the consent string 
%For every logged consent string obtained using the standard API or in the shared cookie, we use a 
with a public script provided by IAB% to decode it
~\cite{iab-cssdk}.
The decoded array contains the CMP identifier or ID (see \texttt{cmpId} in an example of decoded 
consent string in Fig.~\ref{fig:consent-string}). We map the CMP ID to the public CMP list~\cite{iab-cmplist}
to retrieve the CMP company's name. 
%\footnote{We obtained different CMPs across different page loads on two websites. As this may be attributed to the time difference between both crawls, we tested these 2 websites again on October 24, 2019, and still obtained two different CMPs on one website. The reason is that the consent string stored before user action contains a different CMP ID than a consent string stored after the user action. For this particular website, we have kept the CMP in the consent string after the user action.}

\subsection{Detecting Suspected GDPR and ePD Violations}
\label{sec:methodology:violations}

%We
In this section, we first explain what information we extract from the consent strings, and then explain how we detect suspected violations. 
%
%\textbf{Only the purposes stored in a consent string.}
Each consent string contains two arrays: an array of allowed advertisers, 
and an array of accepted purposes. The TCF indicates~\cite{iab-csusecase} that advertisers are supposed to verify that their identifier is allowed in the advertisers array, and that the purposes they use is allowed in the purposes array. 
% should they check whether their identifier is in the vendors array, or should they check if the purpose of the data processing they want to perform is allowed in the array of purposes?
%
As purposes for data processing have a strong legal meaning (see  Section~\ref{sec:background:legal}), we focus on the \textit{purposes} stored in a consent string, and do not analyze the array of allowed {advertisers}. We do, however, remove %the few 
%cases 
websites where a consent string contains %a non-empty array of purposes but 
an empty array of advertisers and a non-empty array of purposes.
We also remove 2 %cases 
websites where the vendors remaining in the consent string based their processing 
on legitimate interests only. We leave the discussion of such cases to the Data Protection Authorities (DPAs).

By using \scname{}, we detect the four major suspected GDPR and ePD violations 
presented in Section~\ref{sec:violations}.
We explain below what methods we used to detect each violation.
 
\textbf{\preaction{}}: 
\scname{} logs all the consent strings by using the standard APIs and reading the shared cookie 
(methods \one, \two\ and \three\ in Figure~\ref{fig:detect_consent}). 
\scname{} is able to detect \preaction\ violation fully automatically while crawling 
\nbasuccess\  websites without performing any user action. 
If the consent string stored in the user's browser is empty (has no accepted purposes), 
then we do not label it as a violation. 
%We do not consider a violation to store an empty consent string in the user's browser 
%before user action and update it after the user has made their choice in the banner. 
We therefore consider a suspected violation only if % report only on positive consent stored without any user action, we
the consent string has \emph{one or more accepted purposes}
%remove all consent strings that do not have any allowed purpose 
(out of five possible purposes in the TCF, see Appendix~\ref{app:purposes}) 
even though no action was performed on the visited website. 
 
\textbf{\nooption{}}: we detect this suspected violation manually by visiting the websites in a clean Chromium session
and using \scname{} to assign the corresponding label. 
To identify whether there is an option to refuse consent, we click on every button and link to verify whether they lead to a second layer of the banner. In the second layer we refuse consent by deselecting all purposes and advertisers. Notice that 
%containing options for selecting purposes and vendors. These 
the second layer is often hidden behind a misleading terminology (e.g. ``learn more''), which does not indicate that refusal is possible. 
%In other words, we check every possible action on the banner, to make sure that our analysis is not missing anything.
%To detect \textbf{\tracking}, we have manually refused consent in the cookie banner and then analyzed the third party trackers still present on the page. 

\textbf{\preticked{}}: 
we detect this suspected violation manually by visiting the websites in a clean Chromium session
and using \scname{} to assign the corresponding label.
We label a website as violating if it has a "parameters" option, and if in the "parameters" page 
there is at least one pre-selected checkbox or enabled slider for at least one purpose. 

{\bf \nonrespect}: to detect this violation, we perform a semi-automatic crawl 
based on a human operator and \scname{} on 
\nbsa{} websites from French-, Italian- or English-speaking countries: 
France, UK, Belgium, Ireland and Italy, and \texttt{.com} websites. 
We only consider banners written in a language that at least one of the authors speak
to ensure that we understand the actions we perform.
First, in a clean Chromium session, a human operator manually refuses consent on the cookie banner (following the procedure detailed in Appendix~\ref{sec:human_operator}).
Then, \scname{} logs all the consent strings by using the standard APIs and reading the shared cookie. 
%In this case, we present a lower bound on violations: we report only on a positive consent that contains all five purposes that exist in the framework.
%We do it because some of the purposes are sometimes pre-selected in the cookie banner and we do not quantify for how many purposes we could effectively opt-out in the manual evaluation. 
%

\iffalse
We consider only consent strings that have \emph{all five accepted purposes}. 
With this approach, we ensure that a suspected violation indeed took place. 
Notice that with this restrictive approach, we miss potential violations for TCF banners that do not allow users to refuse all purposes because in these cases a consent string would contain less than five purposes. 
\fi

Some TCF banners may display purposes that users cannot refuse,
considering that such purposes do not require consent. 
Such decision can be criticized by legal experts and policy makers, 
however we exclude such discussions from our work.
Instead, we further consider only consent strings that have \emph{all five accepted purposes},
to ensure that a violation indeed took place, even if the user couldn't refuse some 
of the purposes of data processing in the TCF banner. Thus, we avoid the case where setting a purpose in a consent string may be considered legal because this purpose can be relied on using legitimate interests.

We underline that the responsibility for suspected violations is joint between publishers and CMPS 
and discuss the difficulty to attribute responsibility %to only one actor 
in Section~\ref{sec:discussion}.

%\NB{We should give explanations on how we detect shared cookies and why and how we detect trackers. Otherwise they appear out of nowhere in the description of the automatic crawl.}
%\subsection{Shared cookies}
%
%\subsection{Trackers}

\section{Two Measurement Campaigns}
\label{sec:campaigns}

%Although the GDPR's scope is worldwide as long as a user located in the EU accesses a service,
We perform a large-scale study of websites registered in the EU because the TCF is likely to be observed more often on European websites. 
We perform two measurement campaigns with \scname{}: a large-scale automatic crawl and a 
smaller-scale semi-automatic crawl, both conducted in September 2019 from France. 
The source code of \scname{} and all the extensions is publicly available so that 
end users and DPAs can test compliance of publishers and CMPs by themselves~\cite{Cookinspect}.

\begin{figure}[t]
  \center
  \includegraphics[width=0.47\textwidth]{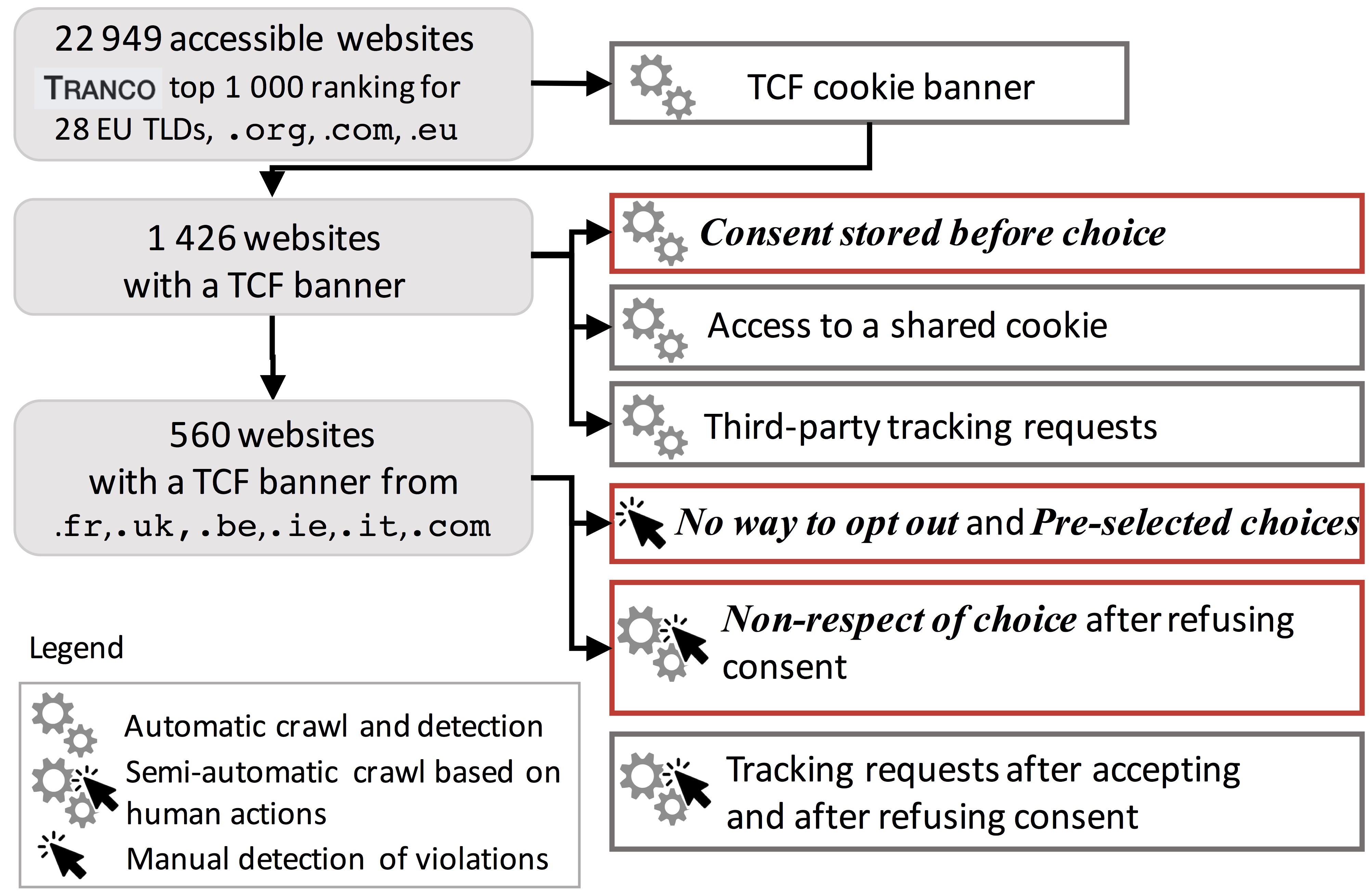}
  \caption{Overview of the website-auditing process combining automatic crawling, semi-automatic 
  crawling with human interaction  and manual analysis to detect suspected GDPR and ePD violations.}
  \SHORTEN
  \label{fig:pipeline}
\end{figure}

%\nc{An arrow is missing in figure~\ref{fig:pipeline}}

%Figure~\ref{fig:pipeline} provides an overview of the main components of our website auditing process. 
%Table~\ref{tab:violations} presents an overview of suspected violations we detect using %our 
%automatic and semi-automatic %crawls. 
%crawling campaigns with \scname.
%For each violation, we show what crawl was used for its detection %(column ``Crawl'') 
%and which component of the Web application allowed us to detect it.
%%(column ``Component'').

\begin{table}[t]
\caption{Overview of methods to detect the suspected GDPR and the ePD violations in TCF banners.}
\label{tab:violations}
\center
\begin{tabular}{|p{2.3cm}|p{1.8cm}|p{1.4cm}|p{1.6cm}|}
\hline
\textbf{Short name} &  \textbf{Type of crawl}  & %\textbf{Analyzed component} & 
\textbf{Analyzed component} & 
\textbf{Number of tested websites} \\\hline
%\multicolumn{4}{l}{\bf \em GDPR violations} \\\hline
\preaction{} &  automatic  & consent string & \nbbanner{} \\\hline
\nooption{} &  semi-automatic  & banner & \nbsa{} \\\hline
\preticked{} &  semi-automatic (when opting out is possible) &  banner & \nbsarefusal{} \\\hline
\nonrespect{} & semi-automatic (when opting out is possible) &  consent string & \nbsarefusal{} \\\hline
% \multicolumn{4}{l}{\bf \em  TCF violations} \\\hline
% Broken banner &  semi-automatic &  banner & \nbsa{} \\\hline
% No banner &  semi-automatic &  banner & \nbsa{} \\\hline
% Consent to nonexistent vendors &  both &  consent string & \nbbanner{} \\\hline
% Wrong CMP id & both &  consent string & \nbbanner{} \\\hline
% Non-registered tracker verifying consent  & both &  third party & \nbbanner{} \\\hline
% Non-registered tracker not verifying consent  & automatic & third party & \nbbanner{} \\\hline
% Unknown third-party verifying consent &  both & third party & \nbbanner{} \\\hline
\end{tabular}
%\SHORTENTAB
\end{table}

Figure~\ref{fig:pipeline} provides an overview of the main components of our website auditing process. 
Table~\ref{tab:violations} presents an overview of suspected violations we detect using %our 
automatic and semi-automatic %crawls. 
crawling campaigns with \scname.
For each violation, we show %what crawl was 
the crawl used for its detection %(column ``Crawl'') 
and which component of the Web application allowed us to detect it.
%(column ``Component'').

\subsection{Automatic Crawl}
First, we perform a stateless \emph{automatic crawl} of \nbasuccess{} selected websites 
(see Section~\ref{sec:methodology:websites} for websites selection) to perform website 
auditing without human intervention to detect: 
%detect the presence of TCF banners and 
%\textbf{\preaction} violation automatically. 
(1) %\begin{enumerate}
%\item 
whether the website contains a TCF banner\footnote{We do not test for the \cmplocator{} iframe presence, as we found only 1 website among 21~000 on which we could find a \cmplocator{} iframe but no \cmp{} function during a test run.},
(2) %\item 
whether positive consent is stored without any user action ({\bf \preaction{}} violation),
(3) %\item 
whether the website accesses the consent string in a shared cookie that we set in the browser,
(4) %\item 
presence of third-party tracking requests prior to any user consent.
%\end{enumerate}

\emph{Procedure:} In a first browser session, we detect if the website implements the TCF by verifying whether a \cmp{} function is defined after load and a 3s delay. If so, we attempt to obtain the consent string %via the standard APIs and the shared cookie (using browser extensions) 
using \scname{} to detect the {\bf \preaction{}} violation.
\scname{} also logs all requests made to third parties.
In a new clean browser session, we place a shared cookie in the browser, and attempt to get it back using a direct \cmp{} call and a postMessage, to measure whether the CMP on the page uses the shared cookie.
%This time, we override the \cmp{} function and collect domains of scripts calling this function.
%In a third clean browser session, we monitor the passing of consent string using remaining methods described in section~\ref{sec:methodology:methods}: postMessage, obtaining shared cookie, and monitoring GET and POST requests.

This crawl takes an average of 11.04s per website. Crawling 28~000 websites takes more than 24 hours. 
This crawl was made from France on September 20\textsuperscript{th} and 21\textsuperscript{st} 2019.

\subsection{Semi-Automatic Crawl}

From the resulting  \nbbanner{} websites that contain
a TCF banner, we select \nbsa{} websites of French, Italian or English-speaking countries (the languages 
that the authors speak fluently) from  \texttt{.uk}, \texttt{.fr},  \texttt{.it},  \texttt{.be}, \texttt{.ie} and \texttt{.com} TLDs to perform a \emph{semi-automatic crawl}.

This crawl performs tests requiring interaction with the cookie banner\footnote{As the TCF does not specify anything regarding the user's interface, we don't have a way to locate a banner and its different elements, and \textit{a fortiori} to automate banner interaction.}. Upon different browser sessions, we both give a positive consent and refuse consent, to make tests regarding respect of given consent, and setting of the shared cookie. We also note whether banners propose an option to refuse consent, and whether specific parameters are pre-selected in favor of consent-sharing. 

%We perform a \textit{semi-automatic crawl} on the selected \nbsa\ websites and
\emph{Procedure:} On a clean browser session, we load the website.
If there is no banner or a broken banner, we stop there.
We manually label when the \textbf{\preticked} or \textbf{\nooption} suspected violation are observed (see the 
complete procedure describing the labelling process in Section~\ref{sec:methodology:violations}).
We then refuse consent on the banner, and observe with \scname{} what consent is stored by the CMP (including the shared cookie) to detect the \textbf{\nonrespect} violation.
After 3 seconds, we refresh the page to log all network requests (to quantify the amount of trackers).
We also attempt to obtain the consent string again.
Then, on a new session, we repeat this procedure, this time giving a positive consent to the banner.
%, and attempt to obtain the consent string using the same methods again (including reloading the page).
%
%We open a new session, load the website and prompt the human operator to refuse consent on the banner. %If there is no option to refuse tracking, they have to validate instead.
%Once this is done, we attempt to obtain the consent string using several methods: direct call to the \cmp{} function, sending a postMessage, and reading the shared cookie. 
%After 3 seconds, we refresh the page and attempt to obtain the consent string again, this time by monitoring all requests, and once again using direct calls to the \cmp{} function and a postMessage.
%Then, on a new session, we prompt the human operator to accept consent, and attempt to obtain the consent string using the same methods again (including reloading the page). %, in order to be able to identify the CMP using the CMP id included in it.
%In both sessions, we keep track of domains trying to obtain the consent string, and of trackers present on the page. We also note whether the shared cookie is set after each decision of the user.
%
%Tests related to the verification of information present in the consent string 
%and verification of consent by third parties
% are performed in both crawls, because different behaviours may be exhibited prior to and after user giving consent.
%
We give the detailed procedure for the human operator to refuse and accept consent on the banner in Appendix~\ref{sec:human_operator}.

Each website takes around 30-40s given a reactive human manipulation.
We performed this stateless crawl from France from September 23\textsuperscript{rd} to October 1\textsuperscript{st} 2019.

%In total, we load the website in 2 successive browser sessions in the automatic crawl, and 2 sessions each reloading the page once in the semi-automatic crawl.

\subsection{Verification Procedure}
\label{sec:verification}

For the \nonrespect{} and the \nooption{} violations, we cross-checked choices made in the banner and whether the banner allows to refuse consent by two human operators to limit errors. The three operators are computer scientists working on web tracking. 
%, and 2 of them are co-authors of this paper.
The semi-automatic crawl is first entirely done by one of the operators.
For each CMP on which we observe a \nonrespect{} on at least one website, we select one of these violating websites. We add an equal number of websites on which the violation is not seen in the pool of sites to be verified. Then, a second human operator refuses consent on all of these websites, unknowingly of which website was considered a violation by the previous tester. We do the same for the \nooption{} violation, this time testing for the possibility to refuse consent.
Then, the third human operator repeats the procedure on a new pool of websites.
%In case of a conflict, 
In case we obtain different results from different operators, we keep the results of the test returning the least violations on the concerned website, and retest all websites of the considered CMP to take into account the fact that some option of the banner may have been missed.

\subsection{Ethical Considerations}
\label{sec:ethics}

Our data collection process does not involve any human subject.
Our study of websites is mostly passive: the only action we perform on websites is %for some of them, 
%to 
clicking on manually-selected cookie acceptance buttons. Hence, we do not tamper with any distant system.
Our large-scale measurement does not present any potential harm for websites, while presenting societal benefits.
Moreover, we respect publishers' consent regarding bots that they express in a \texttt{robots.txt} file. % (see website selection process in Section~\ref{sec:methodology}). 

%==========================================================
\section{Prevalence of TCF Banners %from the TCF 
in Europe}
\label{sec:prevalence}

\newcommand\setrow[1]{\gdef\rowmac{#1}#1\ignorespaces}
\newcommand\clearrow{\global\let\rowmac\relax}

\def\purposestable{
  \begin{table}[htbp]
    \center
    \setlength\tabcolsep{2.5pt}
    \caption{Number of websites seen with the different violations, w.r.t. the maximum number of purposes in observed consent strings. Considered violating cases are shown in bold.}
    \label{tab:res:purposes}
    \begin{tabular}{|p{1.5cm}|p{1.85cm}|p{1.50cm}|p{1.70cm}|p{1.50cm}|}
      \hline
      \textbf{Number of purposes} & \textbf{\preaction{}} & \textbf{\nooption{}} & \textbf{\preticked{}} & \textbf{\nonrespect{}} \\\hline
      1 to 4 purposes & \textbf{2.1\%} (30/1426) & - & - & 6.7\% (34/508) \\\hline
5 purposes & \textbf{7.8\%} (111/1426) & - & - & \textbf{5.3\%} (27/508) \\\hline
\textbf{Total number of violations} & 9.9\% (141/1426) & 6.8\% (38/560) & 46.5\% (236/508) & 5.3\% (27/508) \\\hline

      \addlinespace[1mm]
      \hline
      \textbf{Any violation} & \multicolumn{4}{c|}{54.29\% 304/560} \\
      \hline
    \end{tabular}
  \end{table}
}

\def\percountrytable{
  \begin{table}[t]
    \def\arraystretch{0.95}% default: 1.0
    \center
    \caption{Distribution of websites with a TCF banner across European (and 3 international) TLDs, computed with an automatic crawl.}
    \label{tab:tlds}
    \begin{tabular}{|>{\rowmac}l|>{\rowmac}p{2.1cm}|>{\rowmac}p{2.25cm}|>{\rowmac}p{2.15cm}<{\clearrow}|}
      \hline
      \textbf{TLD} & \textbf{Number of domains in the Tranco top-1 million list} & \textbf{Number of reachable and allowed (for bots) domains} & \textbf{Number of domains with a TCF-related cookie banner} \\\hline
      .uk & 1 000 & 788 (78.8\%) &  \cellcolor{gray!56} 149 (18.9\%) \\
.fr & 1 000 & 815 (81.5\%) &  \cellcolor{gray!51} 139 (17.1\%) \\
.pl & 1 000 & 858 (85.8\%) &  \cellcolor{gray!45} 129 (15.0\%) \\
.it & 1 000 & 824 (82.4\%) &  \cellcolor{gray!44} 123 (14.9\%) \\
.es & 1 000 & 800 (80.0\%) &  \cellcolor{gray!42} 113 (14.1\%) \\
.nl & 1 000 & 838 (83.8\%) &  \cellcolor{gray!23} 65 (7.8\%) \\
.gr & 1 000 & 720 (72.0\%) &  \cellcolor{gray!22} 53 (7.4\%) \\
.pt & 1 000 & 793 (79.3\%) &  \cellcolor{gray!19} 52 (6.6\%) \\
.de & 1 000 & 881 (88.1\%) &  \cellcolor{gray!19} 56 (6.4\%) \\
.ro & 1 000 & 787 (78.7\%) &  \cellcolor{gray!19} 50 (6.4\%) \\
.bg & 547 & 449 (82.1\%) &  \cellcolor{gray!17} 26 (5.8\%) \\
.fi & 1 000 & 824 (82.4\%) &  \cellcolor{gray!17} 47 (5.7\%) \\
.no & 1 000 & 862 (86.2\%) &  \cellcolor{gray!16} 48 (5.6\%) \\
.dk & 1 000 & 824 (82.4\%) &  \cellcolor{gray!14} 41 (5.0\%) \\
.be & 1 000 & 863 (86.3\%) &  \cellcolor{gray!13} 38 (4.4\%) \\
.at & 1 000 & 873 (87.3\%) &  \cellcolor{gray!11} 33 (3.8\%) \\
.ie & 1 000 & 769 (76.9\%) &  \cellcolor{gray!9} 25 (3.3\%) \\
.cz & 1 000 & 916 (91.6\%) &  \cellcolor{gray!9} 28 (3.1\%) \\
.ch & 1 000 & 849 (84.9\%) &  \cellcolor{gray!9} 26 (3.1\%) \\
.se & 1 000 & 787 (78.7\%) &  \cellcolor{gray!8} 21 (2.7\%) \\
.sk & 1 000 & 879 (87.9\%) &  \cellcolor{gray!4} 14 (1.6\%) \\
.hr & 627 & 513 (81.8\%) &  \cellcolor{gray!4} 8 (1.6\%) \\
.hu & 1 000 & 794 (79.4\%) &  \cellcolor{gray!2} 6 (0.8\%) \\
.lu & 186 & 147 (79.0\%) &  \cellcolor{gray!2} 1 (0.7\%) \\
.lt & 745 & 605 (81.2\%) &  \cellcolor{gray!1} 4 (0.7\%) \\
.lv & 537 & 420 (78.2\%) &  \cellcolor{gray!1} 2 (0.5\%) \\
.si & 514 & 426 (82.9\%) &  \cellcolor{gray!1} 2 (0.5\%) \\
.is & 358 & 248 (69.3\%) &  \cellcolor{gray!1} 1 (0.4\%) \\
.ee & 468 & 373 (79.7\%) &  \cellcolor{gray!0} 1 (0.3\%) \\
.li & 105 & 62 (59.0\%) &  \cellcolor{gray!0} 0 (0.0\%) \\
.cy & 108 & 76 (70.4\%) &  \cellcolor{gray!0} 0 (0.0\%) \\
.mt & 62 & 37 (59.7\%) &  \cellcolor{gray!0} 0 (0.0\%) \\
\hline
.com & 1 000 & 711 (71.1\%) &  \cellcolor{gray!40} 97 (13.6\%) \\
.org & 1 000 & 735 (73.5\%) &  \cellcolor{gray!6} 16 (2.2\%) \\
.eu & 1 000 & 803 (80.3\%) &  \cellcolor{gray!4} 12 (1.5\%) \\
\hline
\setrow{\bfseries} all & 28 257 & 22 949 (81.2\%) &  \cellcolor{gray!18} 1 426 (6.2\%) \\
\hline

    \end{tabular}
  %\SHORTENTAB
  \end{table}
}

\clearrow
\def\percountrypreaction{
  \begin{table}[htbp]
    \center
    \caption{Results of the \preaction{} violation on \nbbanner{} websites via an automatic crawl.}
    \label{tab:tlds_preaction}
%    \parbox{.45\linewidth}{
      \begin{tabular}{|>{\rowmac}l|>{\rowmac}p{2.2cm}<{\clearrow}|}
        \hline
        \textbf{TLD} & \textbf{Number of websites} \\\hline
        .uk & \cellcolor{BrickRed!11} 11.4\% (17/149) \\
.fr & \cellcolor{BrickRed!7} 7.2\% (10/139) \\
.pl & \cellcolor{BrickRed!17} 17.8\% (23/129) \\
.it & \cellcolor{BrickRed!9} 9.8\% (12/123) \\
.es & \cellcolor{BrickRed!7} 7.1\% (8/113) \\
.nl & \cellcolor{BrickRed!0} 0.0\% (0/65) \\
.gr & \cellcolor{BrickRed!3} 3.8\% (2/53) \\
.pt & \cellcolor{BrickRed!1} 1.9\% (1/52) \\
.de & \cellcolor{BrickRed!28} 28.6\% (16/56) \\
.ro & \cellcolor{BrickRed!2} 2.0\% (1/50) \\
.bg & \cellcolor{BrickRed!3} 3.8\% (1/26) \\
.fi & \cellcolor{BrickRed!10} 10.6\% (5/47) \\
.no & \cellcolor{BrickRed!2} 2.1\% (1/48) \\
.dk & \cellcolor{BrickRed!4} 4.9\% (2/41) \\
.be & \cellcolor{BrickRed!42} 42.1\% (16/38) \\
.at & \cellcolor{BrickRed!12} 12.1\% (4/33) \\
.ie & \cellcolor{BrickRed!12} 12.0\% (3/25) \\
\hline

      \end{tabular}
%    }
    ~
%    \parbox{.45\linewidth}{
      \begin{tabular}{|>{\rowmac}l|>{\rowmac}p{2.2cm}<{\clearrow}|}
        \hline
        \textbf{TLD} & \textbf{Number of websites} \\\hline
        .cz & \cellcolor{BrickRed!3} 3.6\% (1/28) \\
.ch & \cellcolor{BrickRed!0} 0.0\% (0/26) \\
.se & \cellcolor{BrickRed!9} 9.5\% (2/21) \\
.sk & \cellcolor{BrickRed!7} 7.1\% (1/14) \\
.hr & \cellcolor{BrickRed!0} 0.0\% (0/8) \\
.hu & \cellcolor{BrickRed!0} 0.0\% (0/6) \\
.lu & \cellcolor{BrickRed!0} 0.0\% (0/1) \\
.lt & \cellcolor{BrickRed!25} 25.0\% (1/4) \\
.lv & \cellcolor{BrickRed!0} 0.0\% (0/2) \\
.si & \cellcolor{BrickRed!0} 0.0\% (0/2) \\
.is & \cellcolor{BrickRed!0} 0.0\% (0/1) \\
.ee & \cellcolor{BrickRed!0} 0.0\% (0/1) \\
\hline
.com & \cellcolor{BrickRed!11} 11.3\% (11/97) \\
.org & \cellcolor{BrickRed!12} 12.5\% (2/16) \\
.eu & \cellcolor{BrickRed!8} 8.3\% (1/12) \\
\hline
\setrow{\bfseries} all & \cellcolor{BrickRed!9} 9.9\% (141/1426) \\
\hline

      \end{tabular}
%    }
    \SHORTENTAB
  \end{table}
}

\def\cmptablegdpr{
  \begin{table*}[htpb]
    \center
    \caption{Quantification of suspected violations of the GDPR and the ePD encountered in the different CMPs seen at least 5 times during the semi-automatic crawl (on \texttt{.fr}, \texttt{.uk}, \texttt{.it}, \texttt{.be}, \texttt{.ie} and \texttt{.com} websites), by CMP. The \nonrespect{} and \preticked{} columns display results w.r.t. the number of websites on which refusing consent was possible.}
    \label{tab:quantification_violations_cmp_gdpr}
    \begin{tabular}{|>{\rowmac}l|>{\rowmac}c|>{\rowmac}c|>{\rowmac}c|>{\rowmac}c|>{\rowmac}c|>{\rowmac}c<{\clearrow}|}
      \hline
      & \textbf{Number of} & \multicolumn{4}{c|}{\textbf{Violations}} \\
      \cline{3-6}
      \textbf{CMP} & \textbf{websites} & \textbf{\preaction{}} & \textbf{\nooption{}} & \textbf{\preticked{}} & \textbf{\nonrespect{}} \\
      \hline
      Quantcast & 174 & \cellcolor{BrickRed!3} 3.4\% (6/174) & \cellcolor{BrickRed!5} 5.2\% (9/174) & \cellcolor{BrickRed!37} 37.8\% (62/164) & \cellcolor{BrickRed!0} 0.6\% (1/164) \\\hline
OneTrust & 50 & \cellcolor{BrickRed!74} 74.0\% (37/50) & \cellcolor{BrickRed!4} 4.0\% (2/50) & \cellcolor{BrickRed!83} 83.3\% (40/48) & \cellcolor{BrickRed!8} 8.3\% (4/48) \\\hline
Didomi & 41 & \cellcolor{BrickRed!0} 0.0\% (0/41) & \cellcolor{BrickRed!0} 0.0\% (0/41) & \cellcolor{BrickRed!39} 39.0\% (16/41) & \cellcolor{BrickRed!0} 0.0\% (0/41) \\\hline
Sourcepoint & 34 & \cellcolor{BrickRed!2} 2.9\% (1/34) & \cellcolor{BrickRed!0} 0.0\% (0/34) & \cellcolor{BrickRed!64} 64.7\% (22/34) & \cellcolor{BrickRed!2} 2.9\% (1/34) \\\hline
Evidon & 22 & \cellcolor{BrickRed!0} 0.0\% (0/22) & \cellcolor{BrickRed!22} 22.7\% (5/22) & \cellcolor{BrickRed!25} 25.0\% (4/16) & \cellcolor{BrickRed!25} 25.0\% (4/16) \\\hline
iubenda & 20 & \cellcolor{BrickRed!0} 0.0\% (0/20) & \cellcolor{BrickRed!0} 0.0\% (0/20) & \cellcolor{BrickRed!0} 0.0\% (0/20) & \cellcolor{BrickRed!0} 0.0\% (0/20) \\\hline
Clickio & 14 & \cellcolor{BrickRed!0} 0.0\% (0/14) & \cellcolor{BrickRed!0} 0.0\% (0/14) & \cellcolor{BrickRed!0} 0.0\% (0/14) & \cellcolor{BrickRed!0} 0.0\% (0/14) \\\hline
Oath & 12 & \cellcolor{BrickRed!0} 0.0\% (0/12) & \cellcolor{BrickRed!0} 0.0\% (0/12) & \cellcolor{BrickRed!16} 16.7\% (2/12) & \cellcolor{BrickRed!0} 0.0\% (0/12) \\\hline
Triboo Media & 10 & \cellcolor{BrickRed!0} 0.0\% (0/10) & \cellcolor{BrickRed!0} 0.0\% (0/10) & \cellcolor{BrickRed!0} 0.0\% (0/10) & \cellcolor{BrickRed!0} 0.0\% (0/10) \\\hline
Commanders Act & 10 & \cellcolor{BrickRed!40} 40.0\% (4/10) & \cellcolor{BrickRed!0} 0.0\% (0/10) & \cellcolor{BrickRed!80} 80.0\% (8/10) & \cellcolor{BrickRed!0} 0.0\% (0/10) \\\hline
Axel Springer & 10 & \cellcolor{BrickRed!60} 60.0\% (6/10) & \cellcolor{BrickRed!70} 70.0\% (7/10) & \cellcolor{BrickRed!100} 100.0\% (3/3) & \cellcolor{BrickRed!33} 33.3\% (1/3) \\\hline
OneTag & 9 & \cellcolor{BrickRed!0} 0.0\% (0/9) & \cellcolor{BrickRed!0} 0.0\% (0/9) & \cellcolor{BrickRed!100} 100.0\% (9/9) & \cellcolor{BrickRed!0} 0.0\% (0/9) \\\hline
Cookie Trust WG. & 8 & \cellcolor{BrickRed!25} 25.0\% (2/8) & \cellcolor{BrickRed!25} 25.0\% (2/8) & \cellcolor{BrickRed!60} 60.0\% (3/5) & \cellcolor{BrickRed!0} 0.0\% (0/5) \\\hline
Conversant Europe & 7 & \cellcolor{BrickRed!0} 0.0\% (0/7) & \cellcolor{BrickRed!0} 0.0\% (0/7) & \cellcolor{BrickRed!100} 100.0\% (7/7) & \cellcolor{BrickRed!0} 0.0\% (0/7) \\\hline
Ensighten & 7 & \cellcolor{BrickRed!0} 0.0\% (0/7) & \cellcolor{BrickRed!0} 0.0\% (0/7) & \cellcolor{BrickRed!100} 100.0\% (7/7) & \cellcolor{BrickRed!0} 0.0\% (0/7) \\\hline
SIRDATA & 5 & \cellcolor{BrickRed!0} 0.0\% (0/5) & \cellcolor{BrickRed!0} 0.0\% (0/5) & \cellcolor{BrickRed!0} 0.0\% (0/5) & \cellcolor{BrickRed!0} 0.0\% (0/5) \\\hline
Chandago & 5 & \cellcolor{BrickRed!0} 0.0\% (0/5) & \cellcolor{BrickRed!0} 0.0\% (0/5) & \cellcolor{BrickRed!0} 0.0\% (0/5) & \cellcolor{BrickRed!0} 0.0\% (0/5) \\\hline
\textbf{incorrect CMP ID} & 9 & \cellcolor{BrickRed!11} 11.1\% (1/9) & \cellcolor{BrickRed!11} 11.1\% (1/9) & \cellcolor{BrickRed!62} 62.5\% (5/8) & \cellcolor{BrickRed!12} 12.5\% (1/8) \\\hline
\textbf{others} & 73 & \cellcolor{BrickRed!10} 11.0\% (8/73) & \cellcolor{BrickRed!6} 6.8\% (5/73) & \cellcolor{BrickRed!54} 54.4\% (37/68) & \cellcolor{BrickRed!22} 22.1\% (15/68) \\\hline
\textbf{No consent string found} & 40 & \cellcolor{BrickRed!0} 0.0\% (0/40) & \cellcolor{BrickRed!17} 17.5\% (7/40) & \cellcolor{BrickRed!50} 50.0\% (11/22) & \cellcolor{BrickRed!0} 0.0\% (0/22) \\\hline
\setrow{\bfseries} all & 560 & \cellcolor{BrickRed!11} 11.6\% (65/560) & \cellcolor{BrickRed!6} 6.8\% (38/560) & \cellcolor{BrickRed!46} 46.5\% (236/508) & \cellcolor{BrickRed!5} 5.3\% (27/508) \\\hline

    \end{tabular}
    \SHORTENTAB
  \end{table*}
}

\def\size{1.65cm}
\def\tablegdprviolation[#1]#2{
  \begin{subfigure}{\columnwidth}
  %\center
    \setlength\tabcolsep{2pt}
    \begin{minipage}[c]{0.27\textwidth}
      \caption{#2}
    \end{minipage}\hfill
    \begin{minipage}[c]{0.65\textwidth}
      \begin{tabular}{|>{\rowmac}L{2.5cm}|>{\rowmac}R{1.75cm}|>{\rowmac}R{\size{}}|>{\rowmac}R{\size{}}|>{\rowmac}R{\size{}}|>{\rowmac}R{\size{}}|>{\rowmac}R{\size{}}|>{\rowmac}R{\size{}}<{\clearrow}|}
        \hline
        \textbf{CMP} & \textbf{total} & \textbf{.uk} & \textbf{.fr} &  \textbf{.it} & \textbf{.be} & \textbf{.ie} & \textbf{.com} \\\hline
        \input{tables/cmp_table_gdpr_#1}
      \end{tabular}
      \vspace{0.2cm}
      \end{minipage}
  \end{subfigure}
}

\def\tablesperviolation{
  \begin{table*}[htbp]
    % \center
    \caption{Quantification of violations of the GDPR encountered in the different CMPs for which the considered violations has been seen at least 3 times during the semi-automatic crawl (on \texttt{.fr}, \texttt{.uk}, \texttt{.it}, \texttt{.be}, \texttt{.ie} and \texttt{.com} websites), by CMP. The \nonrespect{} and \preticked{} tables only consider websites on which refusing consent was possible.}
    \label{tab:quantification_violations_cmp_gdpr_per_violation}

    \tablegdprviolation[preaction]{\preaction{}}

    \tablegdprviolation[nooption]{\nooption{}}

    \tablegdprviolation[preticked]{\preticked{}}

    \tablegdprviolation[nonrespect]{\nonrespect{}}

    \SHORTEN
  \end{table*}
}

\def\tableviolatingwebsites[#1]#2{
  \begin{subfigure}{0.50\columnwidth}
  %\center
    \caption*{#2}
    \begin{tabular}{|r|l|}
      \hline
      \textbf{Tranco rank} & \textbf{domain} \\\hline
      \input{tables/websites_violation_#1}
    \end{tabular}
  \end{subfigure}
}

\def\tablesviolatingwebsite{
  \begin{table*}[htbp]
    \def\arraystretch{0.8}% default: 1.0
    % \center
    \caption{Top 20 websites where we observe each suspected violation, ordered by their rank in the Tranco list. See complete lists of websites for each suspected violation in attachment~\cite{attachments}.}
    \label{tab:violating_websites}
    \vspace{-0.5cm}
    \tableviolatingwebsites[preaction]{\preaction{}}
    \tableviolatingwebsites[nooption]{\nooption{}}
    \tableviolatingwebsites[preticked]{\preticked{}}
    \tableviolatingwebsites[nonrespect]{\nonrespect{}}
    \SHORTENTAB
  \end{table*}
}

\def\tablegdprtld#1{
  \begin{subfigure}{\columnwidth}
    \center
    \begin{minipage}[c]{0.45\textwidth}
      \caption{.#1}
    \end{minipage}
    \begin{minipage}[c]{0.54\textwidth}
      \begin{tabular}{|>{\rowmac}l|>{\rowmac}C{1.0cm}|>{\rowmac}p{1.75cm}|>{\rowmac}p{1.75cm}|>{\rowmac}p{1.75cm}|>{\rowmac}p{1.75cm}<{\clearrow}|}
        % \begin{tabular}{|l|p{0.7cm}|p{0.7cm}|p{0.7cm}|p{0.7cm}|p{0.7cm}|}
        \hline
        &  & \multicolumn{4}{c|}{\textbf{Violations}} \\
        \cline{3-6}
        % \textbf{CMP} & \textbf{Number of websites} & \thead{\textbf{No}\\\textbf{option}} & \thead{\textbf{Pre-}\\\textbf{ticked}} & \thead{\textbf{Pre-}\\\textbf{action}\\\textbf{consent}} & \thead{\textbf{Non-}\\\textbf{respect}\\\textbf{of de-}\\\textbf{cision}} \\
        \textbf{CMP} & \textbf{Number of websites} & \textbf{\preaction{}} & \textbf{\nooption{}} & \textbf{\preticked{}} & \textbf{\nonrespect{}} \\
        \hline
        \input{tables/cmp_table_gdpr_#1}
      \end{tabular}
      \vspace{0.2cm}
    \end{minipage}
  \end{subfigure}
}

\def\percountryviolationtables{
  \begin{table*}[htbp]
    % \center
    \setlength{\tabcolsep}{4pt}
    \caption{Per-country violation tables. Quantification of suspected violations of the GDPR and the ePD encountered in the different CMPs seen at least 5 times in that country during the semi-automatic crawl (on \texttt{.fr}, \texttt{.uk}, \texttt{.it}, \texttt{.be}, \texttt{.ie} and \texttt{.com} websites), by CMP. The \nonrespect{} and \preticked{} tables only consider websites on which refusing consent was possible.}
    \label{tab:quantification_violations_cmp_gdpr_tlds}
    % \fbox{
    
  \begin{subfigure}{\columnwidth}
    \center
    \begin{minipage}[c]{0.45\textwidth}
      \caption{.uk}
    \end{minipage}
    \begin{minipage}[c]{0.54\textwidth}
      \begin{tabular}{|>{\rowmac}l|>{\rowmac}C{1.0cm}|>{\rowmac}p{1.75cm}|>{\rowmac}p{1.75cm}|>{\rowmac}p{1.75cm}|>{\rowmac}p{1.75cm}<{\clearrow}|}
        % \begin{tabular}{|l|p{0.7cm}|p{0.7cm}|p{0.7cm}|p{0.7cm}|p{0.7cm}|}
        \hline
        &  & \multicolumn{4}{c|}{\textbf{Violations}} \\
        \cline{3-6}
        % \textbf{CMP} & \textbf{Number of websites} & \thead{\textbf{No}\\\textbf{option}} & \thead{\textbf{Pre-}\\\textbf{ticked}} & \thead{\textbf{Pre-}\\\textbf{action}\\\textbf{consent}} & \thead{\textbf{Non-}\\\textbf{respect}\\\textbf{of de-}\\\textbf{cision}} \\
        \textbf{CMP} & \textbf{Number of websites} & \textbf{\preaction{}} & \textbf{\nooption{}} & \textbf{\preticked{}} & \textbf{\nonrespect{}} \\
        \hline
        Quantcast & 60 & \cellcolor{BrickRed!0} 0.0\% (0/60) & \cellcolor{BrickRed!0} 0.0\% (0/60) & \cellcolor{BrickRed!55} 55.0\% (33/60) & \cellcolor{BrickRed!0} 0.0\% (0/60) \\\hline
Sourcepoint & 21 & \cellcolor{BrickRed!0} 0.0\% (0/21) & \cellcolor{BrickRed!0} 0.0\% (0/21) & \cellcolor{BrickRed!100} 100.0\% (21/21) & \cellcolor{BrickRed!0} 0.0\% (0/21) \\\hline
OneTrust & 16 & \cellcolor{BrickRed!81} 81.2\% (13/16) & \cellcolor{BrickRed!6} 6.2\% (1/16) & \cellcolor{BrickRed!93} 93.3\% (14/15) & \cellcolor{BrickRed!6} 6.7\% (1/15) \\\hline
Evidon & 10 & \cellcolor{BrickRed!0} 0.0\% (0/10) & \cellcolor{BrickRed!10} 10.0\% (1/10) & \cellcolor{BrickRed!11} 11.1\% (1/9) & \cellcolor{BrickRed!22} 22.2\% (2/9) \\\hline
Ensighten & 7 & \cellcolor{BrickRed!0} 0.0\% (0/7) & \cellcolor{BrickRed!0} 0.0\% (0/7) & \cellcolor{BrickRed!100} 100.0\% (7/7) & \cellcolor{BrickRed!0} 0.0\% (0/7) \\\hline
Oath & 5 & \cellcolor{BrickRed!0} 0.0\% (0/5) & \cellcolor{BrickRed!0} 0.0\% (0/5) & \cellcolor{BrickRed!0} 0.0\% (0/5) & \cellcolor{BrickRed!0} 0.0\% (0/5) \\\hline
\textbf{others} & 18 & \cellcolor{BrickRed!22} 22.2\% (4/18) & \cellcolor{BrickRed!16} 16.7\% (3/18) & \cellcolor{BrickRed!33} 33.3\% (5/15) & \cellcolor{BrickRed!0} 0.0\% (0/15) \\\hline
\textbf{No consent string found} & 12 & \cellcolor{BrickRed!0} 0.0\% (0/12) & \cellcolor{BrickRed!8} 8.3\% (1/12) & \cellcolor{BrickRed!45} 45.5\% (5/11) & \cellcolor{BrickRed!0} 0.0\% (0/11) \\\hline
\setrow{\bfseries} all & 149 & \cellcolor{BrickRed!11} 11.4\% (17/149) & \cellcolor{BrickRed!4} 4.0\% (6/149) & \cellcolor{BrickRed!60} 60.1\% (86/143) & \cellcolor{BrickRed!2} 2.1\% (3/143) \\\hline

      \end{tabular}
      \vspace{0.2cm}
    \end{minipage}
  \end{subfigure}

  \begin{subfigure}{\columnwidth}
    \center
    \begin{minipage}[c]{0.45\textwidth}
      \caption{.fr}
    \end{minipage}
    \begin{minipage}[c]{0.54\textwidth}
      \begin{tabular}{|>{\rowmac}l|>{\rowmac}C{1.0cm}|>{\rowmac}p{1.75cm}|>{\rowmac}p{1.75cm}|>{\rowmac}p{1.75cm}|>{\rowmac}p{1.75cm}<{\clearrow}|}
        % \begin{tabular}{|l|p{0.7cm}|p{0.7cm}|p{0.7cm}|p{0.7cm}|p{0.7cm}|}
        \hline
        &  & \multicolumn{4}{c|}{\textbf{Violations}} \\
        \cline{3-6}
        % \textbf{CMP} & \textbf{Number of websites} & \thead{\textbf{No}\\\textbf{option}} & \thead{\textbf{Pre-}\\\textbf{ticked}} & \thead{\textbf{Pre-}\\\textbf{action}\\\textbf{consent}} & \thead{\textbf{Non-}\\\textbf{respect}\\\textbf{of de-}\\\textbf{cision}} \\
        \textbf{CMP} & \textbf{Number of websites} & \textbf{\preaction{}} & \textbf{\nooption{}} & \textbf{\preticked{}} & \textbf{\nonrespect{}} \\
        \hline
        Quantcast & 34 & \cellcolor{BrickRed!0} 0.0\% (0/34) & \cellcolor{BrickRed!0} 0.0\% (0/34) & \cellcolor{BrickRed!32} 32.4\% (11/34) & \cellcolor{BrickRed!0} 0.0\% (0/34) \\\hline
Didomi & 29 & \cellcolor{BrickRed!0} 0.0\% (0/29) & \cellcolor{BrickRed!0} 0.0\% (0/29) & \cellcolor{BrickRed!24} 24.1\% (7/29) & \cellcolor{BrickRed!0} 0.0\% (0/29) \\\hline
Commanders Act & 10 & \cellcolor{BrickRed!40} 40.0\% (4/10) & \cellcolor{BrickRed!0} 0.0\% (0/10) & \cellcolor{BrickRed!80} 80.0\% (8/10) & \cellcolor{BrickRed!0} 0.0\% (0/10) \\\hline
Sourcepoint & 8 & \cellcolor{BrickRed!0} 0.0\% (0/8) & \cellcolor{BrickRed!0} 0.0\% (0/8) & \cellcolor{BrickRed!0} 0.0\% (0/8) & \cellcolor{BrickRed!0} 0.0\% (0/8) \\\hline
OneTrust & 6 & \cellcolor{BrickRed!66} 66.7\% (4/6) & \cellcolor{BrickRed!0} 0.0\% (0/6) & \cellcolor{BrickRed!83} 83.3\% (5/6) & \cellcolor{BrickRed!0} 0.0\% (0/6) \\\hline
SIRDATA & 5 & \cellcolor{BrickRed!0} 0.0\% (0/5) & \cellcolor{BrickRed!0} 0.0\% (0/5) & \cellcolor{BrickRed!0} 0.0\% (0/5) & \cellcolor{BrickRed!0} 0.0\% (0/5) \\\hline
Chandago & 5 & \cellcolor{BrickRed!0} 0.0\% (0/5) & \cellcolor{BrickRed!0} 0.0\% (0/5) & \cellcolor{BrickRed!0} 0.0\% (0/5) & \cellcolor{BrickRed!0} 0.0\% (0/5) \\\hline
\textbf{others} & 39 & \cellcolor{BrickRed!5} 5.1\% (2/39) & \cellcolor{BrickRed!12} 12.8\% (5/39) & \cellcolor{BrickRed!70} 70.6\% (24/34) & \cellcolor{BrickRed!29} 29.4\% (10/34) \\\hline
\textbf{No consent string found} & 3 & \cellcolor{BrickRed!0} 0.0\% (0/3) & \cellcolor{BrickRed!0} 0.0\% (0/3) & \cellcolor{BrickRed!50} 50.0\% (1/2) & \cellcolor{BrickRed!0} 0.0\% (0/2) \\\hline
\setrow{\bfseries} all & 139 & \cellcolor{BrickRed!7} 7.2\% (10/139) & \cellcolor{BrickRed!3} 3.6\% (5/139) & \cellcolor{BrickRed!42} 42.1\% (56/133) & \cellcolor{BrickRed!7} 7.5\% (10/133) \\\hline

      \end{tabular}
      \vspace{0.2cm}
    \end{minipage}
  \end{subfigure}

  \begin{subfigure}{\columnwidth}
    \center
    \begin{minipage}[c]{0.45\textwidth}
      \caption{.it}
    \end{minipage}
    \begin{minipage}[c]{0.54\textwidth}
      \begin{tabular}{|>{\rowmac}l|>{\rowmac}C{1.0cm}|>{\rowmac}p{1.75cm}|>{\rowmac}p{1.75cm}|>{\rowmac}p{1.75cm}|>{\rowmac}p{1.75cm}<{\clearrow}|}
        % \begin{tabular}{|l|p{0.7cm}|p{0.7cm}|p{0.7cm}|p{0.7cm}|p{0.7cm}|}
        \hline
        &  & \multicolumn{4}{c|}{\textbf{Violations}} \\
        \cline{3-6}
        % \textbf{CMP} & \textbf{Number of websites} & \thead{\textbf{No}\\\textbf{option}} & \thead{\textbf{Pre-}\\\textbf{ticked}} & \thead{\textbf{Pre-}\\\textbf{action}\\\textbf{consent}} & \thead{\textbf{Non-}\\\textbf{respect}\\\textbf{of de-}\\\textbf{cision}} \\
        \textbf{CMP} & \textbf{Number of websites} & \textbf{\preaction{}} & \textbf{\nooption{}} & \textbf{\preticked{}} & \textbf{\nonrespect{}} \\
        \hline
        Quantcast & 39 & \cellcolor{BrickRed!12} 12.8\% (5/39) & \cellcolor{BrickRed!23} 23.1\% (9/39) & \cellcolor{BrickRed!20} 20.0\% (6/30) & \cellcolor{BrickRed!3} 3.3\% (1/30) \\\hline
iubenda & 20 & \cellcolor{BrickRed!0} 0.0\% (0/20) & \cellcolor{BrickRed!0} 0.0\% (0/20) & \cellcolor{BrickRed!0} 0.0\% (0/20) & \cellcolor{BrickRed!0} 0.0\% (0/20) \\\hline
Clickio & 14 & \cellcolor{BrickRed!0} 0.0\% (0/14) & \cellcolor{BrickRed!0} 0.0\% (0/14) & \cellcolor{BrickRed!0} 0.0\% (0/14) & \cellcolor{BrickRed!0} 0.0\% (0/14) \\\hline
Triboo Media & 10 & \cellcolor{BrickRed!0} 0.0\% (0/10) & \cellcolor{BrickRed!0} 0.0\% (0/10) & \cellcolor{BrickRed!0} 0.0\% (0/10) & \cellcolor{BrickRed!0} 0.0\% (0/10) \\\hline
OneTag & 9 & \cellcolor{BrickRed!0} 0.0\% (0/9) & \cellcolor{BrickRed!0} 0.0\% (0/9) & \cellcolor{BrickRed!100} 100.0\% (9/9) & \cellcolor{BrickRed!0} 0.0\% (0/9) \\\hline
Axel Springer & 6 & \cellcolor{BrickRed!83} 83.3\% (5/6) & \cellcolor{BrickRed!83} 83.3\% (5/6) & \cellcolor{BrickRed!100} 100.0\% (1/1) & \cellcolor{BrickRed!0} 0.0\% (0/1) \\\hline
\textbf{others} & 22 & \cellcolor{BrickRed!9} 9.1\% (2/22) & \cellcolor{BrickRed!4} 4.5\% (1/22) & \cellcolor{BrickRed!66} 66.7\% (14/21) & \cellcolor{BrickRed!9} 9.5\% (2/21) \\\hline
\textbf{No consent string found} & 3 & \cellcolor{BrickRed!0} 0.0\% (0/3) & \cellcolor{BrickRed!33} 33.3\% (1/3) & \cellcolor{BrickRed!0} 0.0\% (0/1) & \cellcolor{BrickRed!0} 0.0\% (0/1) \\\hline
\setrow{\bfseries} all & 123 & \cellcolor{BrickRed!9} 9.8\% (12/123) & \cellcolor{BrickRed!13} 13.0\% (16/123) & \cellcolor{BrickRed!28} 28.3\% (30/106) & \cellcolor{BrickRed!2} 2.8\% (3/106) \\\hline

      \end{tabular}
      \vspace{0.2cm}
    \end{minipage}
  \end{subfigure}

  \begin{subfigure}{\columnwidth}
    \center
    \begin{minipage}[c]{0.45\textwidth}
      \caption{.be}
    \end{minipage}
    \begin{minipage}[c]{0.54\textwidth}
      \begin{tabular}{|>{\rowmac}l|>{\rowmac}C{1.0cm}|>{\rowmac}p{1.75cm}|>{\rowmac}p{1.75cm}|>{\rowmac}p{1.75cm}|>{\rowmac}p{1.75cm}<{\clearrow}|}
        % \begin{tabular}{|l|p{0.7cm}|p{0.7cm}|p{0.7cm}|p{0.7cm}|p{0.7cm}|}
        \hline
        &  & \multicolumn{4}{c|}{\textbf{Violations}} \\
        \cline{3-6}
        % \textbf{CMP} & \textbf{Number of websites} & \thead{\textbf{No}\\\textbf{option}} & \thead{\textbf{Pre-}\\\textbf{ticked}} & \thead{\textbf{Pre-}\\\textbf{action}\\\textbf{consent}} & \thead{\textbf{Non-}\\\textbf{respect}\\\textbf{of de-}\\\textbf{cision}} \\
        \textbf{CMP} & \textbf{Number of websites} & \textbf{\preaction{}} & \textbf{\nooption{}} & \textbf{\preticked{}} & \textbf{\nonrespect{}} \\
        \hline
        OneTrust & 9 & \cellcolor{BrickRed!100} 100.0\% (9/9) & \cellcolor{BrickRed!0} 0.0\% (0/9) & \cellcolor{BrickRed!77} 77.8\% (7/9) & \cellcolor{BrickRed!11} 11.1\% (1/9) \\\hline
Didomi & 6 & \cellcolor{BrickRed!0} 0.0\% (0/6) & \cellcolor{BrickRed!0} 0.0\% (0/6) & \cellcolor{BrickRed!100} 100.0\% (6/6) & \cellcolor{BrickRed!0} 0.0\% (0/6) \\\hline
Quantcast & 5 & \cellcolor{BrickRed!20} 20.0\% (1/5) & \cellcolor{BrickRed!0} 0.0\% (0/5) & \cellcolor{BrickRed!60} 60.0\% (3/5) & \cellcolor{BrickRed!0} 0.0\% (0/5) \\\hline
\textbf{others} & 3 & \cellcolor{BrickRed!66} 66.7\% (2/3) & \cellcolor{BrickRed!33} 33.3\% (1/3) & \cellcolor{BrickRed!0} 0.0\% (0/1) & \cellcolor{BrickRed!0} 0.0\% (0/1) \\\hline
\textbf{No consent string found} & 4 & \cellcolor{BrickRed!0} 0.0\% (0/4) & \cellcolor{BrickRed!50} 50.0\% (2/4) & \cellcolor{BrickRed!100} 100.0\% (1/1) & \cellcolor{BrickRed!0} 0.0\% (0/1) \\\hline
\setrow{\bfseries} all & 27 & \cellcolor{BrickRed!44} 44.4\% (12/27) & \cellcolor{BrickRed!11} 11.1\% (3/27) & \cellcolor{BrickRed!77} 77.3\% (17/22) & \cellcolor{BrickRed!4} 4.5\% (1/22) \\\hline

      \end{tabular}
      \vspace{0.2cm}
    \end{minipage}
  \end{subfigure}

  \begin{subfigure}{\columnwidth}
    \center
    \begin{minipage}[c]{0.45\textwidth}
      \caption{.ie}
    \end{minipage}
    \begin{minipage}[c]{0.54\textwidth}
      \begin{tabular}{|>{\rowmac}l|>{\rowmac}C{1.0cm}|>{\rowmac}p{1.75cm}|>{\rowmac}p{1.75cm}|>{\rowmac}p{1.75cm}|>{\rowmac}p{1.75cm}<{\clearrow}|}
        % \begin{tabular}{|l|p{0.7cm}|p{0.7cm}|p{0.7cm}|p{0.7cm}|p{0.7cm}|}
        \hline
        &  & \multicolumn{4}{c|}{\textbf{Violations}} \\
        \cline{3-6}
        % \textbf{CMP} & \textbf{Number of websites} & \thead{\textbf{No}\\\textbf{option}} & \thead{\textbf{Pre-}\\\textbf{ticked}} & \thead{\textbf{Pre-}\\\textbf{action}\\\textbf{consent}} & \thead{\textbf{Non-}\\\textbf{respect}\\\textbf{of de-}\\\textbf{cision}} \\
        \textbf{CMP} & \textbf{Number of websites} & \textbf{\preaction{}} & \textbf{\nooption{}} & \textbf{\preticked{}} & \textbf{\nonrespect{}} \\
        \hline
        Quantcast & 12 & \cellcolor{BrickRed!0} 0.0\% (0/12) & \cellcolor{BrickRed!0} 0.0\% (0/12) & \cellcolor{BrickRed!33} 33.3\% (4/12) & \cellcolor{BrickRed!0} 0.0\% (0/12) \\\hline
\textbf{others} & 7 & \cellcolor{BrickRed!42} 42.9\% (3/7) & \cellcolor{BrickRed!14} 14.3\% (1/7) & \cellcolor{BrickRed!40} 40.0\% (2/5) & \cellcolor{BrickRed!0} 0.0\% (0/5) \\\hline
\textbf{No consent string found} & 6 & \cellcolor{BrickRed!0} 0.0\% (0/6) & \cellcolor{BrickRed!16} 16.7\% (1/6) & \cellcolor{BrickRed!50} 50.0\% (1/2) & \cellcolor{BrickRed!0} 0.0\% (0/2) \\\hline
\setrow{\bfseries} all & 25 & \cellcolor{BrickRed!12} 12.0\% (3/25) & \cellcolor{BrickRed!8} 8.0\% (2/25) & \cellcolor{BrickRed!36} 36.8\% (7/19) & \cellcolor{BrickRed!0} 0.0\% (0/19) \\\hline

      \end{tabular}
      \vspace{0.2cm}
    \end{minipage}
  \end{subfigure}

  \begin{subfigure}{\columnwidth}
    \center
    \begin{minipage}[c]{0.45\textwidth}
      \caption{.com}
    \end{minipage}
    \begin{minipage}[c]{0.54\textwidth}
      \begin{tabular}{|>{\rowmac}l|>{\rowmac}C{1.0cm}|>{\rowmac}p{1.75cm}|>{\rowmac}p{1.75cm}|>{\rowmac}p{1.75cm}|>{\rowmac}p{1.75cm}<{\clearrow}|}
        % \begin{tabular}{|l|p{0.7cm}|p{0.7cm}|p{0.7cm}|p{0.7cm}|p{0.7cm}|}
        \hline
        &  & \multicolumn{4}{c|}{\textbf{Violations}} \\
        \cline{3-6}
        % \textbf{CMP} & \textbf{Number of websites} & \thead{\textbf{No}\\\textbf{option}} & \thead{\textbf{Pre-}\\\textbf{ticked}} & \thead{\textbf{Pre-}\\\textbf{action}\\\textbf{consent}} & \thead{\textbf{Non-}\\\textbf{respect}\\\textbf{of de-}\\\textbf{cision}} \\
        \textbf{CMP} & \textbf{Number of websites} & \textbf{\preaction{}} & \textbf{\nooption{}} & \textbf{\preticked{}} & \textbf{\nonrespect{}} \\
        \hline
        Quantcast & 24 & \cellcolor{BrickRed!0} 0.0\% (0/24) & \cellcolor{BrickRed!0} 0.0\% (0/24) & \cellcolor{BrickRed!21} 21.7\% (5/23) & \cellcolor{BrickRed!0} 0.0\% (0/23) \\\hline
OneTrust & 12 & \cellcolor{BrickRed!58} 58.3\% (7/12) & \cellcolor{BrickRed!8} 8.3\% (1/12) & \cellcolor{BrickRed!72} 72.7\% (8/11) & \cellcolor{BrickRed!18} 18.2\% (2/11) \\\hline
Evidon & 8 & \cellcolor{BrickRed!0} 0.0\% (0/8) & \cellcolor{BrickRed!12} 12.5\% (1/8) & \cellcolor{BrickRed!42} 42.9\% (3/7) & \cellcolor{BrickRed!28} 28.6\% (2/7) \\\hline
Sourcepoint & 5 & \cellcolor{BrickRed!20} 20.0\% (1/5) & \cellcolor{BrickRed!0} 0.0\% (0/5) & \cellcolor{BrickRed!20} 20.0\% (1/5) & \cellcolor{BrickRed!20} 20.0\% (1/5) \\\hline
\textbf{others} & 36 & \cellcolor{BrickRed!8} 8.3\% (3/36) & \cellcolor{BrickRed!5} 5.6\% (2/36) & \cellcolor{BrickRed!58} 58.8\% (20/34) & \cellcolor{BrickRed!14} 14.7\% (5/34) \\\hline
\textbf{No consent string found} & 12 & \cellcolor{BrickRed!0} 0.0\% (0/12) & \cellcolor{BrickRed!16} 16.7\% (2/12) & \cellcolor{BrickRed!60} 60.0\% (3/5) & \cellcolor{BrickRed!0} 0.0\% (0/5) \\\hline
\setrow{\bfseries} all & 97 & \cellcolor{BrickRed!11} 11.3\% (11/97) & \cellcolor{BrickRed!6} 6.2\% (6/97) & \cellcolor{BrickRed!47} 47.1\% (40/85) & \cellcolor{BrickRed!11} 11.8\% (10/85) \\\hline

      \end{tabular}
      \vspace{0.2cm}
    \end{minipage}
  \end{subfigure}

  \end{table*}
}

We conducted an automatic crawl of \nba{} websites from 1~000 top Tranco websites for 
31 European TLDs and from \texttt{.com}, \texttt{.org} and \texttt{.eu} domains 
between September 20\textsuperscript{th} and September 23\textsuperscript{rd} 2019.
Among  reachable and authorized websites, \nbbanner{} (6.2\%) had a TCF banner (cookie banner of a CMP implementing the TCF). We show per-TLD details in Table~\ref{tab:tlds}. The \nbbanner{} websites that have a TCF banner are the target of the following automatic crawls.

%From this dataset, we selected \nbsa{} websites that belong to \texttt{.fr},  \texttt{.it}, \texttt{.uk}, \texttt{.be}, \texttt{.ie} or \texttt{.com} TLD and have a TCF banner and performed a semi-automatic crawl on them. \nbsarefusal{} websites (90.56 \% of them) had a banner offering a way to refuse per-purpose consent.

  \begin{table}[t]
    \def\arraystretch{0.95}% default: 1.0
    \center
    \caption{Distribution of websites with a TCF banner across European (and 3 international) TLDs, computed with an automatic crawl.}
    \label{tab:tlds}
    \begin{tabular}{|>{\rowmac}l|>{\rowmac}p{2.1cm}|>{\rowmac}p{2.25cm}|>{\rowmac}p{2.15cm}<{\clearrow}|}
      \hline
      \textbf{TLD} & \textbf{Number of domains in the Tranco top-1 million list} & \textbf{Number of reachable and allowed (for bots) domains} & \textbf{Number of domains with a TCF-related cookie banner} \\\hline
      .uk & 1 000 & 788 (78.8\%) &  \cellcolor{gray!56} 149 (18.9\%) \\
.fr & 1 000 & 815 (81.5\%) &  \cellcolor{gray!51} 139 (17.1\%) \\
.pl & 1 000 & 858 (85.8\%) &  \cellcolor{gray!45} 129 (15.0\%) \\
.it & 1 000 & 824 (82.4\%) &  \cellcolor{gray!44} 123 (14.9\%) \\
.es & 1 000 & 800 (80.0\%) &  \cellcolor{gray!42} 113 (14.1\%) \\
.nl & 1 000 & 838 (83.8\%) &  \cellcolor{gray!23} 65 (7.8\%) \\
.gr & 1 000 & 720 (72.0\%) &  \cellcolor{gray!22} 53 (7.4\%) \\
.pt & 1 000 & 793 (79.3\%) &  \cellcolor{gray!19} 52 (6.6\%) \\
.de & 1 000 & 881 (88.1\%) &  \cellcolor{gray!19} 56 (6.4\%) \\
.ro & 1 000 & 787 (78.7\%) &  \cellcolor{gray!19} 50 (6.4\%) \\
.bg & 547 & 449 (82.1\%) &  \cellcolor{gray!17} 26 (5.8\%) \\
.fi & 1 000 & 824 (82.4\%) &  \cellcolor{gray!17} 47 (5.7\%) \\
.no & 1 000 & 862 (86.2\%) &  \cellcolor{gray!16} 48 (5.6\%) \\
.dk & 1 000 & 824 (82.4\%) &  \cellcolor{gray!14} 41 (5.0\%) \\
.be & 1 000 & 863 (86.3\%) &  \cellcolor{gray!13} 38 (4.4\%) \\
.at & 1 000 & 873 (87.3\%) &  \cellcolor{gray!11} 33 (3.8\%) \\
.ie & 1 000 & 769 (76.9\%) &  \cellcolor{gray!9} 25 (3.3\%) \\
.cz & 1 000 & 916 (91.6\%) &  \cellcolor{gray!9} 28 (3.1\%) \\
.ch & 1 000 & 849 (84.9\%) &  \cellcolor{gray!9} 26 (3.1\%) \\
.se & 1 000 & 787 (78.7\%) &  \cellcolor{gray!8} 21 (2.7\%) \\
.sk & 1 000 & 879 (87.9\%) &  \cellcolor{gray!4} 14 (1.6\%) \\
.hr & 627 & 513 (81.8\%) &  \cellcolor{gray!4} 8 (1.6\%) \\
.hu & 1 000 & 794 (79.4\%) &  \cellcolor{gray!2} 6 (0.8\%) \\
.lu & 186 & 147 (79.0\%) &  \cellcolor{gray!2} 1 (0.7\%) \\
.lt & 745 & 605 (81.2\%) &  \cellcolor{gray!1} 4 (0.7\%) \\
.lv & 537 & 420 (78.2\%) &  \cellcolor{gray!1} 2 (0.5\%) \\
.si & 514 & 426 (82.9\%) &  \cellcolor{gray!1} 2 (0.5\%) \\
.is & 358 & 248 (69.3\%) &  \cellcolor{gray!1} 1 (0.4\%) \\
.ee & 468 & 373 (79.7\%) &  \cellcolor{gray!0} 1 (0.3\%) \\
.li & 105 & 62 (59.0\%) &  \cellcolor{gray!0} 0 (0.0\%) \\
.cy & 108 & 76 (70.4\%) &  \cellcolor{gray!0} 0 (0.0\%) \\
.mt & 62 & 37 (59.7\%) &  \cellcolor{gray!0} 0 (0.0\%) \\
\hline
.com & 1 000 & 711 (71.1\%) &  \cellcolor{gray!40} 97 (13.6\%) \\
.org & 1 000 & 735 (73.5\%) &  \cellcolor{gray!6} 16 (2.2\%) \\
.eu & 1 000 & 803 (80.3\%) &  \cellcolor{gray!4} 12 (1.5\%) \\
\hline
\setrow{\bfseries} all & 28 257 & 22 949 (81.2\%) &  \cellcolor{gray!18} 1 426 (6.2\%) \\
\hline

    \end{tabular}
  %\SHORTENTAB
  \end{table}
{}

\begin{figure}[t]
  \center
  \includegraphics[width=0.45\textwidth]{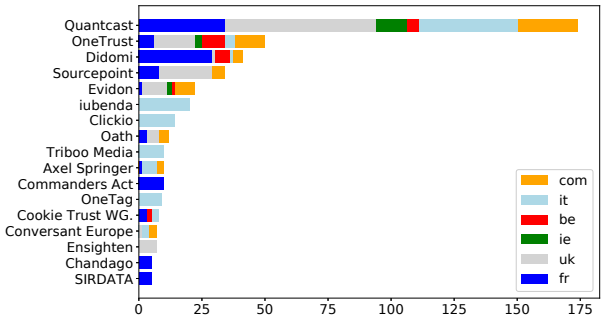}
  \caption{Distribution of the identified CMPs seen at least 5 times during the semi-automatic crawl.}
  \label{fig:cmps_distribution}
  \SHORTEN
\end{figure}

We extract information from the consent strings to identify the CMP present on a website.
As not all websites were setting up a consent string upon our visit (see our methodology in Section~\ref{sec:methodology}), and some consent strings contain an incorrect CMP ID, we have been able to identify the CMP company behind a TCF banner for 
%707 (53.32\%)
298 (20.9\%)
websites in the automatic crawl, and
%409 (94.676 \%)
511 (92.9\%)
websites in the semi-automatic crawl.
We represent the distribution of identified CMPs in the semi-automatic crawl in Figure~\ref{fig:cmps_distribution}.
The most encountered CMP is Quantcast, far beyond OneTrust, Didomi and Sourcepoint.

We have not found any implementation of the version 2 of TCF that came out in August 20\textsuperscript{th} 2019.

%==========================================================
\section{Quantification of Suspected Violations}
\label{sec:results}

In this section, we comment on the results regarding the main suspected violations of the GDPR and the ePD described in section~\ref{sec:violations}. These suspected violations concern consent strings obtained using the standard API and shared cookie (see section~\ref{sec:methodology:methods}). 
We note that there is a joint responsibility between publishers and CMPs for these suspected violations.
Violations related to consent strings seen in GET and POST requests are shown in section~\ref{sec:results:queries}, because 
%we cannot attribute their responsibility to the CMP or the publisher.
we cannot attribute CMPs to these cases.

\subsection{Detected GDPR and ePD Suspected Violations}
\label{sec:results:violations}
\subsubsection{Overview of Suspected Violations}
%\label{sec:results:violations}

We show a summary of the main suspected violations' prevalence, depending on the number of purposes in the consent strings, in Table~\ref{tab:res:purposes}. As a reminder, we consider violations of \preaction{} when we find a consent string with 1 to 5 purposes set, but only when 5 purposes are set for \nonrespect{}.

  \begin{table}[htbp]
    \center
    \setlength\tabcolsep{2.5pt}
    \caption{Number of websites seen with the different violations, w.r.t. the maximum number of purposes in observed consent strings. Considered violating cases are shown in bold.}
    \label{tab:res:purposes}
    \begin{tabular}{|p{1.5cm}|p{1.85cm}|p{1.50cm}|p{1.70cm}|p{1.50cm}|}
      \hline
      \textbf{Number of purposes} & \textbf{\preaction{}} & \textbf{\nooption{}} & \textbf{\preticked{}} & \textbf{\nonrespect{}} \\\hline
      1 to 4 purposes & \textbf{2.1\%} (30/1426) & - & - & 6.7\% (34/508) \\\hline
5 purposes & \textbf{7.8\%} (111/1426) & - & - & \textbf{5.3\%} (27/508) \\\hline
\textbf{Total number of violations} & 9.9\% (141/1426) & 6.8\% (38/560) & 46.5\% (236/508) & 5.3\% (27/508) \\\hline

      \addlinespace[1mm]
      \hline
      \textbf{Any violation} & \multicolumn{4}{c|}{54.29\% 304/560} \\
      \hline
    \end{tabular}
  \end{table}
{}

We find examples of websites for all considered violations. We find that 38 (6.8\%) websites do no provide any way to refuse consent. 236 (46.5\%) websites pre-tick the purpose or vendor options. 141 websites (9.9\%) set a consent string with 1 to 5 purposes before any user action. 27 websites (5.3\%) set a consent string with 5 purposes even though the user refused consent.

\subsubsection{Quantification per Suspected Violation}

Table~\ref{tab:quantification_violations_cmp_gdpr_per_violation} shows the results for each suspected violation, grouped by CMP seen performing a violation at least 3 times in the semi-automatic crawl. For websites for which we found a suspected violation, we provide their global Tranco rank.
To compute the results of the \preticked{} and \nonrespect{} violations, we only consider websites having a banner proposing a way to refuse consent (\nbsarefusal{} websites), i.e. we exclude banners having the \nooption{} violations (38 websites), and broken/missing banners (14 websites).
%Percentages in the different tables are calculated out of this number.
%For the \preaction{} violation, we show per-TLD results in the automatic crawl in Table~\ref{tab:tlds_preaction}.

\tablesperviolation{}

  \begin{table}[htbp]
    \center
    \caption{Results of the \preaction{} violation on \nbbanner{} websites via an automatic crawl.}
    \label{tab:tlds_preaction}
%    \parbox{.45\linewidth}{
      \begin{tabular}{|>{\rowmac}l|>{\rowmac}p{2.2cm}<{\clearrow}|}
        \hline
        \textbf{TLD} & \textbf{Number of websites} \\\hline
        .uk & \cellcolor{BrickRed!11} 11.4\% (17/149) \\
.fr & \cellcolor{BrickRed!7} 7.2\% (10/139) \\
.pl & \cellcolor{BrickRed!17} 17.8\% (23/129) \\
.it & \cellcolor{BrickRed!9} 9.8\% (12/123) \\
.es & \cellcolor{BrickRed!7} 7.1\% (8/113) \\
.nl & \cellcolor{BrickRed!0} 0.0\% (0/65) \\
.gr & \cellcolor{BrickRed!3} 3.8\% (2/53) \\
.pt & \cellcolor{BrickRed!1} 1.9\% (1/52) \\
.de & \cellcolor{BrickRed!28} 28.6\% (16/56) \\
.ro & \cellcolor{BrickRed!2} 2.0\% (1/50) \\
.bg & \cellcolor{BrickRed!3} 3.8\% (1/26) \\
.fi & \cellcolor{BrickRed!10} 10.6\% (5/47) \\
.no & \cellcolor{BrickRed!2} 2.1\% (1/48) \\
.dk & \cellcolor{BrickRed!4} 4.9\% (2/41) \\
.be & \cellcolor{BrickRed!42} 42.1\% (16/38) \\
.at & \cellcolor{BrickRed!12} 12.1\% (4/33) \\
.ie & \cellcolor{BrickRed!12} 12.0\% (3/25) \\
\hline

      \end{tabular}
%    }
    ~
%    \parbox{.45\linewidth}{
      \begin{tabular}{|>{\rowmac}l|>{\rowmac}p{2.2cm}<{\clearrow}|}
        \hline
        \textbf{TLD} & \textbf{Number of websites} \\\hline
        .cz & \cellcolor{BrickRed!3} 3.6\% (1/28) \\
.ch & \cellcolor{BrickRed!0} 0.0\% (0/26) \\
.se & \cellcolor{BrickRed!9} 9.5\% (2/21) \\
.sk & \cellcolor{BrickRed!7} 7.1\% (1/14) \\
.hr & \cellcolor{BrickRed!0} 0.0\% (0/8) \\
.hu & \cellcolor{BrickRed!0} 0.0\% (0/6) \\
.lu & \cellcolor{BrickRed!0} 0.0\% (0/1) \\
.lt & \cellcolor{BrickRed!25} 25.0\% (1/4) \\
.lv & \cellcolor{BrickRed!0} 0.0\% (0/2) \\
.si & \cellcolor{BrickRed!0} 0.0\% (0/2) \\
.is & \cellcolor{BrickRed!0} 0.0\% (0/1) \\
.ee & \cellcolor{BrickRed!0} 0.0\% (0/1) \\
\hline
.com & \cellcolor{BrickRed!11} 11.3\% (11/97) \\
.org & \cellcolor{BrickRed!12} 12.5\% (2/16) \\
.eu & \cellcolor{BrickRed!8} 8.3\% (1/12) \\
\hline
\setrow{\bfseries} all & \cellcolor{BrickRed!9} 9.9\% (141/1426) \\
\hline

      \end{tabular}
%    }
    \SHORTENTAB
  \end{table}
{}

\textbf{\preaction{}}: Table~\ref{tab:tlds_preaction} shows  results of the automatic crawl per TLD. 
We observe 141 websites registering a consent string that contains a positive consent even though the user did not perform any action. 111 of them contain all of the TCF's purposes. This is a striking abuse of the framework, happening on more than 1 in 10 websites using it. 
Interestingly, according to the TCF specification, the APIs we have used to detect consent string  
\emph{should not return the consent string before the user gives their decision on consent} 
(or consent is retrieved from existing cookies)~\cite{iab-jsapi}.

\textbf{\nooption{}}: We observe 38 websites offering no option to refuse consent. These website take part in a framework about user's consent collection, but do not actually offer a way to refuse consent. Collected consent cannot be considered \textit{free}, as required by the GDPR.

\textbf{\preticked{}}: Almost half of tested websites (236 out of 508) pre-select choices. In the Planet49 case~\cite{Planet49} announced few days after we finished the crawling campaigns, % (settled after the crawls), 
the European court of Justice decided that such pre-selected choices 
%do not lead to a valid consent.
lead to an invalid consent.

\textbf{\nonrespect{}}: 27 websites register a positive consent even though the user refused consent. This strikingly violates user's choice, the framework, and the GDPR.

We observe a variety of suspected violations among the different
CMPs. Interestingly, violations are often seen on a partial number of
websites. This shows that CMPs offer several versions of their banners
that behave differently. We further discuss the shared
responsibility of violations between CMPs and publishers in
section~\ref{sec:discussion}.

%Some violations are more prevalent in countries where pre-GDPR laws were different. For instance, more Italian websites offer no way to opt out than the average, but less of them pre-select purposes.
%Many CMPs are in a grey area, i.e., only part of their banners perform violations. We discuss the responsibility of CMPs and publisher in Section~\ref{sec:discussion}.

%\subsubsection{Quantification per CMP}

%In Table~\ref{tab:quantification_violations_cmp_gdpr}, we show the quantification of violations grouped by CMP. This table displays CMPs seen at least 5 times during the semi-automatic crawl.

%\cmptablegdpr{}

%The attribution of the responsibility of the remaining violations to CMPs only is up to interpretation 
%using table~\ref{tab:quantification_violations_cmp_gdpr}, showing the amount of violating websites for each CMP.

We give additional presentations of the results (per-country and per-CMPs views) in Appendix~\ref{app:altpres}.

\subsubsection{Quantification per Publisher}

We observe suspected violations on a wide range of websites. For each suspected violation, we display the lists of top 10 violating websites, ordered by their rank in the Tranco list in Table~\ref{tab:violating_websites}.
\texttt{msn.com}, a web portal ranked 48 in the Tranco list, stores a positive consent before any user choice, then offers no way to opt out.
\texttt{medicalnewstoday.com}, a website about health, does the same, even though medical information is a sensitive category of data.
\texttt{w3schools.com}, a popular website providing web development tutorials, displays a banner with pre-selected choices, but registers a positive consent even if the user goes to the trouble of deselecting them.
\texttt{softonic.com}, website of a major software developer, registers a positive consent before user action, then displays a banner with pre-selected choices, and finally does not respect the user's decision.

\tablesviolatingwebsite{}

\subsection{Escalation of Suspected Violations with the Shared Consent Mechanism}
\label{sec:results:shared_cookie}

Setting a violating consent string in a cookie shared among all TCF websites would constitute an escalation of the problem. We investigate the question: to what extent do websites use the shared cookie?
As explained in section~\ref{sec:methodology:methods}, we try to read it using a browser extension after both giving a positive consent and refusing consent in the semi-automatic crawl. We observe 126 (22.9\%) websites setting the shared cookie.

We then estimate how many websites access and reuse the shared cookie.
We place a custom cookie (respecting the specification) in the browser, query the CMP using the standard APIs, and see if the CMP returns the exact same consent string (with no banner interaction).
Using this protocol, 62 (4.3\%) websites return the same consent strings. This means that CMPs on these websites reuse the shared cookie, even if it has been created by another CMP. This constitutes a lower bound, because CMPs can return another consent string than the one stored in the cookie, and may ignore ours for various reasons (e.g. an unexpected vendor list version).
%We cannot know how many websites read this cookie without a debugger on the browser level, which is out of scope of the paper. However, we can quantify one way to access this cookie by third parties: CMP consent redirecting~\cite{iab-urlbased}.

We also estimate how many websites access the shared cookie by studying how many of them use the HTTP redirect mechanism described in section~\ref{sec:background:consentsharing} to do so.
We first observe that many consent redirecting domains do not respect the specification. Indeed, during manual inquiry, we find redirecting schemes using different values for the GET parameter specifying the redirection URL. For example, on \texttt{mirror.co.uk} we observed a GET request with a \texttt{gdpr\_consent\_string} parameter instead of \texttt{gdpr\_consent}. As we cannot cover these cases exhaustively, we focus on those respecting the specification.
The only domain we observe doing so (\texttt{sddan.consensu.org}, owned by the SIRDATA CMP) is used on 53 (9.5\%) websites during the semi-automatic crawl.
%We observe disparities between countries: 31 \texttt{.fr} (22.3\%) websites in the semi-automatic crawl, 21 (14.1\%) \texttt{.uk} domains and 1 (0.8\%) \texttt{.it} domain (and none for \texttt{.be}, \texttt{.ie} and \texttt{.com} domains).
This hints that the practice of reading the shared consent cookie is quite common. % at least for French websites.
% found on 22 fr websites (17.60%), 2 uk (1.43%), 1 it (0.82%)

%3 websites (0.60 \%) set a fully positive consent (all purposes set) in the shared cookie when the user explicitly refuses consent, and 19 (3.81 \%) sites do so with 1 to 5 purposes. Moreover, 3 websites (0.21 \%) set a consent string with all purposes before any user interaction with the cookie banner.

%Having a look at the way consent is passed to advertisers, the \textit{cookie} case (website setting the consent string in a cookie shared across CMP and publishers) is particularly troubling in terms of privacy: since this consent is reused among different publishers, it constitutes an escalation of the problem.
We observe 3 websites setting the shared cookie in the \preaction{} case, 3 in the \nonrespect{} case with 5 purposes, and 20 (3.9\%) with 1 to 5 purposes.

Visiting one of the 3 websites on which the cookie is set before any user action on the banner will automatically set a global positive consent cookie. Visiting one of the 20 websites that do not respect user decision will set a global positive consent cookie against the user's decision. 
This is particularly troubling in terms of privacy: since this consent is reused among different publishers, it constitutes an escalation of the problem.
We discuss this further in Section~\ref{sec:discussion}.

%\nc{todo: check next paragraph (test websites)}
%Even more troublesome, we observe one website (\texttt{fcinternews.it}) setting a positive consent in the shared cookie only when the user responds with a positive consent, and not when they refuse consent. In other words, this website only propagates positive consents and ignore negative ones.

%\section{Additional results}
\section{Measuring Third-Party Requests: Presence of Consent Strings and Third-Party Trackers}
\label{sec:additional}

In previous sections, we studied violations in consent strings obtained via the standard API and shared cookies, 
as described in Section~\ref{sec:background:consentsharing}. Responsibility of such violations can be 
attributed to CMPs and publishers (see the discussion in Section~\ref{sec:discussion}). 
However, when we find a non-compliant consent string via a URL-based method, 
%we cannot assign the responsibility of this request on any CMP, as such request could be emitted by any third-party script present on the website.
we have no way to know whether that consent string was legitimately transmitted %to the script 
by the CMP or any other third party present on the page. %and not a creation of the script itself. 
%Hence, we separate this case from previous violations, so as not to attribute violations to CMPs and publishers incorrectly.

In this section, we study third-party requests observed in the two crawls. 
We first analyse the consent strings transmitted via URL-based methods, and 
then measure how many third-party trackers are present on the page before user actions, after acceptance and after refusal of consent. 

\subsection{%Consent strings transmitted by third parties in GET and POST requests}
Third-Party Requests with Consent Strings}
\label{sec:results:queries}

%In previous sections, we studied violations in consent strings obtained via the standard API, and shared cookies, as described in Section~\ref{sec:background:consentsharing}. Responsibility of such violations can be attributed to CMPs and publisher (see the discussion in Section~\ref{sec:discussion}). However, when we find a non-compliant consent string in a request,
%%we cannot assign the responsibility of this request on any CMP, as such request could be emitted by any third-party script present on the website.
%we have no way to know whether that consent string was legitimately transmitted to the script by the CMP and not a creation of the script itself. Hence, we separate this case from previous violations, so as not to attribute violations to CMPs and publishers incorrectly.

In this section, we detect the four suspected GDPR and ePD violations 
%(see Section~\ref{sec:violations} for description of violations and Section~\ref{sec:methodology} describing how we detect them) 
by analyzing consent strings that we observed in GET and POST requests to third parties.
%We then quantify violations based on these consent strings. 
%To detect \preaction{}, we collect consent strings without performing any action on the website. To detect \nonrespect{}, we collect consent strings after manually giving a negative consent on the banner.
%

We observed consent strings with positive consent (1 to 5 allowed purposes) 
in GET or POST requests before any user action on
%in the \preaction{} case on 
151 (10.6\%) websites out of \nbasuccess{} websites in the automatic crawl -- 
this indicates websites with a \preaction{} violation. 
For the \nonrespect{} violation, we intercepted consent strings in GET or POST requests 
with 5 purposes on 63 websites (12.4\%).
To evaluate whether these results are complementary to our previous findings, 
we count the number of websites in which we see a violating consent string in GET and POST requests,
 but do not obtain a violating consent string via intercepting the standard APIs or in 
 the shared cookie.
 % or using methods to interrogate the CMP (using direct calls to the \cmp{} function or postMessages, 
% (see section~\ref{sec:background:consentsharing}).
%(see Section~\ref{sec:methodology:methods}). 

%It appears that in the \preaction{} violation, %69 of these websites (4.84 \% of the TCF websites) 
{\bf \preaction{}:} In addition to 66 websites where we observed this violation while intercepting 
consent strings using the standard APIs and the shared cookie, we observed it also 
on additional 69 websites, where GET or POST requests send consent strings with a positive consent (1 to 5 purposes).
% that we do not %otherwise obtain. 
%obtain with standard APIs and share cookie. 
It means that 
%In other words, 
requests containing violating consent strings are sent %by unknown scripts 
while the CMP has not provided a consent string yet.

{\bf \nonrespect{}: } In addition to 27 websites where we observed this violation while intercepting 
consent strings with the standard APIs and the shared cookie, we observed it also 
on additional 26 websites where we obtain consent strings with all 5 purposes 
in GET and POST requests.

We further investigated whether the identifiers of the responsible CMPs (CMP ID) for each consent string 
obtained via GET and POST requests match the CMP IDs obtained from consent strings 
with the standard APIs and the shared cookie.  
%and compared it against CMP IDs 
%As already mentioned in section~\ref{sec:methodo:cmpid}, we strangely 
We found CMP IDs in GET and POST requests %which were not coherent with 
different from the ones found using the standard APIs on 48 websites. 
In 37 of them, both CMP IDs found were from valid CMPs, 
while in the remaining 11 websites, 
CMP IDs were set to either 0, 1 or 4095, which do not exist in the CMP public list~\cite{iab-cmplist}.
%In of them, CMP IDs in both cases are valid (correspond to actual CMPs).
It seems suspicious that consent strings not created by the website's actual CMP 
(or even non-existent CMPs) are sent to third parties.

\subsection{Third-party trackers} 
We %quantify the 
measure the number of third-party trackers on %TCF-
websites with TCF banners depending on user consent: before any user action, after refusing consent and after a positive consent.
To do so, we logged every request to third-party domains with \scname{}. From this, we extract domains which are considered trackers in the Disconnect list~\cite{disconnectlist}.

We first %report on 
measure the number of %trackers
third-party tracking requests without responding to the cookie banner or doing any other action on the website. %Secondly, 
Then we count %trackers 
third-party tracking requests after both giving a positive consent and refusing consent to the cookie banner 
(for websites on which it is possible), and reloading the page.
Each measurement of trackers is done in a single browser session, on a single page load. 
These tests are done on the \nbsarefusal{} websites on which refusing consent is
possible in the semi-automatic crawl. 

\begin{table}[t]
  \center
  \caption{Average number of third-party tracking requests per website before user action, 
  after a positive and after refusing consent.}
  \label{tab:nb_trackers}
  \begin{tabular}{|l|p{2.5cm}|p{2.5cm}|}
    \hline
    \textbf{User action} & \textbf{Number of third party tracking requests}  & \textbf{Total number of third party requests} \\
    \hline
    Before user action & \cellcolor{BrickRed!23} 22.54 & 35.04 \\
    After refusing consent & \cellcolor{BrickRed!29} 28.78 & 42.50 \\
    After a positive consent & \cellcolor{BrickRed!40} 39.59 & 56.75 \\
    \hline
  \end{tabular}
  \SHORTENTAB
\end{table}

%Before any user action, we observe an average of 22.54 trackers per website, out of 35.04 third-party domains.
%After refusing consent, we observe an average of 28.78 trackers, for a total of 42.50 third-party domains. We observe at least one tracker in 96.26 \% of domains after a negative consent.
%After a positive consent, we observe an average of 39.59 trackers, out of 56.75 third-party domains.
%In total (during all tests, i.e. across several browser sessions), we observe an average of 45.09 trackers, out of 65.94 third-party domains.
Table~\ref{tab:nb_trackers} summarizes the results. We observe that giving consent on TCF banners, whether it's a positive or a refusal, has an effect on the number of included %trackers and third parties. 
third-party trackers.
Surprisingly, even refusing consent increases the number of %trackers. 
tracking requests. The number of websites having the \nonrespect{} violation 
(and hence setting a positive consent even if the user refused)
 is not sufficient to explain this increase. 
 We estimate that some scripts, in order to execute and include content, wait for the \cmp{} function to be defined, which should only happen after the user has given their choice to the banner~\cite{iab-jsapi}.
% first figure is obtained using --automatic-crawl option

\begin{table}[t]
  \center
  \caption{Top 20 tracking companies %present on websites 
  observed on \nbsarefusal\ websites
  after refusing consent and a page reload.}
  \label{tab:top_trackers}
  \begin{tabular}{|l|c|c|}
    \hline
    \textbf{Tracking company} & \textbf{Number of websites} & \textbf{TCF Vendor?} \\
    \hline
    Google & \cellcolor{BrickRed!96} 491 (96.7\%) &  \\\hline
AppNexus & \cellcolor{BrickRed!70} 356 (70.1\%) & \ding{51} \\\hline
Facebook & \cellcolor{BrickRed!66} 337 (66.3\%) &  \\\hline
RubiconProject & \cellcolor{BrickRed!58} 299 (58.9\%) & \ding{51} \\\hline
comScore & \cellcolor{BrickRed!55} 280 (55.1\%) & \ding{51} \\\hline
Integral Ad Science & \cellcolor{BrickRed!50} 258 (50.8\%) & \ding{51} \\\hline
Amazon.com & \cellcolor{BrickRed!47} 239 (47.0\%) &  \\\hline
Casale Media & \cellcolor{BrickRed!46} 237 (46.7\%) & \ding{51} \\\hline
Criteo & \cellcolor{BrickRed!45} 232 (45.7\%) & \ding{51} \\\hline
Adform & \cellcolor{BrickRed!45} 230 (45.3\%) & \ding{51} \\\hline
Yahoo! & \cellcolor{BrickRed!43} 221 (43.5\%) & \ding{51} \\\hline
OpenX & \cellcolor{BrickRed!42} 217 (42.7\%) & \ding{51} \\\hline
The Trade Desk & \cellcolor{BrickRed!42} 217 (42.7\%) & \ding{51} \\\hline
Quantcast & \cellcolor{BrickRed!39} 202 (39.8\%) & \ding{51} \\\hline
MediaMath & \cellcolor{BrickRed!39} 199 (39.2\%) & \ding{51} \\\hline
DataXu & \cellcolor{BrickRed!37} 192 (37.8\%) & \ding{51} \\\hline
Adobe & \cellcolor{BrickRed!37} 190 (37.4\%) & \ding{51} \\\hline
PubMatic & \cellcolor{BrickRed!36} 186 (36.6\%) & \ding{51} \\\hline
SmartAdServer & \cellcolor{BrickRed!35} 179 (35.2\%) & \ding{51} \\\hline
SiteScout & \cellcolor{BrickRed!32} 165 (32.5\%) & \ding{51} \\\hline

  \end{tabular}
  \SHORTENTAB
\end{table}

Table~\ref{tab:top_trackers} shows 
%We display 
the top 10 companies that own tracking domains present on websites after refusing consent 
(and a page reload). % in Table~\ref{tab:top_trackers}. 
We matched %tracker 
tracking domains to company names using the Disconnect list~\cite{disconnectlist}. 
We find whether they are part of the TCF by checking if any company name linked to a 
tracker domain in WebXRay's database~\cite{libert2015exposing} is present in the Global Vendor List (version 168).
Some top trackers belong to vendors which are not part of the IAB framework 
(Google, Facebook or Amazon), but the rest of them are (eg. AppNexus, The Rubicon Project, comScore, etc.).

%\begin{table}[h]
%  \center
%  % \setlength{\tabcolsep}{4pt}
%  \caption{Top 20 tracking companies present on websites after a negative consent and a page reload.}
%  \label{tab:top_trackers}
%  \begin{tabular}{|l|c|c|}
%    \hline
%    \textbf{Tracking company} & \textbf{Number of websites} & \textbf{TCF Vendor?} \\
%    \hline
%    \input{tables/top_trackers}
%  \end{tabular}
%\end{table}

During our study, we encountered many unusual cases, detailed in appendix~\ref{app:unusual_cases}.

%==========================================================
\section{Browser Extension}

\begin{figure}[t]
  \center
  \caption{Interface of our browser extension, \extname{}, displaying the consent transmitted by CMPs to advertisers.}
  \label{fig:cookieglasses}
  \includegraphics[width=0.32\textwidth]{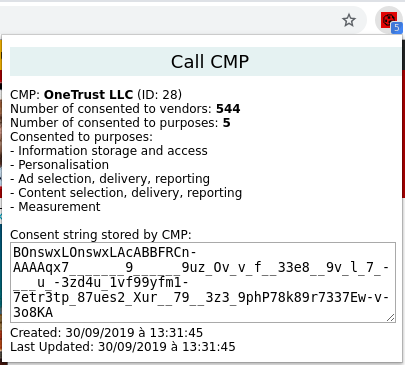}
  \SHORTEN
\end{figure}

We publish a browser extension called \extname{}~\cite{Cookie-Glasses}, to enable users to see if consent stored by CMPs corresponds to their choice.
Users can read information stored in the consent string provided by the CMP in a simple interface.
%, visible in Figure~\ref{fig:cookieglasses}.
The most important pieces of information present in the consent string are decoded and displayed in a readable format (see Figure~\ref{fig:cookieglasses}).

%This extension pretends to be an advertiser and queries the CMP for the user's consent. This makes it possible for user to easily check whether consent transmitted to third parties is consistent with the choice they made on the cookie banner.

Technically, the extension uses postMessages from the standard APIs (see \two{} in Figure~\ref{fig:detect_consent}). It is not possible to use direct calls to the \cmp{} function because of browser security mechanisms. Our tests show that 79\% of TCF websites use the postMessage API. Our extension therefore works on a majority of websites. For the remaining websites, we propose a workaround to manually execute the JS code to obtain the consent string, and decode it with the browser extension.
%Thus, only one of the methods described in section~\ref{sec:methodology:methods} to obtain consent strings is available.

%\nc{(We could read the cookie though)}

%==========================================================
%==========================================================
\section{Limitations}
\label{sec:limitations}

%In this section, we list limitations to our study.

%\subsubsection{Limited to banners that implement IAB Europe TCF}
Our work has some limitations.
First, our scope is limited to banners of IAB Europe's TCF.
Since we do not detect other cookie banners, we only observe a subset of all cookie banners.
Besides, our results on the prevalence of TCF banners represent a lower bound on the actual usage of TCF banners, due to a variety of implementations of the TCF.
For instance, some banners do not define the \cmp{} function on the first page load. In one of its banners (e.g. on \texttt{aol.com}), the Oath CMP redirects the user to another website (of a different domain) to display a consent wall. On this page, the \cmp{} function is not displayed. 
We do not detect such cases in our study. 
%For example, website \texttt{notizie.it} does not load the \cmp{} function on the main page, but displays the banner. We observe queries with consent strings that contain a positive consent before the user has made any choice in the cookie banner.
While we detect TCF banners on 17.1\% of \texttt{.fr} websites, van Eijk et al.~\cite{eijk2019impact} found that 40.2\% of European websites have a cookie banner. 
%, and about 60\% for \texttt{.fr} websites alone. 

%\subsubsection{Websites that do not tamper with JavaScript APIs}
%Secondly, our method to detect violations actively interacts with websites and might introduce changes in their original functionality.
%\NB{I don't find where in the methodology we described this "active interaction".}
%CMPs willing to detect and tamper with our auditing scripts %tests or hide results from our tests could do so. In our experiments, observed a website \texttt{kayak.fr} overriding the \texttt{console.log()} function, which made our script fail to catch the consent string automatically. 

Secondly, results of our semi-automatic crawl are prone to errors due to dark patterns.
Most banners we encountered %presented dark patterns, nudging 
nudge users towards accepting consent: % Some of these patterns 
some of them make it particularly difficult to opt out%refuse tracking%, adding potential human errors when responding to a banner
\footnote{For instance,  an uncolored link is hidden in the middle of a 28~000-word-long privacy policy on \texttt{liberation.fr}, accessed on October 25\textsuperscript{th} 2019).}.
%(for instance, hiding an uncolored link in the middle of a 28~000-word-long privacy policy on \texttt{liberation.fr}, accessed on October 25\textsuperscript{th} 2019).
As a consequence, results of our semi-automatic crawls are prone to errors due to these dark patterns.
To limit such errors, we cross-checked answers to banners by three human operators (see Section~\ref{sec:verification}). Nonetheless, it is still possible that some banners are designed in such a confusing way that %these 
the three persons failed to find the proper way to opt out. %refuse tracking. 
We argue that banners %giving 
that give a technical mean to refuse consent, but %making 
make it so difficult that three computer science researchers do not find it, 
%cannot collect a free and unambiguous consent, and are therefore 
are still in violation with the GDPR.
\ifextended
Additionally, we have not measured the usability aspect of refusing consent for all the banners in our study. We detail the human operator procedures in the Appendix~\ref{sec:human_operator}.
\fi

%Third, in our results (Tables~\ref{tab:quantification_violations_cmp_gdpr_per_violation} and~\ref{tab:tlds_preaction}) we have not separated the cases  when an advertiser relies solely on the legal basis of legitimate interest. We underline that such practices are questionable as discussed in Section~\ref{sec:legal} and leave it for further analysis to the Data Protection Authorities. For every CMP we report in Table~\ref{tab:quantification_violations_cmp_gdpr_per_violation}, we have found at least one publisher for which violations are found independently from this corner case.

%Fourth, our semi-automatic crawl involves human operators, however we have not evaluated the usability aspect of refusing consent for all the banners in our study. We detail the human operator procedures in the Appendix~\ref{sec:human_operator}.

Finally, we only detect violation in client-side consent strings. 
Yet, %out-of-browser 
exchanges of consent strings can also happen %as well. 
outside of the browser. 
IAB %also 
provides extension fields~\cite{iab-extensionfields} for exchanging consent string in its OpenRTB protocol~\cite{OpenRTB}. This protocol is used between ad exchanges and advertisers for Real-Time Bidding. % (RTB). 
As such communication happens server-to-server, we cannot observe it with a client-side approach.

\section{Discussion}
\label{sec:discussion}

%In our work, we analyzed the legal sources and defined four violations in cookie banners.
%We then technically detected and measured  prevalence  of such violations in websites that contain TCF banners. 
In this section, we reflect on our experiments and our results and comment on open problems that can 
be addressed by legal professionals or DPAs. 

%In our work, we analyzed the legal sources and defined four violations in cookie banners. 
%We then technically detected and measured  prevalence  of such violations in websites 
%that contain IAB Europe TCF banners. 
%%
%%Our results are purely technical. 
%%During our experiments, we encountered a large ``grey area'' of cases open for interpretation and discussion.
%%We present these cases in this section.
%We reflect on our experiments and our results,  discuss limitations (Section~\ref{sec:limitations}) 
%and comment on open problems that can 
%be addressed by the researchers in Law or Data Protection Authorities (Section~\ref{sec:discussion}). 

\subsection{Who is responsible for violations in cookie banners?}
%\textbf{Shared responsibility of non-compliant banners between CMPs and publishers}
%When we obtain a consent string via direct calls to the CMP and shared cookies that belong to \texttt{consensu.org} domain, we attribute the responsibility for setting such consent string to either the CMP or the publisher.
It is a complex task to attribute the responsibility of a non-compliant cookie banner on a website to either the CMP or the publisher.
CMPs often propose different versions of their banners that have different legal implications, and provide a documentation on customizing the banner. For instance, OneTrust, on its webpage presenting its CMP solution~\cite{OneTrust}, proposes publishers to ``maximize user opt-in with customizable publisher-specific cookie banners [...] to optimize consent collection''.
We argue that CMPs providing non-compliant cookie banners cannot exonerate themselves and delegate responsibility to the publishers that include them, especially when they claim to provide GDPR- and ePrivacy-compliant consent collection solutions. Conversely, publishers have a part of responsibility if they choose non-compliant banners.
Hence, the responsibility of non-compliant cookie banners is shared between CMPs and publishers.
CMPs and publishers might even be considered co-controllers, but we leave this discussion to lawyers.

Moreover,
it is possible that publishers %make modifications to 
customize the banner when they host the CMP script in their website, % as a first-party script, disrupting 
modifying the original behaviour %planned 
offered by the CMP. In such a case, the responsibility of a violating banner %goes 
should be attributed to the publisher. 
%We have no way to detect such a case without 
Such cases can be detected with extensive case-by-case manual inspection.

%\textbf{Global shared consent} — 
\subsection{Problem of shared consent across publishers}
The TCF defines a ``global'' cookie that is writable and readable by all CMPs (see section~\ref{sec:background:consentstorage}).
We found such an example on \texttt{letudiant.fr}: it obtains the consent string set 
on the website \texttt{senscritique.com} previously visited by the user\footnote{We show a video of this in attachment~\cite{attachments}.}.
This behaviour may not be a violation of the GDPR in itself: consent must be specific to a given purpose, 
not to a publisher. 
However, it seems suspicious that, even while obtaining the consent string invisibly 
for the end user, \texttt{letudiant.fr} still displays the cookie banner to the user. 
This may be considered as an excessive request (publishers ask for a response on consent they already have) 
and a lack of transparency or a deception (user is tricked into thinking \texttt{letudiant.fr} 
does not have their consent)\footnote{%This could also simply be a bug.
However, the appearance of the banner on \texttt{letudiant.fr} could be a mistake and not an explicit deception technique.}. 
In fact, in a report about dark patterns, the CNIL already uses the terms ``bordering on harassment'' to describe the repetitive %asking of 
request for consent on every website, even without any shared consent consideration~\cite{cnil-dp2019}.
The global %cookie situation 
consent has been criticized by the Privacy International NGO~\cite{privacyinternational}, which denounced the lack of users' knowledge that consent is global to websites, and that opting out is near impossible.
This concept of global consent requires further analysis by legal experts.

%Moreover, it seems risky to have code written by many different possible actors taken as a safe source of consent. What if one publisher or CMP sets this cookie incorrectly, violating user's consent?
Additionally, such a design in the TCF assumes that all the CMPs who use the global consent mechanism trust each other on setting the consent string accordingly to the choices made by the user.
But a TCF-wide problem would arise if one publisher or CMP set this cookie incorrectly, violating user's consent.
We found that this was not a hypothetical scenario: we detected 3
websites that set a positive consent in the shared cookie 
before the user makes any choice in the banner and 20 websites doing so after the user 
explicitly refused consent
%both 
%when the user explicitly refuses consent and before the user gives their choice (see 
(more details in Section~\ref{sec:results:shared_cookie}).
%Conversely, we also observe a publisher (\texttt{eklablog.fr}) doing so with a negative consent.
%

\subsection{Unclear purposes in IAB Europe's TCF}
%\textbf{Purposes in IAB Europe's TCF} — 
The TCF proposes %Regarding 
five pre-defined purposes (reproduced in Table~\ref{tab:purposes} in the Appendix): 
we leave for discussion to %lawyers 
legal professionals whether 
defined purposes are explicit and specific.
%Even for the authors of this paper (computer science %as 
%researchers on web tracking and legal scholars), it is difficult to understand the exact 
%meaning of each purpose. 
The CNIL has already pronounced that the TCF %consent framework 
defined its purposes in a vague, imprecise way, in the decision against the Vectaury company~\cite{CNIL-Vectaury-2018}.

\subsection{IAB Europe TCF version 2}

We have not observed any application of the IAB Europe TCF version 2 that was announced in August 2019. 
This new version introduces 12 purposes for data processing, and adds more flexibility 
to choose a legal basis (consent or legitimate interest). Since the implementation of the framework by CMPs and publishers is responsible for these violations, they might still occur on websites with CMPs that implement TCF v2.

\iffalse
\subsection{Unclear semantics of consent string format}
\nc{Todo: update}

%\textbf{Ambiguity of consent string format} —
Each consent string contains a list of  
% both have a 
\textit{purposes} and a list of  \textit{vendors}. 
%array. 
The specification does not 
clearly say how advertisers are supposed to verify their consent, nor how the CMP should populate 
these fields. What happens if one of the two % arrays 
lists is empty? Should vendors assume they have 
consent if their identifier is in the \textit{vendors} array, or should they also check the purposes? 
%Since we cannot know 
The TCF doesn't specify 
how vendors should interpret the consent string. 
%, we cannot know 
%what happens in this ambiguous case. 
Moreover, should the \textit{vendors} %array 
list only be populated 
with vendors declaring to use consent for the purposes set in the \textit{purposes} %array?
list?
Such questions could be an interesting terrain of investigation for DPAs.
%For example, consider a user that consented to purposes number 4 %(see Figure~\ref{fig:consent_string}) 
%and to one vendor ``Oath (EMEA) Limited'' (see Figure~\ref{fig:gvl}).
%This vendor uses only purposes 1 and 2 with consent and purposes 3 and 5 with legitimate interest, 
%however it never uses purpose 4. The framework is unclear on who verifies that Oath is indeed not supposed 
%to process personal data under such consent.
\fi

\subsection{Consent strings can be created by anyone}
As shown in Section~\ref{sec:results:queries}, we observe on 37 websites requests to third parties containing consent strings that we suspect are being forged by non-CMP scripts running on the page, because they contain a CMP identifier that doesn't correspond to the CMP present on the page.
Even though the whole purpose of the TCF is to provide a way for actors in the advertisement industry to prove that they received consent from the user, we state that this proof is weak. The consent string's format does not contain any cryptographic proof that it was created by a given CMP, on a given website, in concordance with the user's choice. Consent strings can be forged by anyone, as our observation shows.
Such consent strings have been flagged as ``consent fraud'' by rogue third parties by an actor of the online advertising ecosystem~\cite{grutchfield-consentfraud2019}.

\ifextended
\section{Impact}

After the release of our study, NOYB, a EU privacy organization, has filed 
 three complaints to the CNIL (the French Data Protection Authority) 
 requiring to investigate practices regarding consent and setting of cookies 
 by three websites: \texttt{cdiscount.com}, \texttt{allocine.fr} and \texttt{vanityfair.fr}~\cite{noyb-cookies}. 
 
 Interestingly, we made a manual inspection of the three websites with the help of \extname{}~\cite{Cookie-Glasses} on February 14, 2020, namely 9 weeks after the complaint was issued.
\texttt{cdiscount.fr} changed the CMP to ``Commanders Act'' and on February 14, 2020, upon the user's refusal 
of consent, the CMP stores zero purpose and zero advertiser in the consent string. 
\texttt{allocine.fr} kept the same CMP (``WEBEDIA''), however on February 14, 2020, upon the user's refusal 
of consent, the CMP stores zero purpose and zero advertiser in the consent string. 
\texttt{vanityFair.fr} also kept the same CMP (``OneTrust LLC'') and on February 14, 2020, upon the user's refusal 
of consent, the CMP stores zero purpose and 413 advertisers in the consent string. 
%Nevertheless, 
%such consent string is not valid since no purpose is allowed.
Such consent string cannot be used by advertisers since no purpose is allowed.
\fi
%==========================================================
\section{Related Work}
\label{sec:related_work}

%We first discuss previous works on cookie banners before GDPR. 
%Previous work on cookie banners can be decomposed into pre- and post-GDPR. As pre-GDPR 
%These works 
The first lines of research on cookie banners published before the GDPR 
laid on the legal basis of the ePD and its implementation in various European countries, 
%these works are 
and were very country-specific. As the GDPR changed behaviour regarding 
cookies~\cite{libert2018changes, sanchez2019can}, trackers and other third-party 
content~\cite{libert2018changes} and cookie banners~\cite{eijk2019impact, degeling2018we}, 
we precise if each work made measurements before or after its enforcement (May 2018).
%Because the ePD had to be implemented by the European countries to be effective, 

%In 2013, Borghi et al. first studied the lawfulness of the collection of consent for commercial communications (advertisement and emails) on 200 websites from the UK. They found that, while 69\% of studied websites asks for consent in some way, only 16.2\% of them obtain a valid consent~\cite{borghi2013online}.

%In early 2014, a Dutch press article mentioned a study regarding the 2012 implementation of the ePD in the Dutch law. Two entrepreneurs tested almost 600~000 websites, and found that 28\% of them installed tracking cookies on first page load, i.e. before consent~\cite{heck-nrc2014}.

%\subsection{Post-GDPR works}
To %our 
the best of our knowledge, %four %post-GDPR 
the following works measured prevalence of cookie banners after GDPR.
%works on cookie banners have been published after GDPR came in force. %so far.
Sanchez et al.~\cite{sanchez2019can} studied the impact of the GDPR on cookie-setting practices. They found that the GDPR had a global impact, influencing even US-based websites. Similarly to our semi-automatic crawl, they manually refused consent on 2~000 cookie banners to extract statistics such as the number of cookies after consent refusal.
Van Eijk et al.~\cite{eijk2019impact} studied the impact on user's location on cookie banners. Using a crowd-sourced list, they automatically detected cookie banners on 40.2\% of European websites. They found that the provenance of the user has not so much impact as the expected audience of a website regarding the prevalence of banners. They also observed important variations between websites of different top-level domains.
Degeling et al.~\cite{degeling2018we} studied characteristics of 31 cookie banner libraries, including several ones%of them 
provided by CMPs of the TCF, by installing them locally. They found that 62.2\% of European websites displayed cookie banners in October 2018. %They 
The authors observed a 16\% increase in cookie banners adoption by website pre- and post-GDPR.
Nouwens et al.~\cite{nouwens2020dark} studied dark patterns in 5 popular CMPs of IAB Europe's TCF. They estimate that only 11.8\% of banners meet some minimum requirements of European law. They also study how banners design affect users' choice, and notably finds that the absence of a ``refuse'' button on the first layer of the banner increases positive consent by about 22\%.
%Finally, Utz et al.\cite{utz2019uninformed} studied the impact of graphical user interfaces parameters on users' decision regarding consent.

%Some works have more generally studied the impact of the GDPR of tracking.
%Libert et al., in a factsheet for the press, studied the impact of the GDPR on the amount of third-party content and cookies on news websites. On about 180 European news sites, they observe a 22\% drop in the amount of third-party cookies before (April 2018) and after (July 2018) the GDPR, but only 2\% drop in third-party content~\cite{libert2018changes}.

%\subsection{Opt out possibility}

Similarly to our \nooption{} violation, some works measured how many banners offer no way to opt out~\cite{degeling2018we}.
In 2015, Leenes and Kosta found  this issue on 25\% of 100 Dutch websites~\cite{leenes2015taming}.
The same year, the Article 29 Working Party found it on 54\% of top 250 websites of 8 EU countries~\cite{wp292015sweep}.
Vallina et al. found cookie banners offering no option to refuse consent on 1.36\% of porn websites~\cite{vallina2019tales}.

%\subsection{Influence of consent on trackers and cookies}

Several works measured the influence of %consent 
cookie banners on the number of trackers or cookies. % after GDPR.
%Pre-GDPR, 
Before the GDPR, Carpineto et al.~\cite{carpineto2016automatic} measured how many websites set cookies without displaying a banner.
Traverso et al.~\cite{traverso2017benchmark} measured the number of trackers before and after giving a positive consent on banners on 100 Italians websites. They found an average of 29.5 trackers prior to giving consent.
In 2016, Englehardt and Narayanan~\cite{englehardt2016online} found 18 third parties per websites prior to any consent.
In 2017, Trevisan et al.~\cite{trevisan2019years} found that 49\% of websites installed profiling cookies before user consent, and that 80.5\% of websites installing tracking cookies did not wait for user's consent to do so.
After the GDPR came in force, %In 2019, 
Sanchez et al.~\cite{sanchez2019can} %counted 
measured the number of cookies after refusing consent on banners.
Instead, in our work, we measure trackers both before and after both giving a positive consent and refusing consent. %, post-GDPR.

\ifextended
Detecting all tracking cookies is still a globally unresolved problem, and ad- and trackers-blockers mostly detect trackers with the help of filter lists (that contain regular expressions or 2\textsuperscript{nd}-level TLDs). Fouad et al.~\cite{fouad2020tracking} showed that such filter lists, EasyList\&EasyPrivacy and Disconnect, respectively miss 25.22\% and 30.34\% of trackers they detected. The reason filter lists miss trackers is because tracking cookies are often tightly integrated with the functional and useful content of the website, such as Google customized search engine. Moreover, the usage of third-party tracking by advertisers is decreasing over time~\cite{libert2018changes}, and browser vendors claim they will stop using such cookies~\cite{schuh-building2020}, which means advertisers will switch to other forms of tracking, such as first-party cookies (that are even harder to block) or browser fingerprinting (which is difficult to detect~\cite{laperdrix2019browser}). Therefore, our work on cookie banners is complementary to the ad- or tracking- blocking tools.
\fi

%\subsection{Legal Work}

On the legal side, some regulators have already been active. The French DPA (CNIL) sued an advertisement company that used the TCF, invoking a lack of informed, free, specific and unambiguous consent~\cite{ryan-brave2018}. %For the CNIL, the consent text was not clear enough regarding the final use of collected data, and the formulation may lead users to incorrectly assume that refusing consent prevents a free access to the website, or lead to more intrusive advertisement.
In early 2020, they published a project document for guidelines on cookie banners~\cite{cnil-2020recommandation} and criticized the TCF in a blog post~\cite{poilve-2020mecanismes}.
They also noted that pre-selecting consent checkboxes was not compliant with the Article 32 of the GDPR.
The European Court of Justice's decision in the Planet49~\cite{Planet49} case recently settled that pre-selecting options was not GDPR-compliant.
%. They required the list of recipient of users' data to appear immediately when consent text is displayed.
Finally, the UK's DPA (ICO)~\cite{ico2019rtb} recently published a report on adtech and Real-Time Bidding, studying both IAB Europe's TCF and Google's framework. Among other considerations, they concluded that the TCF lacked transparency and observed a systemic lack of compliance to their data protection requirements. %of the real-time bidding sector.

%==========================================================

\section{Conclusions}
\label{sec:future_work}

In this paper, we have systematically %studies %analyzed 
studied cookie banners of IAB Europe's Transparency and Consent Framework (TCF). 
By collaborating with a researcher in law (one of the co-authors), we have identified four legal violations 
of both the GDPR and the ePD and we have detected them on 
\nbbanner{} European websites that use TCF cookie banners.
We have detected at least one of these suspected violations in 54\% of websites.
%and found privacy violations on 72.45\% of them. 
%The observed privacy violations are indeed violations of GDPR and ePD, that we 
 %with references to legal documents and decisions provided by a researcher in law. 
 Finally, to help users and Data Protection Authorities (DPAs) further investigate these violations, 
 we provide a browser extension called \extname{}, that is able to detect some of them.
 %detecting some of these violations, so that users and DPAs can investigate these issues by themselves.

Beyond suspected violations in the implementations of cookies banners, 
we believe that the TCF
% TCF 
suffers from several problems open for discussion and improvement. % issues in this framework. Firstly, 
%defined purposes of data processing are unclear. Secondly, 
First, the consent string format has an unclear semantics, which makes it hard to 
interpret and use by the third parties that rely on such consent. 
Second, the TCF does not provide guidelines on which actors who obtain user consent (assuming 
it was obtained in a compliant way) are supposed to respect it: should the publishers, CMPs or 
some other actors ensure that the third parties respect consent they received? 
%. Thirdly, no real guidelines are defined regarding how consent should be checked, and who should change their behaviour in case of a negative consent.
%
%Because of all this unclarity, CMPs implement banners in various ways, which lead to the violations we observe. 
We believe that European regulators should %provide more guidance 
take a more active stand regarding the implementation of cookie banners: % and CMPs: 
either with supportive actions, such as guidance, or %investigations. 
%decisions with penalties. 
regulatory decisions and associated fines. 
%Thus, our paper helps DPAs take both supportive and repressive actions.

%\newpage

%\input{appendix}

\section*{Acknowledgment}
The authors would like to thank Imane Fouad for helping with the verification procedure in cookie banners. 
%The authors also thank SIRDATA CMP for helping us detect a bug in the original implementation of \extname{}.  
This work has been partially supported by the ANR JCJC project PrivaWeb (ANR-18-CE39-0008), the ANSWER project PIA FSN2 No. P159564-2661789/ DOS0060094 between Inria and Qwant, and by the Inria DATA4US Exploratory Action project.

\bibliographystyle{IEEEtranS}
\bibliography{paper}

% Generated by IEEEtranS.bst, version: 1.12 (2007/01/11)
\begin{thebibliography}{10}
\providecommand{\url}[1]{#1}
\csname url@samestyle\endcsname
\providecommand{\newblock}{\relax}
\providecommand{\bibinfo}[2]{#2}
\providecommand{\BIBentrySTDinterwordspacing}{\spaceskip=0pt\relax}
\providecommand{\BIBentryALTinterwordstretchfactor}{4}
\providecommand{\BIBentryALTinterwordspacing}{\spaceskip=\fontdimen2\font plus
\BIBentryALTinterwordstretchfactor\fontdimen3\font minus
  \fontdimen4\font\relax}
\providecommand{\BIBforeignlanguage}[2]{{%
\expandafter\ifx\csname l@#1\endcsname\relax
\typeout{** WARNING: IEEEtranS.bst: No hyphenation pattern has been}%
\typeout{** loaded for the language `#1'. Using the pattern for}%
\typeout{** the default language instead.}%
\else
\language=\csname l@#1\endcsname
\fi
#2}}
\providecommand{\BIBdecl}{\relax}
\BIBdecl

\bibitem{Acar-etal-13-CCS}
G.~Acar, M.~Ju{\'{a}}rez, N.~Nikiforakis, C.~D{\'{\i}}az, S.~F. G{\"{u}}rses,
  F.~Piessens, and B.~Preneel, ``Fpdetective: dusting the web for
  fingerprinters,'' in \emph{Conference on Computer and Communications Security
  ({CCS}'13)}, 2013.

\bibitem{wp292015sweep}
{Article 29 Working Party}, ``Cookie sweep combined analysis - report,''
  \url{https://ec.europa.eu/newsroom/article29/document.cfm?action=display&doc_id=56123},
  accessed on 2019.11.01, 2015.

\bibitem{attachments}
``Attachments to the paper ({Dropbox} repository),''
  \url{https://www.dropbox.com/sh/fw8ubf23z3ai0ei/AABY4qRO3FKXeGELfiPGFHica}.

\bibitem{carpineto2016automatic}
C.~Carpineto, D.~Lo~Re, and G.~Romano, ``Automatic assessment of website
  compliance to the {European} cookie law with {CooLCheck},'' in
  \emph{Proceedings of the 2016 ACM on Workshop on Privacy in the Electronic
  Society}, 2016.

\bibitem{CNIL-deliberation-2019}
{CNIL}, ``Article 2 of the deliberation n2019-093 of july 4, 2019 adopting
  guidelines relating to the application of article 82 of the law of january 6,
  1978 modified to the operations of reading or writing to a user's terminal
  (including cookies and other tracers) (corrigendum),''
  \url{https://www.legifrance.gouv.fr/affichTexte.do?cidTexte=JORFTEXT000038783337},
  accessed on 30 October, 2019.

\bibitem{CNIL-Vectaury-2018}
------, ``Décision n {MED} 2018-042 du 30 octobre 2018 mettant en demeure la
  société {Vectaury},''
  \url{https://www.legifrance.gouv.fr/affichCnil.do?oldAction=rechExpCnil&id=CNILTEXT000037594451&fastReqId=974682228&fastPos=2},
  accessed on 31 October 2019.

\bibitem{cnil-dp2019}
------, ``{IP} report: Shaping choices in the digital world,''
  \url{https://linc.cnil.fr/fr/ip-report-shaping-choices-digital-world},
  accessed on 2019.10.30, 2019.

\bibitem{cnil-2020recommandation}
------, ``Projet de recommandation sur les modalités pratiques de recueil du
  consentement prévu par l'article 82 de la loi du 6janvier 1978 modifiée,
  concernant les opérations d’accès ou d’inscription d’informations
  dans le terminal d'un utilisateur (recommandation "cookies et autres
  traceurs"),''
  \url{https://www.cnil.fr/sites/default/files/atoms/files/projet_de_recommandation_cookies_et_autres_traceurs.pdf},
  accessed on 17 January 2020., 01 2020.

\bibitem{degeling2018we}
M.~Degeling, C.~Utz, C.~Lentzsch, H.~Hosseini, F.~Schaub, and T.~Holz, ``We
  value your privacy... now take some cookies: Measuring the {GDPR}'s impact on
  web privacy,'' in \emph{Network and Distributed System Security Symposium
  ({NDSS})}, 2018.

\bibitem{disconnectlist}
Disconnect, ``disconnect-tracking-protection,''
  \url{https://github.com/disconnectme/disconnect-tracking-protection},
  accessed on 2019.07.16.

\bibitem{eijk2019impact}
R.~v. Eijk, H.~Asghari, P.~Winter, and A.~Narayanan, ``The impact of user
  location on cookie notices (inside and outside of the {European} union),'' in
  \emph{Workshop on Technology and Consumer Protection (ConPro'19)}, 2019.

\bibitem{englehardt2016online}
S.~Englehardt and A.~Narayanan, ``Online tracking: A 1-million-site measurement
  and analysis,'' in \emph{conference on computer and communications security
  ({CCS}'13)}, 2016.

\bibitem{ePD-09}
``{Directive 2009/136/EC of the European Parliament and of the Council of 25
  November 2009},''
  \url{https://eur-lex.europa.eu/legal-content/EN/TXT/?uri=celex\%3A32009L0136},
  accessed on 2019.10.31.

\bibitem{Planet49}
{European Court of Justice}, ``Judgement of the court of justice of the {EU},
  {Case} c-673/17,''
  \url{http://curia.europa.eu/juris/document/document.jsf;?&docid=218462&doclang=EN&cid=8679428},
  accessed on 2019.10.31.

\bibitem{EDPB-4-18}
{European Data Protection Board}, ``Guidelines on consent under regulation
  2016/679” ({WP} 259 rev.01), adopted on 10 april 2018,''
  \url{https://ec.europa.eu/newsroom/article29/item-detail.cfm?item_id=623051}.

\bibitem{EDPB-3-13}
------, ``Opinion 03/2013 on purpose limitation ({WP} 203), adopted on 2 april
  2013,''
  \url{https://ec.europa.eu/justice/article-29/documentation/opinion-recommendation/files/2013/wp203_en.pdf}.

\bibitem{EDPB-6-14}
------, ``Opinion 06/2014 on the notion of legitimate interests of the data
  controller under article 7 of directive 95/46/ec ({WP} 217),''
  \url{https://ec.europa.eu/justice/article-29/documentation/opinion-recommendation/files/2014/wp217_en.pdf}.

\bibitem{EDPB-15-11}
------, ``Opinion 15/2011 on the definition of consent ({WP} 187), adopted on
  13 july 2011,''
  \url{https://ec.europa.eu/justice/article-29/documentation/opinion-recommendation/files/2011/wp187_en.pdf}.

\bibitem{EDPB-2-10}
------, ``Opinion 2/2010 on online behavioural advertising, 22 june 2010, {WP}
  171, p. 10,''
  \url{https://ec.europa.eu/justice/article-29/documentation/opinion
  recommendation/files/2007/wp136_en.pdf}.

\bibitem{EDPB-4-07}
------, ``Opinion 4/2007 on the concept of personal data ({WP} 136), adopted on
  20.06.2007,''
  \url{https://ec.europa.eu/justice/article-29/documentation/opinion
  recommendation/files/2007/wp136_en.pdf}.

\bibitem{EDPB-2-13}
------, ``Working document 02/2013 providing guidance on obtaining consent for
  cookies, adopted on 2 october 2013,''
  \url{https://www.pdpjournals.com/docs/88135.pdf}.

\bibitem{grutchfield-consentfraud2019}
C.~Grutchfield, ``The adtech truth – don’t mess with my consent string!''
  \url{https://newdigitalage.co/2019/10/09/behind-the-curtain-the-adtech-truth-dont-mess-with-my-consent-string/amp/},
  accessed on 2020.02.04, 2019.

\bibitem{iab-csusecase}
IAB, ``Consent string use cases,''
  \url{https://github.com/InteractiveAdvertisingBureau/Consent-String-SDK-JS/blob/master/consent_string_use_cases.md},
  accessed on 2020.02.07.

\bibitem{OpenRTB}
{IAB}, ``Openrtb (real-time bidding),''
  \url{https://www.iab.com/guidelines/real-time-bidding-rtb-project/}, accessed
  on 2019.09.16.

\bibitem{safeFrames}
------, ``Safeframe,'' \url{https://www.iab.com/guidelines/safeframe/},
  accessed on 2019.09.16, 2014.

\bibitem{iab-cmpid1}
{IAB Europe}, ``{CMP} {ID} 1 is not currently assigned to a {Consent}
  {Management} {Provider} ({CMP}),''
  \url{http://advertisingconsent.eu/2019/01/cmp-id-1-is-not-currently-assigned-to-a-consent-management-provider-cmp/},
  accessed on 2019.09.02.

\bibitem{iab-cmplist}
------, ``{CMP} list,'' \url{https://advertisingconsent.eu/cmp-list/},
  downloaded in 2019.04.

\bibitem{iab-policy}
------, ``{IAB} europe transparency \& consent framework policies,''
  \url{https://iabeurope.eu/wp-content/uploads/2019/08/IABEurope_TransparencyConsentFramework_v1-1_policy_FINAL.pdf},
  accessed on 2019.11.20.

\bibitem{iab-jsapi}
{IAB Europe} and {IAB Tech Lab}, ``Consent management provider javascript {API}
  v1.1: Transparency \& consent framework,''
  \url{https://github.com/InteractiveAdvertisingBureau/GDPR-Transparency-and-Consent-Framework/blob/master/CMP\%20JS\%20API\%20v1.1\%20Final.md\#API-provided},
  04 2018.

\bibitem{iab-tcf2018}
------, ``Transparency and consent framework,''
  \url{https://github.com/InteractiveAdvertisingBureau/GDPR-Transparency-and-Consent-Framework},
  accessed on 2019.05.03, 04 2018.

\bibitem{iab-gvl}
------, ``Global vendor list (gvl),''
  \url{https://github.com/InteractiveAdvertisingBureau/GDPR-Transparency-and-Consent-Frameworkhttps://vendorlist.consensu.org/vendorlist.json},
  accessed June 2019, 06 2019.

\bibitem{iab-cssdk}
{IAB Tech Lab}, ``Transparency and consent framework: Consent string {SDK}
  (javascript),''
  \url{https://github.com/InteractiveAdvertisingBureau/Consent-String-SDK-JS}.

\bibitem{iab-extensionfields}
------, ``{OpenRTB} advisory - {GDPR},''
  \url{https://iabtechlab.com/wp-content/uploads/2018/02/OpenRTB_Advisory_GDPR_2018-02.pdf},
  accessed on 2019.10.16, 02 2018.

\bibitem{iab-urlbased}
{IAB Tech Lab} and {IAB Europe}, ``{GDPR} consent passing for {URL}-based
  services: Transparency and consent framework,''
  \url{https://github.com/InteractiveAdvertisingBureau/GDPR-Transparency-and-Consent-Framework/blob/master/URL-based\%20Consent\%20Passing_\%20Framework\%20Guidance.md},
  04 2018.

\bibitem{ICO-cookie-19}
{Information Commissioner's Office}, ``{ICO} guidance on the rules on use of
  cookies and similar technologies,''
  \url{https://ico.org.uk/media/for-organisations/guide-to-pecr/guidance-on-the-use-of-cookies-and-similar-technologies-1-0.pdf}.

\bibitem{ico2019rtb}
------, ``Update report into adtech and real time bidding,''
  \url{https://ico.org.uk/media/about-the-ico/documents/2615156/adtech-real-time-bidding-report-201906.pdf},
  accessed on 2019.07.10, 2019.

\bibitem{le2019tranco}
V.~Le~Pochat, T.~Van~Goethem, S.~Tajalizadehkhoob, M.~Korczy{\'n}ski, and
  W.~Joosen, ``Tranco: a research-oriented top sites ranking hardened against
  manipulation,'' in \emph{Network and Distributed System Security Symposium
  ({NDSS})}, 2019.

\bibitem{leenes2015taming}
R.~Leenes and E.~Kosta, ``Taming the cookie monster with {Dutch} law - a tale
  of regulatory failure,'' \emph{Computer Law \& Security Review}, vol.~31,
  2015.

\bibitem{Lern-etal-16-USENIX}
A.~Lerner, A.~K. Simpson, T.~Kohno, and F.~Roesner, ``Internet {Jones} and the
  raiders of the lost trackers: An archaeological study of web tracking from
  1996 to 2016,'' in \emph{25th USENIX Security Symposium ({USENIX} Security
  16)}, 2016.

\bibitem{libert2015exposing}
T.~Libert, ``Exposing the hidden web: An analysis of third-party http requests
  on 1 million websites,'' \emph{International Journal of Communication}, 2015.

\bibitem{libert2018changes}
T.~Libert, L.~Graves, and R.~K. Nielsen, ``Changes in third-party content on
  european news websites after {GDPR},'' Reuters Institute for the Study of
  Journalism, 2018.

\bibitem{Cookie-Glasses}
C.~Matte, ``Cookie glasses,'' \url{https://github.com/Perdu/Cookie-Glasses},
  2019.

\bibitem{Cookinspect}
------, ``Cookinspect,'' \url{https://github.com/Perdu/Cookinspect}, 2020.

\bibitem{Niki-etal-13-SP}
N.~Nikiforakis, A.~Kapravelos, W.~Joosen, C.~Kruegel, F.~Piessens, and
  G.~Vigna, ``Cookieless monster: Exploring the ecosystem of web-based device
  fingerprinting,'' in \emph{{IEEE} Symposium on Security and Privacy
  ({SP}'13)}, 2013.

\bibitem{nouwens2020dark}
M.~Nouwens, I.~Liccardi, M.~Veale, D.~Karger, and L.~Kagal, ``Dark patterns
  after the gdpr: Scraping consent pop-ups and demonstrating their influence,''
  in \emph{ACM CHI Conference on Human Factors in Computing Systems}, 2020.

\bibitem{Olej-etal-14-NDSS}
L.~Olejnik, M.~Tran, and C.~Castelluccia, ``Selling off user privacy at
  auction,'' in \emph{Network and Distributed System Security Symposium
  ({NDSS}'14)}, 2014.

\bibitem{OneTrust}
OneTrust, ``Consent management publishers advertisers,''
  \url{https://www.onetrust.com/solutions/consent-management-platform/},
  accessed on 2019.10.15.

\bibitem{papa-etal-19-www}
P.~Papadopoulos, N.~Kourtellis, and E.~P. Markatos, ``Cookie synchronization:
  Everything you always wanted to know but were afraid to ask,'' in \emph{The
  World Wide Web Conference ({WWW}'19)}, 2019.

\bibitem{poilve-2020mecanismes}
B.~Poilvé, ``Mécanismes et (r)écueil du consentement,''
  \url{https://linc.cnil.fr/mecanismes-et-recueil-du-consentement}, access on
  17 January 2020, Laboratoire d'Innovation Numérique de la {CNIL}, 01 2020.

\bibitem{privacyinternational}
{Privacy International}, ``Most cookie banners are annoying and deceptive. this
  is not consent.''
  \url{https://privacyinternational.org/explainer/2975/most-cookie-banners-are-annoying-and-deceptive-not-consent},
  accessed on 2019.08.12, 2019.

\bibitem{Roes-etal-12-NSDI}
F.~Roesner, T.~Kohno, and D.~Wetherall, ``Detecting and defending against
  third-party tracking on the web,'' in \emph{{USENIX} Symposium on Networked
  Systems Design and Implementation ({NSDI}'12)}, 2012.

\bibitem{ryan-brave2018}
J.~Ryan, ``French regulator shows deep flaws in {IAB}’s consent framework and
  {RTB},'' \url{https://brave.com/cnil-consent-rtb/}, accessed on 2019.03.28,
  2018.

\bibitem{sanchez2019can}
I.~Sanchez-Rola, M.~Dell'Amico, P.~Kotzias, D.~Balzarotti, L.~Bilge, P.-A.
  Vervier, and I.~Santos, ``Can {I} opt out yet?: {GDPR} and the global
  illusion of cookie control,'' in \emph{Asia Conference on Computer and
  Communications Security ({AsiaCCS}'19)}, 2019.

\bibitem{epd}
{The European Parliament and the Council of the European Union}, ``{Directive
  2002/58/EC of the European Parliament and of the Council of 12 July 2002
  concerning the processing of personal data and the protection of privacy in
  the electronic communications sector (Directive on privacy and electronic
  communications)},'' 2002.

\bibitem{gdpr}
------, ``{Regulation (EU) 2016/679 of the European Parliament and of the
  Council of 27 April 2016 on the protection of natural persons with regard to
  the processing of personal data and on the free movement of such data, and
  repealing Directive 95/46/EC (General Data Protection Regulation)},'' 2016.

\bibitem{traverso2017benchmark}
S.~Traverso, M.~Trevisan, L.~Giannantoni, M.~Mellia, and H.~Metwalley,
  ``Benchmark and comparison of tracker-blockers: Should you trust them?'' in
  \emph{Network Traffic Measurement and Analysis Conference ({TMA}'17)}, 2017.

\bibitem{trevisan2019years}
M.~Trevisan, S.~Traverso, E.~Bassi, and M.~Mellia, ``4 years of {EU} cookie
  law: Results and lessons learned,'' \emph{Proceedings on Privacy Enhancing
  Technologies Symposium ({PETS}'19)}, 2019.

\bibitem{utz2019uninformed}
C.~Utz, M.~Degeling, S.~Fahl, F.~Schaub, and T.~Holz, ``(un)informed consent:
  Studying gdpr consent notices in the field,'' in \emph{Conference on Computer
  and Communications Security ({CCS'19})}, 2019.

\bibitem{vallina2019tales}
P.~Vallina, {\'A}.~Feal, J.~Gamba, N.~Vallina-Rodriguez, and A.~F. Anta,
  ``Tales from the porn: A comprehensive privacy analysis of the web porn
  ecosystem,'' in \emph{Proceedings of the Internet Measurement Conference
  ({ICM}'19)}, 2019.

\end{thebibliography}

\appendices
\section{Purposes Defined in IAB Europe's TCF}
\label{app:purposes}

We reproduce purposes defined in the TCF in Table~\ref{tab:purposes}.

\begin{table*}[t]
\caption{Purposes defined in IAB Europe's TCF, accessible at \url{https://register.consensu.org/}, accessed on May 3\textsuperscript{rd}, 2019.}
\label{tab:purposes}
\begin{tabular}{|p{0.9cm}|p{1.7cm}|p{14.5cm}|}
\hline
\textbf{Purpose number} & \textbf{Purpose name} & \textbf{Purpose description} \\\hline
1 & Information storage and access & The storage of information, or access to information that is already stored, on your device such as advertising identifiers, device identifiers, cookies, and similar technologies. \\\hline
2 & Personalisation & The collection and processing of information about your use of this service to subsequently personalise advertising and/or content for you in other contexts, such as on other websites or apps, over time. Typically, the content of the site or app is used to make inferences about your interests, which inform future selection of advertising and/or content. \\\hline
3 & Ad selection, delivery, reporting & The collection of information, and combination with previously collected information, to select and deliver advertisements for you, and to measure the delivery and effectiveness of such advertisements. This includes using previously collected information about your interests to select ads, processing data about what advertisements were shown, how often they were shown, when and where they were shown, and whether you took any action related to the advertisement, including for example clicking an ad or making a purchase. This does not include personalisation, which is the collection and processing of information about your use of this service to subsequently personalise advertising and/or content for you in other contexts, such as websites or apps, over time. \\\hline
4 & Content selection, delivery, reporting & The collection of information, and combination with previously collected information, to select and deliver content for you, and to measure the delivery and effectiveness of such content. This includes using previously collected information about your interests to select content, processing data about what content was shown, how often or how long it was shown, when and where it was shown, and whether the you took any action related to the content, including for example clicking on content. This does not include personalisation, which is the collection and processing of information about your use of this service to subsequently personalise content and/or advertising for you in other contexts, such as websites or apps, over time. \\\hline
5 & Measurement & The collection of information about your use of the content, and combination with previously collected information, used to measure, understand, and report on your usage of the service. This does not include personalisation, the collection of information about your use of this service to subsequently personalise content and/or advertising for you in other contexts, i.e. on other service, such as websites or apps, over time. \\\hline
\end{tabular}
\SHORTENTAB
\end{table*}

\section{Attachments}
\label{app:attachments}

In a public repository~\cite{attachments}, we provide files that are relevant to this work:
the full list of websites for each suspected violation,
%screenshots of each website mentioned in this paper,
and videos showing examples of them. %suspected GDPR and ePD violations.

\section{Data for Reproducible Research}
\label{sec:reproduce}

For the sake of research reproducibility, we indicate all data relevant to this work in Table~\ref{tab:reproductible}.

For selecting the websites, 
we use Tranco to build lists~\cite{le2019tranco}. %This work 
Within Tranco, we select the following options: Alexa and Majestic lists. We don't use The Cisco Umbrella list because it is DNS-based, and may not be representative of web traffic. Likewise, we exclude the Quantcast list because it is based on US traffic only. We also select the option to remove domains flagged as dangerous by Google Safe Browsing
%, and keep the pre-selected option to keep pay-level TLDs only.

From Tranco's top 1 million list, we extract the first 1~000 websites of the top-level domain (TLD)
of each European country, %plus %some 
and 1~000 websites from country-independant TLDs: \texttt{.com}, \texttt{.eu} and \texttt{.org} on 
September 20\textsuperscript{th} 2019.

\begin{table*}[!htbp]
%\twocolumn[
\caption{Data for reproducible research}
\label{tab:reproductible}
\begin{tabular}{|p{4.5cm}|p{13.0cm}|}
\hline
%\multicolumn{2}{|c|}{Initial crawl} \\\hline
%Crawling dates (5k crawl) & 2019.04.09, 2019.04.12, 2019.04.15, 2019.04.16, 2019.04.18, 2019.04.19 \\\hline
%Crawling dates (top 500 .fr) & 2019.04.24 - 25 \\\hline
%Software - inclusion chains & DeepCrawling - commit 9338810 \\\hline
Software - Selenium & python-selenium 3.141.0-1 \\\hline
Software - Chromium & chromium 76.0.3809.100-1 \\\hline
Operating system & Arch Linux \\\hline
Kernel (result of \texttt{uname -a}) & Linux 5.2.5-arch1-1-ARCH \#1 SMP PREEMPT Wed Jul 31 08:30:34 UTC 2019 x86\_64 GNU/Linux \\\hline
%Number of visits per site & 1 \\\hline
%Timeout & 10s (default) \\\hline
User-Agent & Mozilla/5.0 (X11; Linux x86\_64) AppleWebKit/537.36 (KHTML, like Gecko) HeadlessChrome/73.0.3683.103 Safari/537.36 \\\hline
%Browser dimensions & \\\hline
Location & France \\\hline
%ISP & Free, Renater \\\hline
Tranco list & \url{https://tranco-list.eu/list/4NKX/1000000}, generated on 2019.09.20 \\\hline
Disconnect list commit & eb817fb1 (2019-12-10) \\\hline
WebXRay commit & 04c3c8e8 (2019-06-18) \\\hline
Crawling date (automatic crawl) & 2019-09-20 - 2019-09-21 \\\hline
Crawling date (semi-automatic crawl) & 2019-09-23 - 10-01 \\\hline
\end{tabular}
%]
\SHORTENTAB
\end{table*}

\section{Procedure for the Human Operators}
\label{sec:human_operator}

In this section, we give the precise procedure that human operators had to follow to refuse consent and give a positive consent on the banners during the semi-automatic crawl.

First, we attempt to refuse consent. If there is a ``refuse'' button on the banner, we click it directly; otherwise, we open the banner's ``parameters''. There, we untick any purpose-related option (checkbox or slider), independently from the kind of option (including e.g. functional cookies). If there is a ``refuse all'' button, we click it even if options are unticked by default. When banners propose vendors-related options, we ignore them.
%untick them if a button makes it possible to reject all of them at once (i.e., we don't untick hundreds of boxes when this is the only possibility, such as on \texttt{goodtoknow.co.uk}). \b{(Todo: add reference saying that this is not valid according to some DPA.)}.
When the banner does not possess a ``parameters'' button, but only a link to the privacy policy (such as on \texttt{liberation.fr}), we follow this link and attempt to find a way to refuse consent within a reasonable time for a common website user (10 seconds), then come back to the main page. If options to refuse consent are located on another page linked by the banner, we come back to the main page after refusing consent. We always close the banner after refusing consent, by clicking the button whose terminology least indicates that we allow tracking\footnote{Some banners, such as the one on \texttt{rtl.fr}, have both an ``accept'' button and a small cross to close it. The ``accept'' button is misleading, because it sets a consent string with all purposes set even if the purpose options are unticked, while the small cross does not.}. Once everything is done, we manually label whether we encountered different cases: pre-selected options\footnote{We consider that there is a \preticked{} violation according to the visual representation of options (pre-ticked boxes, sliders set to acceptance). Ambiguous cases such as \texttt{lefigaro.fr}, where neither ``accept'' nor ``deny'' is set by default are not considered a violation.}, banner not appearing, non-functional banner, banner proposing no way to refuse consent (considering links inside the banner). In one case (\texttt{healthline.com}), the banner proposed a way to refuse consent, but access to the website was then refused. We mark such a case as \nooption{}.

Secondly, on a second browser session (or directly if there is no option to refuse consent on the banner), we accept tracking by clicking on the ``accept'' button, or close the banner when it is the only option (we close the banner in all cases).

If the banner does not appear on first load, we reload the website until the banner appears, up to 3 times.

\section{Alternative Presentations of the Results}
\label{app:altpres}

We display results of violations observed in the semi-automatic crawl organized by country in Table~\ref{tab:quantification_violations_cmp_gdpr_tlds}, and organized by CMP in Table~\ref{tab:quantification_violations_cmp_gdpr} (for CMPs seen at least 5 times). This is interesting for DPAs, who can then see which CMPs to investigate in priority.
We do not display results for the automatic crawl because we can only identify CMPs providing consent strings before consent in this case (which would introduce a bias, and only concern 21\% of websites).

\percountryviolationtables{}

  \begin{table*}[htpb]
    \center
    \caption{Quantification of suspected violations of the GDPR and the ePD encountered in the different CMPs seen at least 5 times during the semi-automatic crawl (on \texttt{.fr}, \texttt{.uk}, \texttt{.it}, \texttt{.be}, \texttt{.ie} and \texttt{.com} websites), by CMP. The \nonrespect{} and \preticked{} columns display results w.r.t. the number of websites on which refusing consent was possible.}
    \label{tab:quantification_violations_cmp_gdpr}
    \begin{tabular}{|>{\rowmac}l|>{\rowmac}c|>{\rowmac}c|>{\rowmac}c|>{\rowmac}c|>{\rowmac}c|>{\rowmac}c<{\clearrow}|}
      \hline
      & \textbf{Number of} & \multicolumn{4}{c|}{\textbf{Violations}} \\
      \cline{3-6}
      \textbf{CMP} & \textbf{websites} & \textbf{\preaction{}} & \textbf{\nooption{}} & \textbf{\preticked{}} & \textbf{\nonrespect{}} \\
      \hline
      Quantcast & 174 & \cellcolor{BrickRed!3} 3.4\% (6/174) & \cellcolor{BrickRed!5} 5.2\% (9/174) & \cellcolor{BrickRed!37} 37.8\% (62/164) & \cellcolor{BrickRed!0} 0.6\% (1/164) \\\hline
OneTrust & 50 & \cellcolor{BrickRed!74} 74.0\% (37/50) & \cellcolor{BrickRed!4} 4.0\% (2/50) & \cellcolor{BrickRed!83} 83.3\% (40/48) & \cellcolor{BrickRed!8} 8.3\% (4/48) \\\hline
Didomi & 41 & \cellcolor{BrickRed!0} 0.0\% (0/41) & \cellcolor{BrickRed!0} 0.0\% (0/41) & \cellcolor{BrickRed!39} 39.0\% (16/41) & \cellcolor{BrickRed!0} 0.0\% (0/41) \\\hline
Sourcepoint & 34 & \cellcolor{BrickRed!2} 2.9\% (1/34) & \cellcolor{BrickRed!0} 0.0\% (0/34) & \cellcolor{BrickRed!64} 64.7\% (22/34) & \cellcolor{BrickRed!2} 2.9\% (1/34) \\\hline
Evidon & 22 & \cellcolor{BrickRed!0} 0.0\% (0/22) & \cellcolor{BrickRed!22} 22.7\% (5/22) & \cellcolor{BrickRed!25} 25.0\% (4/16) & \cellcolor{BrickRed!25} 25.0\% (4/16) \\\hline
iubenda & 20 & \cellcolor{BrickRed!0} 0.0\% (0/20) & \cellcolor{BrickRed!0} 0.0\% (0/20) & \cellcolor{BrickRed!0} 0.0\% (0/20) & \cellcolor{BrickRed!0} 0.0\% (0/20) \\\hline
Clickio & 14 & \cellcolor{BrickRed!0} 0.0\% (0/14) & \cellcolor{BrickRed!0} 0.0\% (0/14) & \cellcolor{BrickRed!0} 0.0\% (0/14) & \cellcolor{BrickRed!0} 0.0\% (0/14) \\\hline
Oath & 12 & \cellcolor{BrickRed!0} 0.0\% (0/12) & \cellcolor{BrickRed!0} 0.0\% (0/12) & \cellcolor{BrickRed!16} 16.7\% (2/12) & \cellcolor{BrickRed!0} 0.0\% (0/12) \\\hline
Triboo Media & 10 & \cellcolor{BrickRed!0} 0.0\% (0/10) & \cellcolor{BrickRed!0} 0.0\% (0/10) & \cellcolor{BrickRed!0} 0.0\% (0/10) & \cellcolor{BrickRed!0} 0.0\% (0/10) \\\hline
Commanders Act & 10 & \cellcolor{BrickRed!40} 40.0\% (4/10) & \cellcolor{BrickRed!0} 0.0\% (0/10) & \cellcolor{BrickRed!80} 80.0\% (8/10) & \cellcolor{BrickRed!0} 0.0\% (0/10) \\\hline
Axel Springer & 10 & \cellcolor{BrickRed!60} 60.0\% (6/10) & \cellcolor{BrickRed!70} 70.0\% (7/10) & \cellcolor{BrickRed!100} 100.0\% (3/3) & \cellcolor{BrickRed!33} 33.3\% (1/3) \\\hline
OneTag & 9 & \cellcolor{BrickRed!0} 0.0\% (0/9) & \cellcolor{BrickRed!0} 0.0\% (0/9) & \cellcolor{BrickRed!100} 100.0\% (9/9) & \cellcolor{BrickRed!0} 0.0\% (0/9) \\\hline
Cookie Trust WG. & 8 & \cellcolor{BrickRed!25} 25.0\% (2/8) & \cellcolor{BrickRed!25} 25.0\% (2/8) & \cellcolor{BrickRed!60} 60.0\% (3/5) & \cellcolor{BrickRed!0} 0.0\% (0/5) \\\hline
Conversant Europe & 7 & \cellcolor{BrickRed!0} 0.0\% (0/7) & \cellcolor{BrickRed!0} 0.0\% (0/7) & \cellcolor{BrickRed!100} 100.0\% (7/7) & \cellcolor{BrickRed!0} 0.0\% (0/7) \\\hline
Ensighten & 7 & \cellcolor{BrickRed!0} 0.0\% (0/7) & \cellcolor{BrickRed!0} 0.0\% (0/7) & \cellcolor{BrickRed!100} 100.0\% (7/7) & \cellcolor{BrickRed!0} 0.0\% (0/7) \\\hline
SIRDATA & 5 & \cellcolor{BrickRed!0} 0.0\% (0/5) & \cellcolor{BrickRed!0} 0.0\% (0/5) & \cellcolor{BrickRed!0} 0.0\% (0/5) & \cellcolor{BrickRed!0} 0.0\% (0/5) \\\hline
Chandago & 5 & \cellcolor{BrickRed!0} 0.0\% (0/5) & \cellcolor{BrickRed!0} 0.0\% (0/5) & \cellcolor{BrickRed!0} 0.0\% (0/5) & \cellcolor{BrickRed!0} 0.0\% (0/5) \\\hline
\textbf{incorrect CMP ID} & 9 & \cellcolor{BrickRed!11} 11.1\% (1/9) & \cellcolor{BrickRed!11} 11.1\% (1/9) & \cellcolor{BrickRed!62} 62.5\% (5/8) & \cellcolor{BrickRed!12} 12.5\% (1/8) \\\hline
\textbf{others} & 73 & \cellcolor{BrickRed!10} 11.0\% (8/73) & \cellcolor{BrickRed!6} 6.8\% (5/73) & \cellcolor{BrickRed!54} 54.4\% (37/68) & \cellcolor{BrickRed!22} 22.1\% (15/68) \\\hline
\textbf{No consent string found} & 40 & \cellcolor{BrickRed!0} 0.0\% (0/40) & \cellcolor{BrickRed!17} 17.5\% (7/40) & \cellcolor{BrickRed!50} 50.0\% (11/22) & \cellcolor{BrickRed!0} 0.0\% (0/22) \\\hline
\setrow{\bfseries} all & 560 & \cellcolor{BrickRed!11} 11.6\% (65/560) & \cellcolor{BrickRed!6} 6.8\% (38/560) & \cellcolor{BrickRed!46} 46.5\% (236/508) & \cellcolor{BrickRed!5} 5.3\% (27/508) \\\hline

    \end{tabular}
    \SHORTENTAB
  \end{table*}
{}

\section{Unusual Cases}
\label{app:unusual_cases}

We list unusual cases encountered during our whole study.

%On one website, the consent checkbox in the preferences is unticked by default. However, if you don't manually open the preferences and click the ``Save'' button, after browsing the website for a while, the checkbox is automatically ticked and your consent is set in the consent string. & \url{http://www.gazeta.pl} & 1 \\\hline

\textbf{Multiple banners at once} — We observed websites displaying two cookie banners, e.g. \url{psicologiaymente.com} or \url{matchendirect.fr}. On these two sites, each banner seems to follow different regulation (pre- or post-GDPR). Our guess is that publishers forgot to remove the oldest ones.

\textbf{Multiple banners on different loads} — We encountered one specific website (\url{kayak.fr}) displaying 4 different banners under different clean browser sessions. These banners provide different characteristics (consent wall or not, existence of a refuse button, access to more specific configurations). Similarly, \url{public.fr} displays 2 different banners when loaded several times with a clean browser: one allowing parameters configuration, and one only providing an accept button.

\textbf{Specifications not followed} — CMPs on some websites do not respect the TCF's specifications at all. On \url{dominos.fr}, the \cmp{} function is defined, but only ever returns an empty JSON object. \url{express.co.uk} sets 24 purposes in the consent string, even though only 5 of them are defined in the TCF and mentioned on the banner's text.

\textbf{Banner not displayed on front page} — On some websites, such as \url{gamepedia.com}, the banner is not displayed on the front page.

%\textbf{Redirection} — On \url{aol.com}, user is redirected to another website (of different domain) to be showed a consent  wall.

\textbf{Redirections upon refusal} — On some websites, such as \texttt{tvguide.co.uk}, users are redirected to the privacy policy page when (and only if) refusing consent. Even more questionably, \texttt{mon-programme-tv.be} redirects users automatically to Wikipedia if they refuse consent. On \texttt{toro.it}, users are redirected on the privacy policy page of another domain, itself containing a new cookie banner.

\textbf{\texttt{consensu.org}'s page} — While the \texttt{consensu.org} domain is used for global consent cookie sharing across publishers and for consent redirection through its subdomains, its main web pages \url{https://consensu.org} and \url{https://www.consensu.org} display a generic park page.
%, which might be considered a way to dissimulate its activity \b{This is a strong claim and my opinion, I don't know how to say this carefully}.

\textbf{Claiming GDPR does not apply} — The URL-based consent passing method specification~\cite{iab-urlbased} includes a parameter called \texttt{gdpr}, used to indicate whether GDPR applies. We observe many queries setting this parameter to 0, claiming that GDPR does not apply. As there are many reasons for the GDPR not to apply to a given script, we cannot decide whether such claims are founded.

\textbf{Extremely tiresome cases} — During our semi-automatic crawl, some banners were extremely hard to configure. For instance, the one on \texttt{rtl.fr} will display 8 purposes separated by hundreds of vendors, making it hard to disable each purpose. Furthermore, each vendor in each list is preticked, making it extremely tiresome to disable each of them.

\textbf{Unticked options ambiguity} — Some banners, e.g. Quantcast's banner on \texttt{sciencesetavenir.fr}, show unticked options %(sliders in this case)
when parameters are opened. However, a consent refusal is set upon saving, while a positive consent is set if user accepts without opening the parameters. This can lure users into thinking they have nothing to do to refuse consent, while they actually have to open the parameters to do so.

\textbf{No choice before acceptance} — Some banners, e.g. Evidon's banner on \texttt{ticketweb.co.uk}, only give the option to define consent preferences \textit{after} user has accepted tracking: the banner only displays an ``accept'' button, and reveals the parameters button once this accept button has been clicked.

\textbf{Hidden parameters} — On some websites, parameters to refuse consent are hidden into a long cookie policy document linked by the cookie banner. For instance, on \texttt{liberation.fr}, the link to open these parameters is hidden in the middle of a 12~000-word-long policy document and is visually indistinct from the rest of the text.

\textbf{No implementation} — Some websites display a banner of one of the TCF-affiliated CMPs, but do not implement elements from the specification. For instance, \texttt{dominos.fr} displays a classical OneTrust banner, but does not provide a \cmp{} function nor a \cmplocator{} iframe. We cannot detect these cases in our automatic crawl.

%We observed some behaviours which did not respect the specifications of the TCF, making our analysis more difficult.
%\textbf{Consent-string passing} — The specifications of the URL-based method for full-consent string passing~\cite{iab-urlbased} indicates values for ``URL parameters'', suggesting the use of GET queries to send information. In practice, we saw queries on 680 (51.28\%) websites using the GET method and on 330 (24.89\%) using the POST method. Moreover, parameters are sometimes embedded. \b{I need an example}

%\textbf{Consent redirection} — Similarly, we only observed a few queries respecting the CMP consent redirecting scheme~\cite{iab-urlbased} used to share consent string across publishers. While the specification requires the existence of a \texttt{redirect} parameter, we only observe different methods. For instance, on \texttt{barnesandnoble.com}, the parameter is called \texttt{rdct\_url}. %, as shown in Figure~\ref{fig:redirection_barnesandnoble}.
%We only observe one redirector URL respecting the specification, \texttt{sddan.consensu.org}, owned by the CMP called ``SIRDATA''. Strangely, this URL is called from websites having different CMPs.

%\textbf{Wrong response format} — On 2 (0.15\%) websites, the CMP does not respect the format of response to queries specified in the TCF (uses \texttt{VendorConsentData} instead of \texttt{VendorConsents}). Ex: \texttt{healthline.com}.

\textbf{Wrong CMP id} — We observe the following incorrect CMP IDs in consent strings: 1, 0 and 4095 (resp. 155, 45 and 3 websites). As of September 2\textsuperscript{nd} 2019, identifiers in IAB Europe's public CMP list~\cite{iab-cmplist} range from 2 to 265. IAB Europe stated that CMP ID 1 is incorrect and should not be used~\cite{iab-cmpid1}, which indicates that this is clearly a violation of the TCF. While some CMPs always return a consent string containing an invalid CMP ID, some CMPs only do so before users give their consent, e.g. Conversant Europe on \texttt{inc.com}. %Ex: \texttt{heavy.com}

%\textbf{Multiple consent strings} — In our automatic crawl, we catch more than one consent string on 141 (10.63 \%) websites. 55 (4.15 \%) websites have consent strings indicating different CMP identifiers

%\textbf{Multiple CMP ids} — In both crawls, 48 (3.36 \%) websites have consent strings indicating different CMP identifiers. 37 of them have consent strings indicating different valid identifiers.

%\textbf{Updated shared cookie keeps the original CMP} — When updating the shared cookie's consent string, some websites set a new consent string with the CMP indicated in the previous consent string. For instance, if \texttt{letudiant.fr} sets a consent string of the SIRDATA CMP in the shared cookie, the website \texttt{glamour.es} sets a consent string with the same CMP, even though the CMP present on this website is OneTrust.

\textbf{Broken banner} – We observe banners on which either refusing or accepting consent is not possible due to a bug on 6 websites. %(1.07 \% of websites in the semi-automatic crawl).
Ex: \texttt{olympia.ie}

%\textbf{No banner} 8 (1.43 \%)

\textbf{Consent to nonexistent vendors} — Some CMPs set consent for nonexistent vendors in the consent string. For instance, the CMP on \texttt{mycanal.fr} sets vendor IDs from 1 to 2000, even though vendor identifiers go up to a maximum of 670 in the GVL (as of September 2019). We observe this issue on 114 (20\%) websites in the semi-automatic crawl. % (20\% of websites).

\textbf{HTTP only} — 95 (7\%) TCF-websites only provide an HTTP access. It is worrisome that websites using tracking technologies do so on an unencrypted connection.

\textbf{Unusual consent verification} — While monitoring consent verification made by third parties (using browser extensions to override the \cmp{} function to catch direct calls, monitor postMessages, GET and POST requests), we observe third parties unregistered in the TCF doing so.
We detect if third parties are trackers using the Disconnect list. We observe at least one tracker unregistered in the TCF querying the CMP to obtain consent in 44\% of websites, and at least one third-party unregistered in the TCF querying the CMP to obtain consent in 55\% of websites.
It is unclear why vendors would verify consent if they're not registered to the framework.

%\begin{figure}[t]
%  \center
%  \includegraphics[width=0.47\textwidth]{figures/redirection_barnesandnoble}
%  \caption{Request using the CMP consent redirection scheme on \texttt{barneandnobles.com}, as seen in Chromium's Developer Tools.}
%  \label{fig:redirection_barnesandnoble}
%\end{figure}

%\input{vendors_using_features}

\end{document}